\documentclass[twocolumn,tighten]{aastex62}

\shortauthors{French $\&$ Wakker}
\usepackage{graphicx}
\usepackage{subfigure}
\usepackage{amsmath}

\usepackage{breakcites}
\usepackage{microtype}

\newcommand{\degrees}{\ensuremath{^{\circ}}}

\newcommand{\Lstar}{$L^*$}
\newcommand{\kms}{$\rm km\, s^{-1}$}

\newcommand{\HI}{\mbox{H\,{\sc i}} }

\newcommand{\II}{\,{\sc ii}}

\graphicspath{{figures//}}

\submitjournal{ApJ}
\accepted{May 27, 2020}
\begin{document}


\title{Evidence for a Rotational Component in the Circumgalactic Medium of Nearby Galaxies\footnote{Based on observations made with the Southern African Large Telescope (SALT).}}

\author[0000-0003-3681-0016]{David M. French}
\affil{Space Telescope Science Institute, 3700 San Martin Drive, Baltimore, MD 21218}
\affil{Department of Astronomy, University of Wisconsin, Madison, WI 53706, USA}

\author[0000-0002-0507-7096]{Bart P. Wakker}
\affil{Department of Astronomy, University of Wisconsin, Madison, WI 53706, USA}


\cleardoublepage

\begin{abstract}

We present results of a study comparing the relative velocity of $\rm Ly\alpha$ absorbers to the rotation velocity of nearby galaxy disks in the local universe ($z \leq 0.03$). We have obtained rotation curves via long-slit spectroscopy of eight galaxies with the Southern African Large Telescope, and combine this dataset with an additional 16 galaxies with data from the literature. Each galaxy appears within $3R_{\rm vir}$ of a QSO sightline with archival Cosmic Origin Spectrograph (COS) spectra. We study the velocity orientation of absorbers with respect to nearby galaxy's rotation, and compare with results from both the \cite{steidel2002} monolithic halo model and a new cylindrical Navarro-Frenk-White galaxy halo model to interpret these data in the context of probing 3D galaxy halos via 1D QSO absorption-line spectroscopy. Relative to these models we find that up to $59\pm5\%$ of $\rm Ly\alpha$ absorbers have velocities consistent with co-rotation. We find the $\rm Ly\alpha$ co-rotation fraction to decrease with galaxy luminosity (\Lstar) and impact parameter in a model-independent fashion. We report that both anti-rotating absorbers and those found near luminous galaxies ($L \gtrsim 0.5$\Lstar) mostly have low Doppler $b$-parameters ($b \lesssim 50$ \kms). Absorbers consistent with co-rotation show a wide range of Doppler $b$-parameters. Finally, we find a strong anticorrelation between co-rotation fraction and galaxy inclination, which is at odds with recent metal-line kinematic studies and suggests the kinematic and geometric distribution of the circumgalactic medium is complex and multiphase.

\end{abstract}

\keywords{galaxies:intergalactic medium, galaxies:evolution, galaxies:halos, quasars: absorption lines}

\section{Introduction}
The current lambda cold-dark-matter ($\Lambda$CDM) cosmological model describes galaxies forming hierarchically out of over-densities in the underlying dark matter distribution. As the surrounding intergalactic medium (IGM) is funneled toward a growing galaxy, simulations predict the angular momentum of the inflowing gas is redistributed onto the disk and seeds the overall rotation of the galaxy (e.g., \citealt{chen2003, sharma2005, brook2011, kimm2011, pichon2011, stewart2011a, stewart2013, ho2019}). As this infalling gas is responsible for birthing and continuing to feed the galaxies throughout their lifetimes, it is expected that the extended gaseous halos should rotate in the same sense as both the galactic disks and dark matter halos.  

In this $\Lambda$CDM picture, accretion falls broadly into two types. In the so-called ``hot-mode", gas shock-heats at the virial radius as it encounters the galaxy halo. The inner, denser region of this hot gaseous halo then rains down onto the disk as it radiatively cools (e.g., \citealt{fillmore1984, bertschinger1985, danovich2012, shen2013, stevens_2017}). However, most gas arrives cold ($T\sim 10^{4}$ K) from the IGM, and the proposed radiative shock is unstable to cooling. Thus this hot-halo scenario may not actually be created \citep{birnboim2003, keres2005, ocvirk2008, brooks2009, dekel2009}.



In contrast, as part of  the alternative ``cold-mode" accretion model, filaments of gas from the IGM would merge smoothly with the disk, thus converting a significant fraction of their infall velocity to rotational velocity of the galaxy \citep{keres2005, keres2009a, stewart2017, el-badry_2018, ho2019}. Alternatively, if these cold filaments break into cloudlets entrained in the hotter ambient halo they may still accrete and carry angular momentum \citep{keres2009b, wetzel2015, oppenheimer2018, melso2019}.



This cold-mode of accretion likely dominates the global growth of all but the most massive halos at high redshifts ($z \gtrsim 3$), and the growth of lower-mass ($M_{\rm halo} \leq 5 \times 10^{11} ~M_{\rm *}$) objects at late times \citep{dekel2006, vandevoort2011}. Furthermore, cosmological SPH simulations such as those by \cite{stewart2011b, stewart2013} and \cite{ho2019} suggest that halo gas should co-rotate with disk gas out to at least 100 kpc, and that absorption in intervening QSO sightlines should be able to accurately capture this rotation signature. Thus, observing co-rotating gas in the halos of galaxies would provide the most direct evidence of cold-mode accretion.

\begin{deluxetable*}{l l l l l l l c l r}
\vspace{-5pt}
\tablewidth{0pt}
\tabletypesize{\scriptsize}
\tablecaption{SALT Galaxy Observations}
\tablehead{
\colhead{Galaxy}	&  \colhead{R.A.}	&  \colhead{Decl.}  	&  \colhead{$v_{\rm sys}$}& \colhead{Published $v_{\rm sys}$} & \colhead{Type}	&  \colhead{$v_{\rm obs}$}	& \colhead{$v_{\rm rot} = v_{\rm obs} / \sin(\emph{i})$}	& \colhead{Obs. Date} & \colhead{$T_{\rm exp}$}  \\
			  	&          			&  			 	& \colhead{(\kms)}  				& \colhead{(\kms)}  		     	   &					&  \colhead{(\kms)}  		& \colhead{(\kms)}					&					& \colhead{(ks)} \\
\colhead{(1)}  	  	& \colhead{(2)}    	&  \colhead{(3)}   	& \colhead{(4)}  				& \colhead{(5)}    		     	   &	\colhead{(6)}  		& \colhead{(7)}    		&\colhead{(8)}  						& \colhead{(9)}  		& \colhead{(10)}   }
\startdata
 CGCG039-137 	& 11 21 27.0		& +03 26 41.7		& $6918 \pm24$				&	$6902 \pm 52^{a}$	& Scd			& $132 \pm 16$		& $143 \pm 25$			& 05 11 2016		& 700	\\ 
 ESO343-G014	 	& 21 37 45.2		& $-$38 29 33.2	& $9139 \pm32$				&	$9162 \pm 45^{b}$	& S				& $203 \pm 32$		& $203 \pm 32$			& 05 16 2016		& 1000	\\ 
 IC 5325		 	& 23 28 43.4		& $-$41 20 00.5	& $1512 \pm8$					&	$1503 \pm 7^{c}$	& SAB(rs)bc		& \phantom{s}$53 \pm 5$	& $125 \pm 27$			& 05 17 2016		& 600	\\ 
 MCG-03-58-009	& 22 53 40.9		& $-$17 28 44.0	& $9015 \pm19$				&	$9030 \pm 10^{d}$	& Sc				& $150 \pm 11$			& $171 \pm 22$			& 05 16 2016		& 1200	\\ 
 NGC 3633	 	& 11 20 26.2		& +03 35 08.2		& $2587 \pm7$					&	$2600 \pm 2^{f}$	& SAa			& $149 \pm 6$			& $157 \pm 9$				& 05 11 2016		& 1200	\\ 
 NGC 4939		& 13 04 14.4		& $-$10 20 22.6	& $3093 \pm33$				&	$3110 \pm 4^{e}$	& SA(s)bc			& $204 \pm 23$		& $234 \pm 39$			& 05 14 2016		& 500	\\ 
 NGC 5786	 	& 14 58 56.3		& $-$42 00 48.1	& $2975 \pm22$				&	$2998 \pm 5^{h}$	& SAB(s)bc		& $156 \pm 14$		& $172 \pm 25$			& 05 11 2016		& 250	\\ 
 UGC 09760	 	& 15 12 02.4		& +01 41 55.5		& $2094 \pm16$				&	$2023 \pm 2^{i}$	& Sd				& \phantom{s}$46 \pm 10$ & \phantom{s}$46 \pm 16$	& 05 11 2016		& 500	\\
 \hline
\enddata
\vspace{-5pt}
\tablecomments{SALT targeted galaxies. Columns are as follows: 1) the galaxy name, 2) and 3) R.A., Decl., respectively, in J2000, 4) measured galaxy systemic velocity, 5) published galaxy systemic velocity, 6) morphological type (RC3), 7) observed rotation velocity, 8) inclination-corrected rotation velocity, 9) observation date, and 10) exposure time. All observations used the RSS PG2300 grating.}
\tablerefs{a. \cite{sdssDR3}; b. \cite{6dFDR3}; c. \cite{RC3}; d. \cite{mathewson1996}; e. \cite{koribalski2004}; f. \cite{lu1993}; g. \cite{grogin1998}; h. \cite{dinella1996}; i, \cite{giovanelli1997}}
 \label{salt_targets}
 \vspace{-20pt}
\end{deluxetable*}

Some observational evidence of corotating halo gas has been obtained at higher redshifts. In pioneering studies focusing on the Mg\II~absorber kinematics and their connection with neighboring galaxies, \cite{charlton1998} and \cite{steidel2002} (and later \citealt{kacprzak2010, bouche_2013, bouche_2016, ho2017, martin_2019}, see also \citealt{kacprzak2017} for a review) found tantalizing evidence that a significant fraction of Mg\II~absorbers have velocities that can be explained by an extended gaseous disk. However, as noted by \cite{steidel2002} among others, a simple extended disk model is insufficient to explain the observed bulk motion implied by their sample of 5 Mg\II~absorber-galaxy systems, and a rotating \emph{halo} may be a better model.


Approaching the question from a different angle, \cite{bowen2016} probed the halo of a single galaxy, NGC 1097, with four nearby QSO sightlines at impact parameters of 48 - 165 kpc, and suggested that an extended, slowly rotating disk with additional inflowing IGM material best matches observations. Additionally, \cite{diamond-stanic2016} detect co-rotating H$\alpha$ emission and Mg\,{\sc ii} and Fe\,{\sc ii} absorption toward a Milky Way-like galaxy at $z=0.413$.

Additionally, the picture may have changed since $z \sim 0.5$, the epoch 5 Gyr ago that most of these Mg\II~systems are probing. By $z \sim 0$ simulations (e.g., \citealt{keres2005, keres2009a, stewart2017}) predict a drop-off in cold-mode accretion and a decrease in the density of IGM filaments. Observational confirmation has been even more inconclusive in this low-redshift regime, where metal-poor IGM inflows are best traced via the Ly$\alpha$ absorption line. In the largest such study, involving Ly$\rm \alpha$ absorber-galaxy kinematics, \cite{cote2005} probed the halos of nine galaxies using \emph{Hubble Space Telescope} (\emph{HST}) observed background QSOs, and found that large warps would be needed to explain the velocity of \HI~absorbers by an extended rotating disk. Additionally, \cite{wakker2009} compiled a sample of 76 sightlines, which included only four galaxy-QSO systems for which the galaxy's rotation curve was known from the literature, and found that only one-fourth of Ly$\alpha$ absorbers appeared to co-rotate with the associated galaxy disk. Similarly, \cite{kacprzak2011_kinematics} claimed a reduction in Mg\II~co-rotation around $\rm \sim $\Lstar~ galaxies between $z\sim 0.5$ and $z\sim 0.1$. The need remains for a larger-scale, low-redshift study specifically targeting the Ly$\alpha$ absorbers that can directly trace cool, metal-poor IGM inflows.

This current work aims to address many of these observational challenges by (1) significantly increasing the sample size of galaxy-absorber systems, and (2) implementing a 3D rotating halo model to aid in the interpretative challenge that is inherent in a 1D sightline probing a 3D structure. To significantly improve observational statistics in this low-redshift regime, we have obtained rotation curves from the Southern African Large Telescope (SALT) for eight nearby spiral galaxies which are located within $3R_{\rm vir}$ of a background QSO observed by the Cosmic Origins Spectrograph (COS) on board \textit{HST}. A literature search yielded an additional 16 galaxies with published rotation curves and known orientations. Each of these is probed by at least one QSO within $3R_{\rm vir}$.

In Section \ref{data} we describe the selection and reduction of both SALT and COS spectra. In Section \ref{model} we present the rotating halo model we have developed to aid in the interpretation of our observations. In Section \ref{discussion} we discuss the overall results of this exercise and present a physically-motivated interpretation of these results. See Section \ref{summary} for a summary of our results and conclusions. Each galaxy-QSO system is discussed in detail in Appendices \ref{SALT_sample} (SALT-observed galaxies) and \ref{ancillary_data} (galaxies from the literature).

\section{Data and Analysis} \label{data}


\subsection{SALT Data}
Our sample contains eight galaxies observed with the SALT Robert Stobie Spectrograph (RSS) in long-slit mode \citep{burgh2003, kobulnicky2003, buckley2006, odonoghue2006}. Our selection of these galaxies and associated QSO targets included several steps. First we identified galaxies in the NASA Extragalactic Database (NED) that are visible to SALT, within the redshift window of $z \leq 0.33$ ($cz \leq 10,000$ \kms), and within $3 R_{\rm vir}$ of QSOs with existing COS G130M spectra available. From these we selected those with angular sizes less than 6' to enable easy sky subtraction using the outer edges of the slit and thus avoiding additional, off-target exposures, and surface brightnesses sufficient to keep exposure times below $\sim 1300 s$. This resulted in a pool of 48 galaxies, which were submitted to the SALT observing queue with the expectation that SALT would observe as many as possible within our awarded time. We obtained data for 14 galaxies, but two proved to be unusable due to issues with spectral identification and low signal-to-noise. 

Finally, we applied the QSO-galaxy matching scheme outlined in \cite{french2017} to exclude systems for which multiple galaxies could reasonably be matched with a particular absorption line. In short, for each QSO-galaxy system we calculate the likelihood value $\mathcal{L} = e^{-(\rho/R_{\rm vir})^2} e^{-(\Delta v / 200)^2}$, where $\rho$ is the impact parameter, $R_{\rm vir}$ the galaxy virial radius, and $\Delta v$ the difference between the absorption and galaxy system velocities in \kms. We require that $\mathcal{L} \ge 0.01$ for all systems, and for no other nearby galaxies to be within a factor of 5$\times \mathcal{L}$. This requires that the QSO be within approximately $3R_{vir}$ of the galaxy, and be obviously closer than any other neighboring galaxies. After a careful inspection using this scheme we chose to remove an additional four galaxies from the SALT sample. 


All SALT galaxy spectra were reduced and extracted using the standard PySALT reduction package \citep{crawford2010}\footnote{http://pysalt.salt.ac.za/}, which includes procedures to prepare the data, correct for gain, cross-talk, bias, and overscan, and finally mosaic the images from the three CCDs. Next, we rectify the images with wavelength solutions found via Ne and Ar arc lamp spectra. Finally, we perform a basic sky subtraction using an off-target portion of the spectrum, and extract 5-10 pixel wide 1D strips from the reduced 2D spectrum. 

For each resulting 1D spectrum, we identify the H$\alpha$ emission lines and perform a nonlinear least-squares Voigt profile fit using the Python package LMFIT\footnote{\url{http://cars9.uchicago.edu/software/python/lmfit/contents.html}}. While normally Gaussian profiles are used for fitting emission lines, we found that a Voigt profile resulted in a better fit of the peak velocity (which is the measurement of prime importance for this analysis). The line centroid and 1$\sigma$ standard errors are returned, and these fits are then shifted to rest-velocity based on the galaxy systemic redshift with heliocentric velocity corrections calculated via the IRAF \emph{rvcorrect} procedure. The final rotation velocity is calculated by then applying the inclination correction, $v_{\rm rot} = v / \sin(i)$.

\begin{deluxetable*}{l l l l r r r}
\vspace{-5pt}
\tablewidth{0pt}
\tabletypesize{\scriptsize}
\setlength{\tabcolsep}{0.15in}
\tablecolumns{7}
\tablecaption{Summary of QSO Sample}
\tablehead{
\colhead{Target}  		&  \colhead{Galaxy} 			&  \colhead{R.A.}  		&  \colhead{Decl.}		& \colhead{z} 	&  \colhead{Program} &  \colhead{${T_{\rm exp}}$ (ks)} }
\colnumbers
\startdata
2E1530+1511				&	NGC 5951			&	15 33 14.3		&	+15 01 03.0		&   0.09000	& 14071		& 9348	\\
3C 232					&	NGC 3067			&	09 58 20.9		&	+32 24 02.0		&   0.53060	& 8596		& 44662	\\
CSO295					&	NGC3432			&	10 52 05.6		&	+36 40 40.0		&   0.60900	& 14772		& 1088	\\
FBQSJ0908+3246			&	NGC2770			&	09 08 38.8		&	+32 46 20.0		&   0.25989	& 14240		& 7430	\\
MRC2251-178  			&   MCG-03-58-009  	& 	22 54 05.9  	&	$-$17 34 55.0  	&   0.06609  & 12029	& 5515	\\
MRK 335					&	NGC 7817			&	00 06 19.5		&	+20 12 11.0		&   0.02578	& 11524		&  5122	\\
MRK 771					&	NGC 4529			&	12 32 03.6		&	+20 09 30.0		&   0.06301	& 12569		& 1868	\\
MRK 876					&	NGC 6140			&	16 13 57.2		&	+65 43 11.0		&   0.12900	& 11524		& 12579	\\
PG0804+761				&	UGC 04238		&	08 10 58.7		&	+76 02 43.0		&   0.10200	& 11686		& 5510	\\
PG1259+593				&	UGC 08146		&	13 01 12.9		&	+59 02 07.0		&   0.47780	& 11541		& 9200	\\
PG1302-102  			&   NGC 4939  		&   13 05 33.0  	& $-$10 33 19.0  	&   0.27840 & 12038	    & 5979	\\
QSO1500-4140  			&   NGC 5786  		&   15 03 34.0  	& $-$41 52 23.0  	&   0.33500 & 11659		& 9258	\\
RBS1503					&	NGC 5907			&	15 29 07.5		&	+56 16 07.0		&   0.09900	& 12276		& 1964	\\
RBS1768  				&   ESO343-G014  	&   21 38 49.9  	&	$-$38 28 40.0  	&   0.18299 & 12936		& 6962	\\
RBS2000  				&   IC 5325  		&   23 24 44.7  	&	$-$40 40 49.0  	&   0.17359 & 13448		& 5046	\\
RX\_J1017.5+4702		&   NGC 3198			&	10 17 31.0		&	+47 02 25.0  	&   0.33544 & 13314 	& 8655  \\
RX\_J1054.2+3511		&	NGC 3432			&	10 54 16.2		&	+35 11 24.0		&   0.20300	& 14772		& 533	\\
RX\_J1117.6+5301  		&   NGC 3631  		&   11 17 40.5  	&   +53 01 51.0  	&   0.15871 & 14240  	& 4943  \\
RX\_J1121.2+0326  		&   CGCG039-137, NGC 3633  & 11 21 14.0 	&	+03 25 47.0 	&   0.15200 & 12248		& 2695	\\
RX\_J1236.0+2641		&	NGC 4565			&	12 36 04.0		& 	+26 41 36.0		&   0.20920	& 12248		& 4235	\\
SBS1116+523				&	NGC 3631			&	11 19 47.9		&	+52 05 53.0		&   0.35568	& 14240		& 4949	\\
SDSSJ091052.80+333008.0	&	NGC 2770			&	09 10 52.8		&	+33 30 08.0		&   0.11631	& 14240		& 7442	\\
SDSSJ091127.30+325337.0	&	NGC 2770			&	09 11 27.3		&	+32 53 37.0		&   0.29038	& 14240		& 10028	\\
SDSSJ095914.80+320357.0	&	NGC 3067			&	09 59 14.8		&	+32 03 57.0		&   0.56462	& 12603		& 2273	\\
SDSSJ104335.90+115129.0	&	NGC 3351			&	10 43 35.9		&	+11 05 29.0		&   0.79400	& 14071		& 4736	\\
SDSSJ111443.70+525834.0 & 	NGC 3631  		&   11 14 43.7  	&   +52 58 34.0  	&   0.07921 & 14240  	& 13440   \\
SDSSJ112439.50+113117.0	&	NGC 3666			&	11 24 39.4		&	+11 31 17.0		&   0.14300	& 14071		& 10427	\\
SDSSJ112448.30+531818.0	&	UGC 06446, NGC 3631 &	11 24 48.3		&	+53 18 19.0		&   0.53151	& 14240		& 7920	\\
SDSSJ151237.15+012846.0  & 	UGC 09760  		&   15 12 37.2  	&	+01 28 46.0  	&   0.26625 & 12603		& 7590	\\
TON1009					&	NGC 2770			&	09 09 06.2		&	+32 36 30.0		&   0.81028	& 12603		& 4740	\\
TON1015					&	NGC 2770			&	09 10 37.0		&	+33 29 24.0		&   0.35400	& 14240		& 4774	\\
\hline
\enddata

\tablecomments{QSO targets in our sample. Columns are as follows: 1) the QSO target name, 2) names of associated galaxies, 3 and 4) target R.A and Decl. in J2000, 5) target redshift, 6) HST program number for the target data, and 7) the total integration time if multiple exposures were taken.}
\label{COS_targets}
\vspace{-20pt}
\end{deluxetable*}

Final errors are calculated as a quadrature sum of $1\sigma$ fit errors, systemic redshift error, and inclination uncertainty as follows:



\begin{eqnarray} \label{error_calculation}
	\nonumber
	\sigma^2 = \left( \frac{\partial v_{rot}}{\partial \lambda_{obs}} \right)^2 (\Delta \lambda_{obs})^2 + \left(\frac{\partial v_{rot}}{\partial v_{sys}} \right)^2 (\Delta v_{sys})^2 + \\
	\nonumber
	\left( \frac{\partial v_{rot}}{\partial i} \right)^2 (\Delta i)^2,
\end{eqnarray}

\noindent where $\Delta \lambda_{\rm obs}$, $\Delta v_{\rm sys}$, and $\Delta i$ are the errors in observed line center, galaxy redshift, and inclination, respectively. We determine the inclination error by calculating the standard deviation of the set of all axis ratio values available in NED for each galaxy. This inclination component tends to dominate the error for low-inclination galaxies.

The final physical scale is calculated using the SALT image scale of 0.1267 arcsec $\rm pixel^{-1}$, multiplied by the 4 pixel spatial binning, and converted to physical units using a redshift-independent distance if available, and a Hubble flow estimate if not (corrected for Virgocentric flow following \cite{huchra1982}). We adopt a Hubble constant of $H_0= 71$ \kms~$\rm Mpc^{-1}$ throughout.

Finally, we calculate our approaching and receding velocities via a weighted mean of the outer-half of each rotation curve, with errors calculated as weighted standard errors in the mean. Our final redshifts are calculated by forcing symmetric rotation, such that the outer-half average velocity for each side matches in magnitude. The upper-left panel of Figure \ref{figure:model_fits} shows an example of this; the black points and error bars are the observed rotation measurements, the dark green lines show the average rotation velocity for the outer-half edge of each rotation curve, and the green shading shows the $1\sigma$ error for this average value. Table \ref{salt_targets} summarizes the observations for this final sample. See Appendix \ref{SALT_sample} for rotation curves and finder charts for each observed galaxy.

\subsection{Additional Galaxy Data}
We have augmented our observed sample of galaxies with an additional 16 systems from the literature. While rotation curves for many galaxies have been published, a far smaller sample also included the necessary orientation information for our purposes. Of these, only the 18 included here were also relatively isolated based on our likelihood criteria and within $3R_{\rm vir}$ of a background QSO with sufficient spectral coverage and signal-to-noise. The resulting sample comes from a variety of sources, and we have endeavored to keep the galaxy properties (e.g., \emph{i}, $v_{\rm sys}$) adopted by the original data authors. For a small subset of these we have adopted more modern inclination and/or position angle measurements where appropriate (see Appendix \ref{ancillary_data} for details). We used the plot digitization software WebPlotDigitizer\footnote{WebPlotDigitizer; https://automeris.io/WebPlotDigitizer/} to extract rotation curve data from figures when necessary. Rotation velocity errors are calculated with Eq. \ref{error_calculation} as for the SALT-observed sample.

\subsection{COS Spectra}
The Barbara A. Mikulski Archive for Space Telescopes (MAST) archives yield 31 QSO targets observed by COS which lie within $3R_{\rm vir}$ of our SALT galaxies (see Table \ref{COS_targets}). These targets vary widely in signal-to-noise from approximately 5 to 100 due to our choosing them based only on their proximity to galaxies with known rotation. The reduction procedure for these spectra follows those described by \cite{wakker2015} and \cite{french2017}. In short, spectra are processed with CALCOS v3.0 or higher and are aligned using a cross-correlation, and then shifted to make sure that (a) the velocities of the interstellar lines match the 21 cm \HI profile, and (b) the velocities of the lines in a single absorption system line up properly. Multiple exposures are combined by summing total counts per pixel before converting to flux. The COS instrument is described in detail by \cite{green2012}.


Once reduced we fit each absorption system with Voigt profiles using the VoigtFit package \citep{krogager2018}. The VoigtFit routine first fits a continuum around the line region using third-order or lower Chebyshev polynomials. Instrumental broadening is taken into account by convolving COS line-spread function (LSF) tables for the appropriate central wavelength and detector lifetime position with the fitted Voigt profile model. In all cases we use the minimum number of components to obtain a satisfactory fit (i.e., reduced $\chi ^2 \sim 1$). The resulting fits are shown in Figure \ref{figure:line_fits1} and component velocities $v$, column densities $N$, Doppler widths $b$, and associated errors are reported in Table \ref{models}.

\section{Halo Rotation Model} \label{model}
In order to better understand how the potential 3D rotation of galaxy halo gas is mapped onto a 1D QSO sightline, we have developed a simple halo rotation model. This model is seeded by an observed rotation curve, which is then extrapolated out to a radius and height of $3R_{\rm vir}$ to form a coherently rotating cylindrical halo. For each galaxy-QSO pair we create two rotation models: 1) a cylindrical halo model with rotation velocities which smoothly decline as a function of radius based on a Navarro-Frenk-White (NFW) profile fit \citep{navarro1996, navarro1997} to the rotation curve data, and 2) the thick-disk model developed in \cite{steidel2002}. In Section \ref{discussion} we compare the results of these models, as well as to using the simple on-sky apparent rotation velocity of each galaxy.


For our cylindrical halo model, we first fit an NFW rotation velocity profile to the observed rotation curve. The form of this NFW fit is as follows:

\begin{equation}
	V(R) = V_{200} \left [\frac{\ln(1 + c x) - c x / (1 + c x)}{x [ \ln(1 + c) - c / (1 + c)]} \right]^{\frac{1}{2}},
\end{equation}

\noindent where $x = R / R_{200}$, with $R_{200}$ being the radius at which the density contrast with respect to the critical density of the universe exceeds 200, $c = R_{200} / R_s$, with $R_s$ being the characteristic radius of the halo, and $V_{200}$ being the characteristic velocity at $R_{200}$. For all fits $R_{200}$ is set to each galaxy's virial radius, $R_{\rm vir}$. We have taken this form from \cite{deblok2008}. The best-fit concentration parameters, $c$, are mostly in the expected 10-20 range, but get up to $\sim 50$ for several galaxies. This is likely because we are fitting to a rotation profile containing both dark matter and baryons, which results in more contracted profiles with higher concentrations than dark-matter-only halos. The resulting NFW fits tend to be somewhat poor toward the inner parts of the rotation curve (as has been noted by others, e.g., \citealt{cote2005}). Regardless, we are most interested in achieving a physically motivated, declining velocity profile in the outer halo regions where most of our QSO sightlines are located, for which these fits are certainly adequate.

We estimate the virial radius of each galaxy according to the combined ``halo-matching" and constant mass-to-light ratio technique developed in \cite{stocke2013} (see the green line in their Figure 1). This method uses a halo matching with the CfA \emph{B}-band galaxy luminosity function of \cite{marzke1994} with $\alpha = -1.25$ for the faint $L < 0.2 L^{\**}$ end of the luminosity function, and a constant mass-to-light ratio of $M_{\rm halo}/L_{\rm gal} = 50 M_{\odot}/L_{\odot}$ above this (see \citealt{stocke2013} and references therein for further detail).

 \begin{figure*}[ht!]
        \centering
        \vspace{0pt}

        \includegraphics[width=0.46\linewidth]{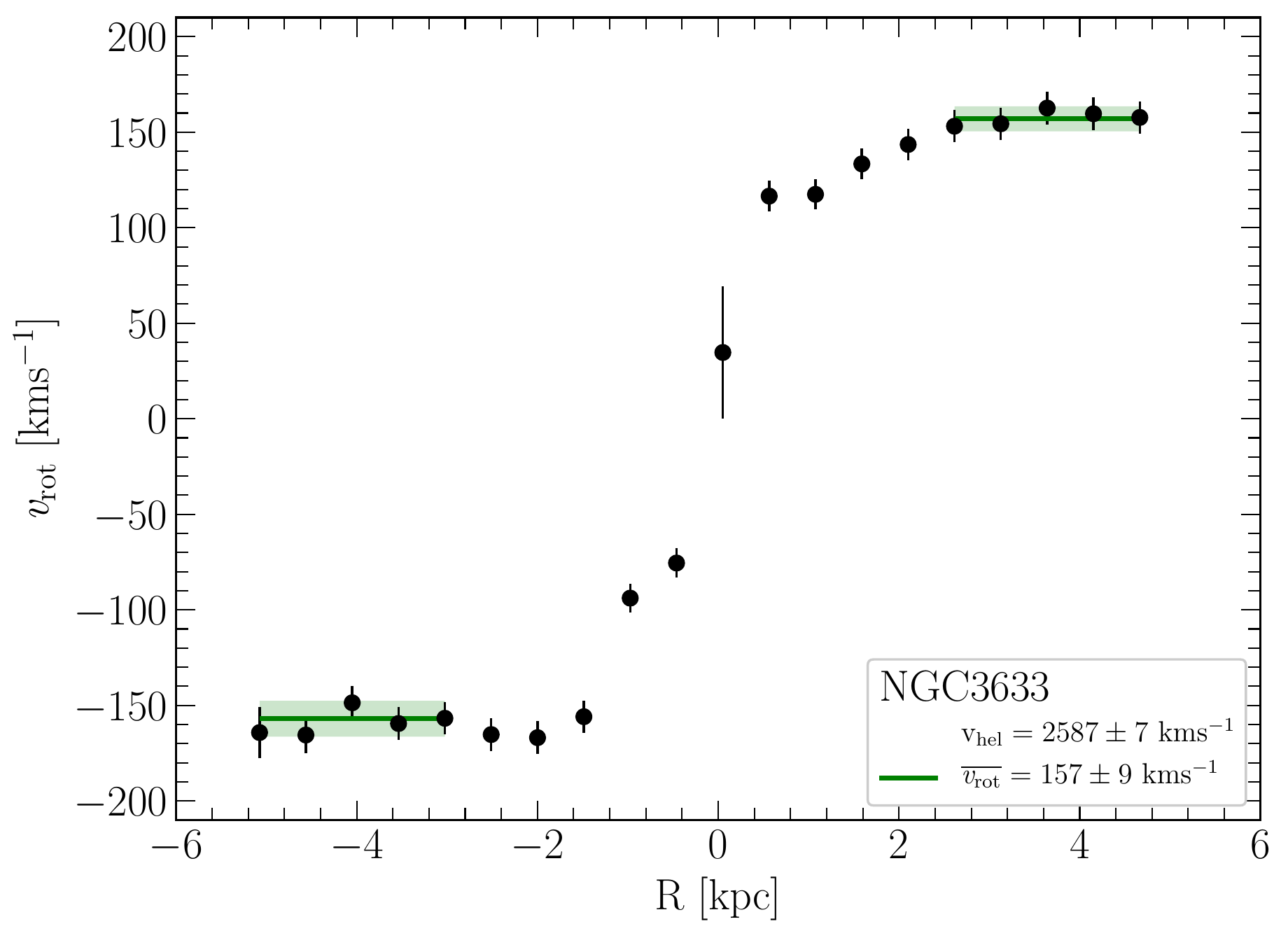}
        \includegraphics[width=0.446\linewidth]{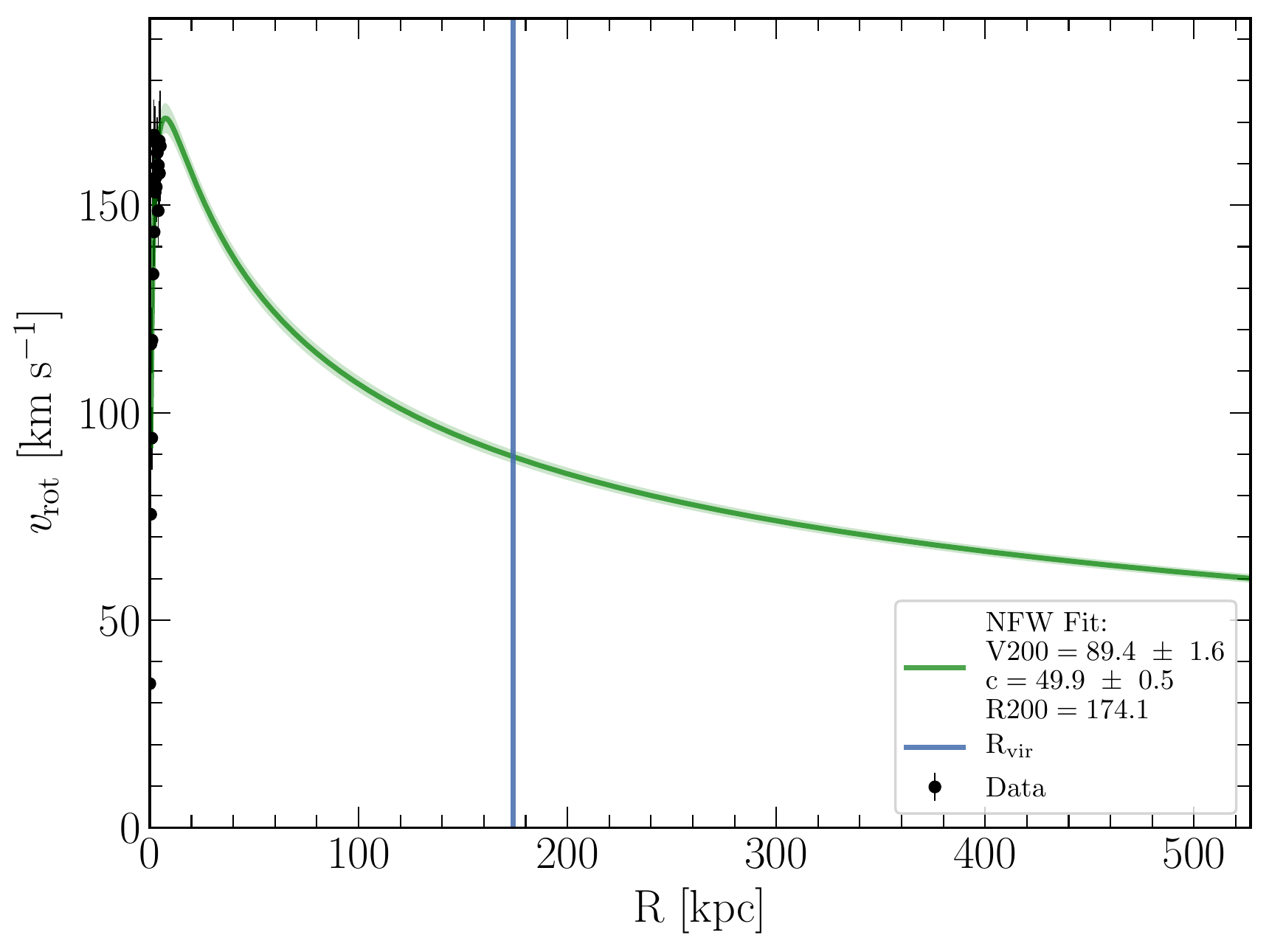}
        
	\vspace{3pt}

        \includegraphics[width=0.455\linewidth]{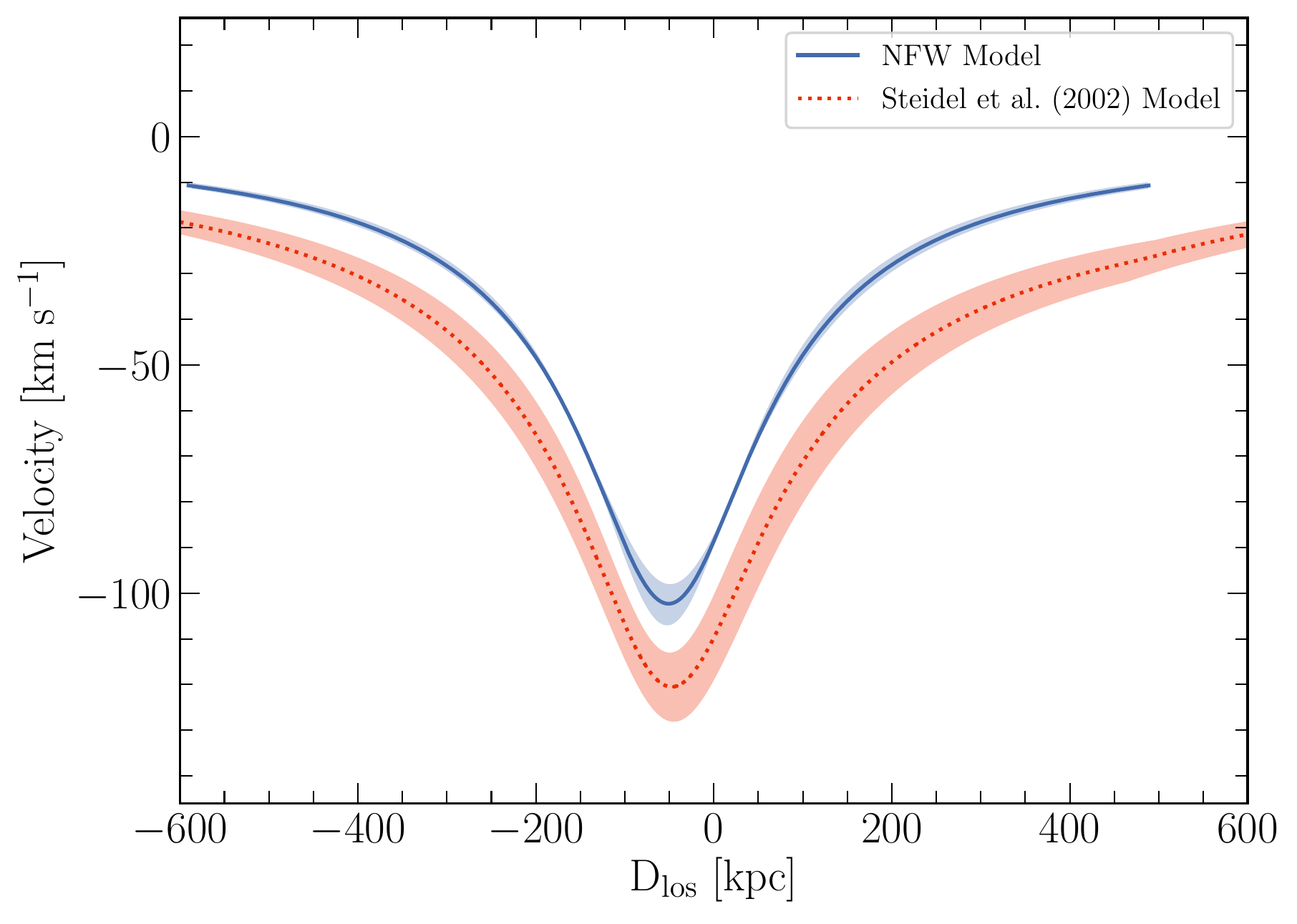}

	\vspace{-4pt}

        \includegraphics[width=0.317\linewidth]{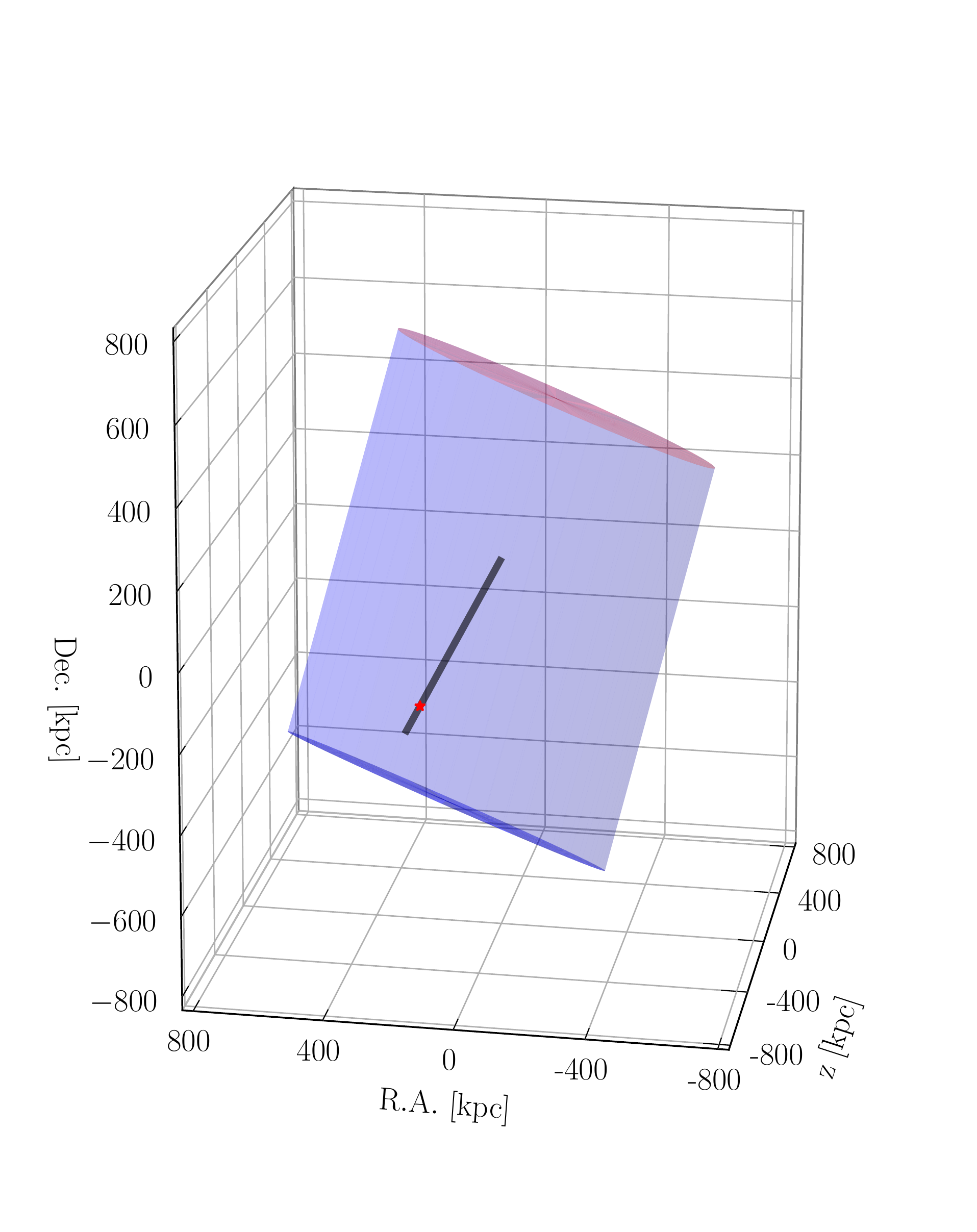}
        \includegraphics[width=0.297\linewidth]{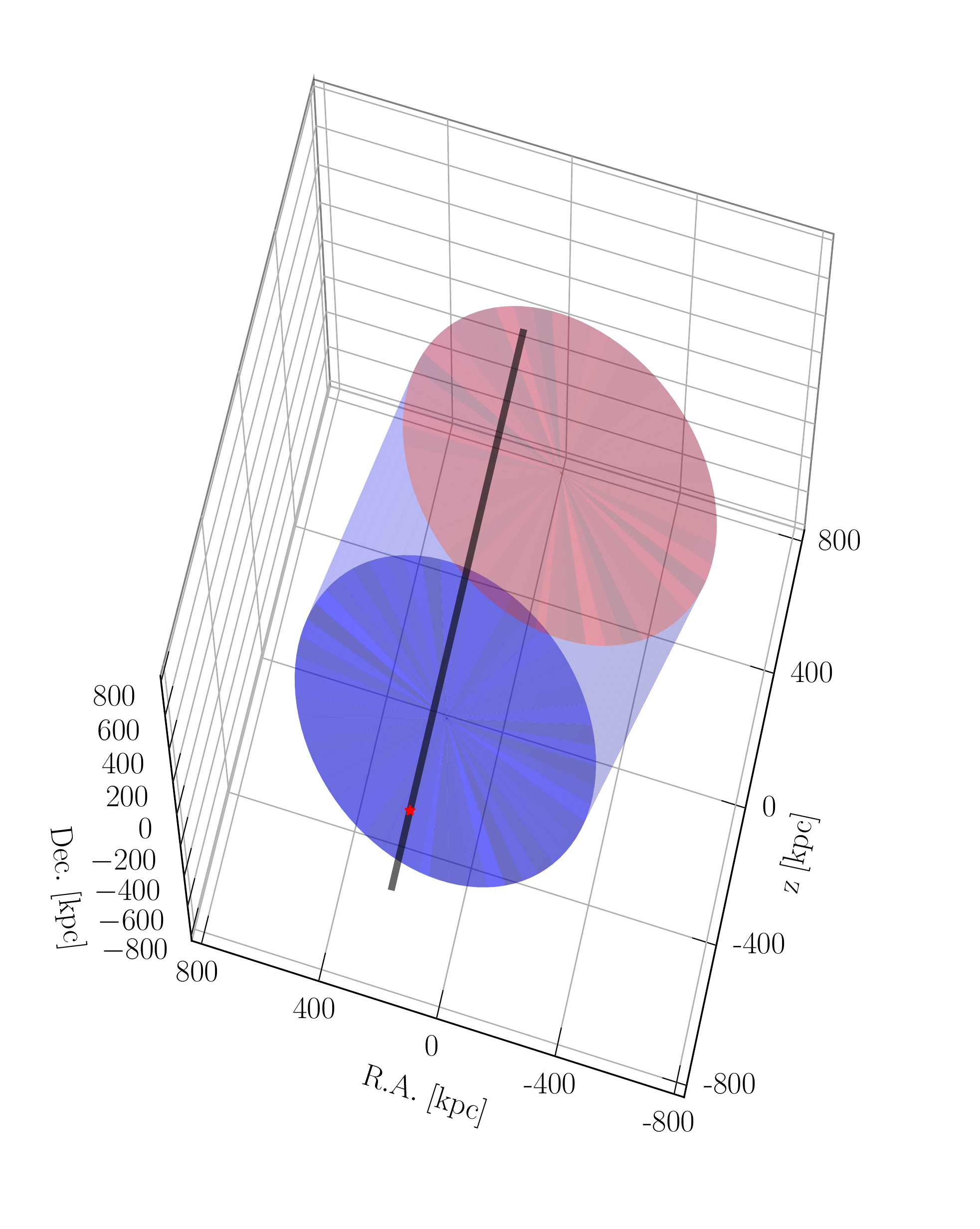}
        \includegraphics[width=0.317\linewidth]{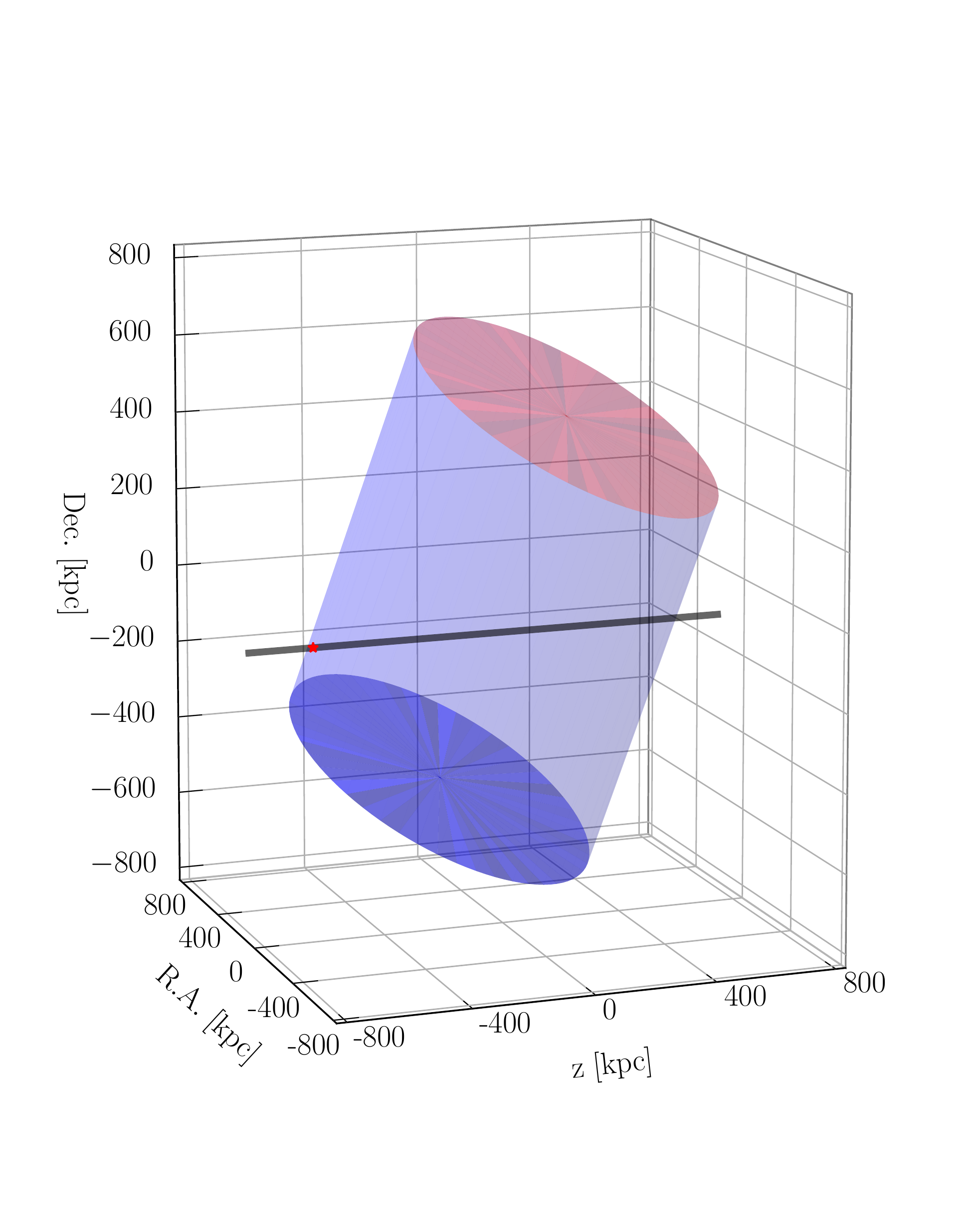}
        \caption{\small{\textbf{Top:} Left: The rotation curve for NGC 3633 is shown in black, with the outer-half mean rotation velocity indicated in green (corresponding to $v_{\rm rot}$). Right:  The observed rotation curve is again shown in black, with an NFW profile fit overlaid in green extending to $3R_{\rm vir}$ ($1 R_{\rm vir}$ is shown by the vertical blue line).
        \textbf{Middle: } The resulting model velocity predictions for the Steidel and our cylindrical NFW models are shown by the dashed-red and solid-blue lines, respectively, as a function of distance along the sightline. 
        \textbf{Bottom:} A 3D example mock-up of our cylindrical NFW halo model showing the orientation and extent of the NGC 3633 model from three different viewing angles. The near-side `top' of the NGC 3633 model halo is shown by the dark-blue oval, with the far side shown in red. The black line shows the location of the sightline toward RX\_J1121.2+0326 as it penetrates the halo, and the red star marks the first intercept point ($\rho = 184$ kpc, Az. = $55^{\circ}$, Inc. = $72^{\circ}$).}}
	\label{figure:model_fits}
        \vspace{6.2pt}
\end{figure*}

\begin{figure*}
\centering
 \includegraphics[width=0.90\linewidth]{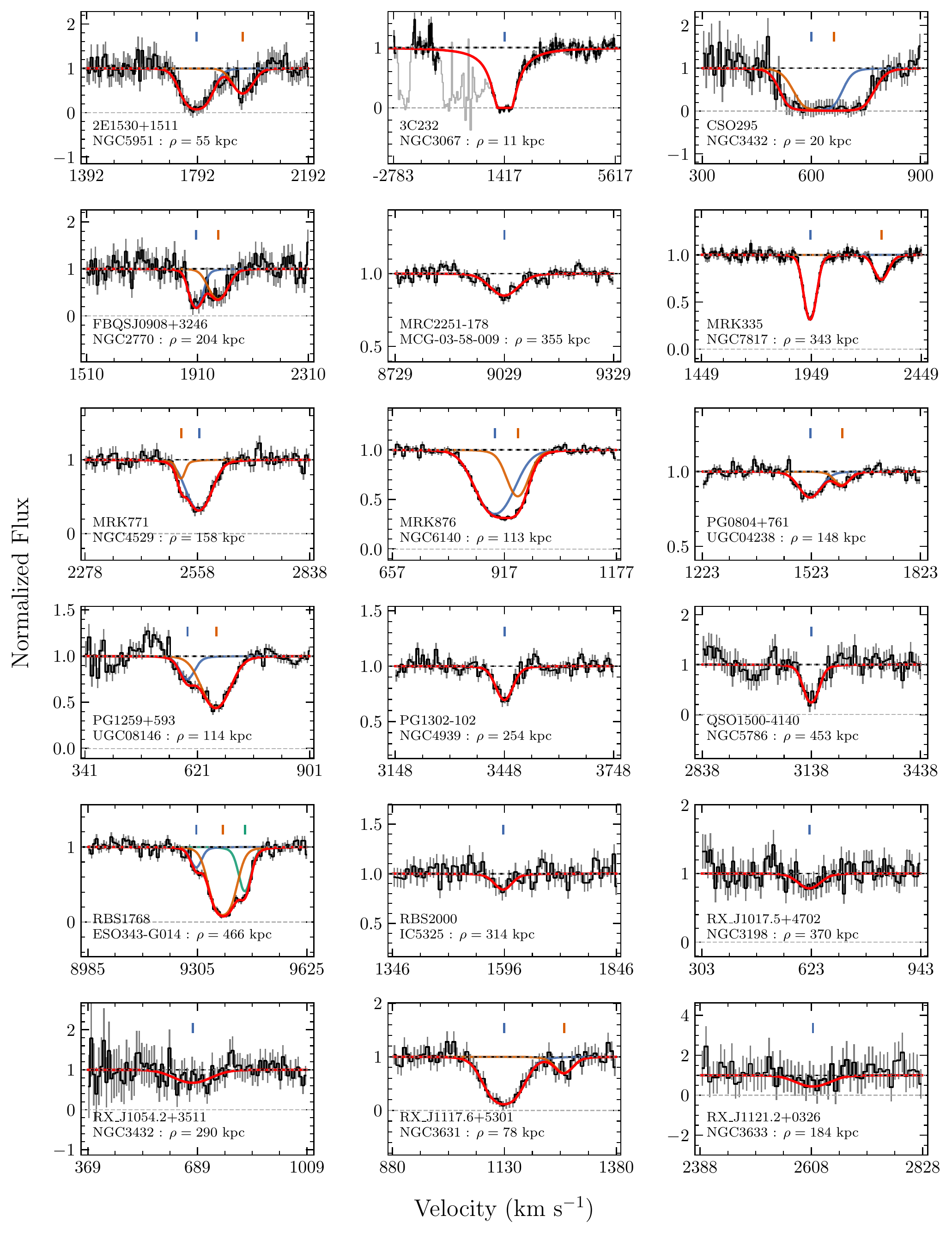}
  \caption{\small{COS spectra of Ly$\alpha$ absorbers. The individual component fits are indicated by colored tick marks and curves, with the heavy red line indicating the composite profile fit. The AGN and associated galaxy names and impact parameters are indicated in the lower-right corner of each profile. The X's above the second and third components toward SDSSJ104335.90+115129.0 indicate interloping absorbers not included in this dataset.}}
  \vspace{5pt}
  \label{figure:line_fits1}
\end{figure*}
\begin{figure*}
\figurenum{\ref{figure:line_fits1}}
\centering
  \includegraphics[width=0.90\linewidth]{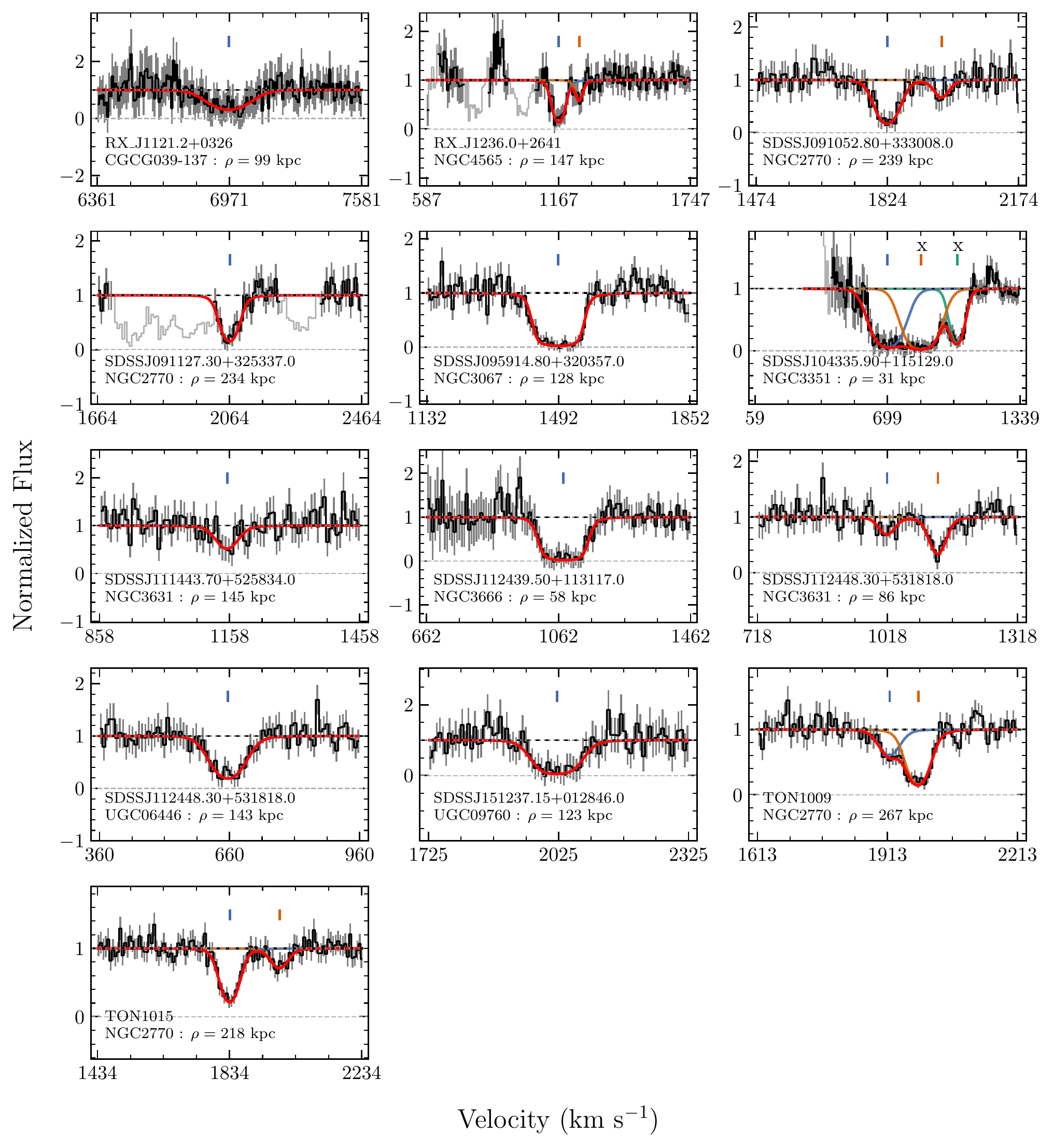}
  \caption{\small{(Continued.)}}
  \vspace{5pt}

%

\end{figure*}
 

Next, we project the NFW fit onto a plane oriented to a mock QSO sightline identically to the input galaxy-QSO pair orientation. By then stacking multiple such rotation planes along the galaxy z-axis direction, we build a cylindrical halo embedded with the rotation curve fit.

Finally, we calculate the projected rotation velocity encountered at each position along the sightline. The result is a function representing the rotation velocity encountered by the sightline as a function of distance along it. Hence, each model produces the velocity a co-rotating absorber would project onto the spectrum as a function of distance from the center of the galaxy (scaled according to $R_{\rm vir}$). 

We also compute the co-rotation velocity ranges produced by the thick-disk model developed by \cite{steidel2002}, which represents a monolithic rotating disk with a flat rotation curve. This model produces a line-of-sight velocity $v_{\rm los}$ as a function of impact parameter $\rho$, galaxy inclination $i$, QSO-galaxy azimuth angle $\Phi$, and maximum projected rotation velocity $v_{\rm max}$ as follows:

\begin{eqnarray}
\nonumber
v_{\rm los} = \frac{- v_{\rm max}}{\sqrt{1 + \left (\frac{y}{p} \right)^2}} e^{-\frac{|y-y_0|}{h_v \tan i}} ~\rm with,\\
y_0 = \frac{\rho \sin \phi}{\cos i}~and~p = \rho \cos \phi.
\end{eqnarray}

Here $h_v$ represents the scale height for the velocity lag in the $z$-direction. We have assumed a thick disk for maximum disk/halo rotation for all galaxies ($h_v$ = 1000 kpc; see, e.g., \citealt{kacprzak2019a}). This effectively maximizes the potential for rotation above and below the disk, resulting in a cylindrical halo of complimentary shape to our NFW model but with constant rotation velocity at all radii (see Fig. 6 in \cite{steidel2002} for a diagram of the resulting halo).

We calculate $1 \sigma$ model errors via a bootstrapping method which reruns the NFW fitting and modeling while resampling the rotation curve data points together with a $3\degrees$ position angle uncertainty. The final model output is the minimum and maximum velocity an absorber would need in order to be consistent with co-rotation within these 1 $\sigma$ error regions.

Figures \ref{figure:model_fits} illustrates an example model for the SALT-observed galaxy NGC 3633, with our observed rotation curve, the NFW fit, and the resulting NFW and Steidel model output velocity distributions from left to right on top, and a 3D halo mock-up from three different viewing angles on the bottom. In most cases, and as seen in this example, the two model outputs have similar \emph{shape}, but vary in maximum predicted velocity.

\begin{longrotatetable}
\begin{deluxetable*}{l l l r r r r r r r r r r r r}
\tabletypesize{\tiny}
\tablewidth{0pt}
\tablecaption{Halo Model Results and Ly$\alpha$ Absorption Properties\label{models}}
\tablehead{
\colhead{$\#$} &\colhead{Galaxy} & \colhead{Target} & \colhead{$\rho$ } & \colhead{Az.} & \colhead{Inc.} & \colhead{$R_{vir}$} &
\colhead{$L$} & \colhead{$v_{\rm sys}$} & \colhead{$v_{\rm rot}$\tablenotemark{a}} &  \colhead{$\Delta v_{\rm Ly\alpha}$} & \colhead{$b$} & \colhead{$log N$} & \colhead{$v_{Steidel}$\tablenotemark{b}}  & \colhead{$v_{NFW}$\tablenotemark{c}}  \\
& &	&  \colhead{(kpc)} & \colhead{(deg)} & \colhead{(deg)} & \colhead{(kpc)} & \colhead{$L_{\**}$} & \colhead{(\kms)}	     & \colhead{(\kms)} & \colhead{(\kms)} & \colhead{(\kms)} & & \colhead{(\kms)} & \colhead{(\kms)} }
\colnumbers
\startdata
1  &        CGCG039-137  &   RX\_J1121.2+0326  &         99  & 86  & 72 &  155  & 0.6  & $6918 \pm 24$ & $136 \pm 24$  &  $53 \pm 27$  &  $112.2 \pm 17.3$ &  $14.27 \pm 0.06$  &  [1.1, 137.2]  &    [1.6, 210.4]   \\
2  &        ESO343-G014  &   RBS1768  &                 466  & 75  & 90 &  187  & 1.1  & $9139 \pm 32$ & $-203 \pm 32$ & $166 \pm 32$  &   $14.4 \pm 5.9$  &    $13.05 \pm 0.08$  &  [-0.1, -0.0]  &    [-158.8, -20.2]   \\
2  &        ESO343-G014  &   RBS1768  &                 466  & 75  & 90 &  187  & 1.1  & $9139 \pm 32$ & $-203 \pm 32$ & $244 \pm 32$  &   $31.6 \pm 2.9$  &    $14.26 \pm 0.03$  &  [-0.1, -0.0]  &    [-158.8, -20.2]   \\
2  &        ESO343-G014  &   RBS1768  &                 466  & 75  & 90 &  187  & 1.1  & $9139 \pm 32$ & $-203 \pm 32$ & $308 \pm 32$  &   $15.3 \pm 3.5$  &    $13.58 \pm 0.06$  &  [-0.1, -0.0]  &    [-158.8, -20.2]   \\
3  &        IC 5325  &        RBS2000  &                 314  & 67  & 25 &  175  & 0.9  & $1512 \pm 8$  &  $-53 \pm 11$ &  $84 \pm 9$  &    $20.9 \pm 9.0$  &    $12.85 \pm 0.1$  &   [-14.6, -1.7]  &   [-30.3, -4.5]   \\
4  &        MCG-03-58-009  & MRC2251-178  &             355  & 74  & 61 &  259  & 2.9  & $9015 \pm 19$ & $150 \pm 19$  &  $14 \pm 19$  &  $48.7 \pm 4.9$  &    $13.08 \pm 0.04$  &  [15.9, 101.6]  &   [8.8, 137.4]   \\
5  &        NGC 2770  &       FBQSJ0908+3246  &          204  & 56  & 80 &  221  & 1.8  & $1948 \pm 4$  &  $-146 \pm 6$  &  $-38 \pm 5$  &  $22.7 \pm 5.0$  &    $13.95 \pm 0.1$  &   [-12.2, -3.8]  &   [-164.9, -14.7]   \\
5  &        NGC 2770  &       FBQSJ0908+3246  &          204  & 56  & 80 &  221  & 1.8  & $1948 \pm 4$  &  $-146 \pm 6$  &  $43 \pm 6$  &   $37.8 \pm 7.1$  &    $13.81 \pm 0.05$  &  [-12.2, -3.8]  &   [-164.9, -14.7]   \\
6  &        NGC 2770  &       TON1009  &                 267  & 38  & 80 &  221  & 1.8  & $1948 \pm 4$  &  $-146 \pm 6$  &  $-35 \pm 7$ &   $26.0 \pm 10.1$  &   $13.38 \pm 0.12$  &  [-23.4, -7.5]  &   [-142.0, -28.2]   \\
6  &        NGC 2770  &       TON1009  &                 267  & 38  & 80 &  221  & 1.8  & $1948 \pm 4$  &  $-146 \pm 6$  &  $32 \pm 5$  &  $24.5 \pm 3.7$  &    $14.06 \pm 0.05$  &  [-23.4, -7.5]  &   [-142.0, -28.2]   \\
7  &        NGC 2770  &       TON1015  &                 218  & 58  & 80 &  221  & 1.8  & $1948 \pm 4$  &  $146 \pm 6$  &   $-114 \pm 4$  &  $28.3 \pm 2.5$  &    $13.96 \pm 0.04$  &  [17.3, 123.8]  &   [14.3, 163.8]   \\
7  &        NGC 2770  &       TON1015  &                 218  & 58  & 80 &  221  & 1.8  & $1948 \pm 4$  &  $146 \pm 6$  &   $36 \pm 6$  &   $29.6 \pm 8.8$  &    $13.24 \pm 0.08$  &  [17.3, 123.8]  &   [14.3, 163.8]   \\
8  &        NGC 2770  &       SDSSJ091052.80+333008.0  & 239  & 63  & 80 &  221  & 1.8  & $1948 \pm 4$  &  $146 \pm 6$  &   $-124 \pm 4$  &  $31.1 \pm 3.8$  &    $14.05 \pm 0.06$  &  [15.2, 120.4]  &   [12.9, 167.7]   \\
8  &        NGC 2770  &       SDSSJ091052.80+333008.0  & 239  & 63  & 80 &  221  & 1.8  & $1948 \pm 4$  &  $146 \pm 6$  &   $21 \pm 6$  &  $20.4 \pm 10.2$  &   $13.22 \pm 0.11$  &  [15.2, 120.4]  &   [12.9, 167.7]   \\
9  &        NGC 2770  &       SDSSJ091127.30+325337.0  & 234  & 33  & 80 &  221  & 1.8  & $1948 \pm 4$  &  $-146 \pm 6$  &  $114 \pm 11$  &  $28.0 \pm 10.0$  &   $14.0 \pm 0.2$  &    [-134.2, -35.8]  & [-141.9, -27.9]   \\
10  &       NGC 3067  &       3C 232  &                    11  & 71  & 71 &  144  & 0.5  & $1465 \pm 5$  &  $135 \pm 9$  &   $-48 \pm 9$  &   $80.8 \pm 6.0$  &    $20.09 \pm 0.02$  &  [-134.4, -0.4]  &  [-136.5, -0.3]   \\
11  &       NGC 3067  &       SDSSJ095914.80+320357.0  & 128  & 40  & 71 &  144  & 0.5  & $1465 \pm 5$  &  $135 \pm 9$  &   $27 \pm 5$  &  $28.4 \pm 10.7$  &   $16.23 \pm 1.43$  &  [14.7, 124.7]  &   [10.2, 86.4]   \\
12  &       NGC 3198  &       RX\_J1017.5+4702  &        370  & 58  & 73 &  217  & 1.7  &  $660 \pm 1$  &   $-145 \pm 5$  &  $-37 \pm 8$  &  $39.2 \pm 15.0$  &   $13.18 \pm 0.12$  &  [-106.0, -30.9]  & [-96.7, -19.0]   \\
13  &       NGC 3351  &       SDSSJ104335.90+115129.0   & 31  & 46  & 42 &  198  & 1.3  &  $778 \pm 4$  &   $-133 \pm 11$  & $-79 \pm 19$  &  $78.6 \pm 14.7$  &   $14.53 \pm 0.12$  &  [-93.5, -2.8]  &   [-128.3, -1.9]   \\
14  &       NGC 3432  &       CSO295  &                   20  & 79  & 90 &  136  & 0.4  &  $616 \pm 4$  &   $119 \pm 8$  &   $-16 \pm 16$  &  $48.1 \pm 12.0$  &   $15.05 \pm 0.37$  &  [0.3, 124.5]  &    [0.3, 141.3]   \\
14  &       NGC 3432  &       CSO295  &                  20  &  79  & 90 &  136  & 0.4  &  $616 \pm 4$  &   $119 \pm 8$  &   $46 \pm 16$  &   $62.1 \pm 9.9$  &    $15.18 \pm 0.32$  &  [0.3, 124.5]  &    [0.3, 141.3]   \\
15  &       NGC 3432  &       RX\_J1054.2+3511  &       290  &  60  & 90 &  136  & 0.4  &  $616 \pm 4$  &   $119 \pm 8$  &   $73 \pm 14$  &  $66.8 \pm 21.7$  &   $13.58 \pm 0.12$  &  [0.0, 0.1]  &      [30.4, 130.7]   \\
16  &       NGC 3631  &       RX\_J1117.6+5301  &        78  &  78  & 17 &  139  & 0.5  &  $1156 \pm 1$ &  $42 \pm 7$  &   $-26 \pm 2$  &  $38.2 \pm 2.6$  &    $14.21 \pm 0.04$  &  [0.4, 11.2]  &     [1.5, 40.2]   \\
16  &       NGC 3631  &       RX\_J1117.6+5301  &        78  &  78  & 17 &  139  & 0.5  &  $1156 \pm 1$  &  $42 \pm 7$  &  $109 \pm 4$  &  $20.9 \pm 9.5$  &    $13.17 \pm 0.1$  &   [0.4, 11.2]  &     [1.5, 40.2]   \\
17  &       NGC 3631  &       SBS1116+523  &             163 &  37  & 17 &  139  & 0.5  &  $1156 \pm 1$  &  $-42 \pm 7$ &   $...$     &       $...$   &  $...$  & [-11.3, -2.3]  &   [-24.0, -7.9]   \\
18  &       NGC 3631  &       SDSSJ111443.70+525834.0  & 145 &  74  & 17 &  139  & 0.5  &  $1156 \pm 1$  &  $42 \pm 7$  &    $2 \pm 5$  & $27.4 \pm 8.7$  &    $13.52 \pm 0.09$  &  [0.7, 9.2]  &      [2.5, 29.5]   \\
19  &       NGC 3631  &       SDSSJ112448.30+531818.0  & 86  &  77  & 17 &  139  & 0.5  &  $1156 \pm 1$  &  $-42 \pm 7$  &   $-138 \pm 5$  &  $18.6 \pm 9.4$  &    $13.18 \pm 0.11$  &  [-10.9, -0.4]  &   [-38.6, -1.8]   \\
19  &       NGC 3631  &       SDSSJ112448.30+531818.0  & 86  &  77  & 17 &  139  & 0.5  &  $1156 \pm 1$  &  $-42 \pm 7$  &   $-21 \pm 3$  &   $17.2 \pm 4.0$  &    $13.7 \pm 0.07$  &   [-10.9, -0.4]  &   [-38.6, -1.8]   \\
20  &       NGC 3633  &       RX\_J1121.2+0326  &       184  &  55  & 72 &  174  & 0.9  &  $2587 \pm 7$  &  $-149 \pm 9$  &  $21 \pm 16$  &   $36.3 \pm 20.1$  &   $13.7 \pm 0.18$  &   [-128.2, -16.1]  & [-107.4, -10.0]   \\
21  &       NGC 3666  &       SDSSJ112439.50+113117.0  & 58  &  86  & 78 &  154  & 0.6  &  $1063 \pm 2$  &  $-124 \pm 6$  &  $-1 \pm 3$  &    $36.0 \pm 8.9$  &    $15.53 \pm 0.67$  &  [-120.0, -0.7]  &  [-129.3, -0.9]   \\
22  &       NGC 4529  &       MRK 771  &                 158  &  26  & 80 &  193  & 1.2  &  $2536 \pm 11$  & $-104 \pm 15$  & $22 \pm 11$  &   $33.0 \pm 3.9$  &    $13.82 \pm 0.04$  &  [-111.2, -17.8]  & [-124.9, -18.4]   \\
22  &       NGC 4529  &       MRK 771  &                 158  &  26  & 80 &  193  & 1.2  &  $2536 \pm 11$  & $-104 \pm 15$  & $-23 \pm 12$  &  $4.0 \pm 12.4$  &    $13.03 \pm 0.49$  &  [-111.2, -17.8]  & [-124.9, -18.4]   \\
23  &       NGC 4565  &       RX\_J1236.0+2641  &       147  &  38  & 86 &  229  & 2.0  &  $1230 \pm 5$  &  $252 \pm 12$  &  $-64 \pm 6$  &   $26.8 \pm 5.5$  &    $14.05 \pm 0.12$  &  [6.3, 14.7]  &     [14.9, 182.2]   \\
23  &       NGC 4565  &       RX\_J1236.0+2641  &       147  &  38  & 86 &  229  & 2.0  &  $1230 \pm 5$  &  $252 \pm 12$  &  $27 \pm 7$  &    $16.6 \pm 10.3$  &   $13.31 \pm 0.14$  &  [6.3, 14.7]  &     [14.9, 182.2]   \\
24  &       NGC 4939  &       PG1302-102  &             254  &  64  & 61 &  320  & 5.5  &  $3093 \pm 33$  & $-205 \pm 34$  & $356 \pm 33$  &  $26.4 \pm 3.6$  &    $13.23 \pm 0.04$  &  [-163.6, -23.3]  & [-162.1, -9.0]   \\
25  &       NGC 5786  &       QSO1500-4140  &           453  &   2  & 65 &  248  & 2.6  &  $2975 \pm 22$  & $156 \pm 23$  &  $163 \pm 22$  &  $18.9 \pm 3.3$  &    $13.85 \pm 0.08$  &  [50.9, 161.2]  &   [33.8, 93.0]   \\
26  &       NGC 5907  &       RBS1503  &                478  &  66  & 90 &  193  & 1.2  &   $667 \pm 3$  &   $-229 \pm 6$  &  $...$      &           $...$   &   $...$  & [-0.1, -0.1]  &    [-138.2, -31.5]   \\
27  &       NGC 5951  &       2E1530+1511  &             55  &  88  & 74 &  203  & 1.4  &  $1780 \pm 1$  &  $127 \pm 7$  &   $12 \pm 3$  &    $48.5 \pm 4.0$  &    $14.36 \pm 0.05$  &  [0.2, 121.4]  &    [0.2, 134.6]   \\
27  &       NGC 5951  &       2E1530+1511  &             55  &  88 &  74 &  203  & 1.4  &  $1780 \pm 1$  &  $127 \pm 7$  &   $179 \pm 4$  &   $39.8 \pm 6.3$  &    $13.73 \pm 0.05$  &  [0.2, 121.4]  &    [0.2, 134.6]   \\
28  &       NGC 6140  &       MRK 876  &                  113 &  18 &  49 &  106  & 0.2  &   $910 \pm 4$  &   $104 \pm 4$  &   $7 \pm 6$  &    $49.2 \pm 4.1$  &    $13.9 \pm 0.06$  &   [9.0, 78.2]  &     [24.1, 103.8]   \\
28  &       NGC 6140  &       MR K876  &                  113 &  18 &  49 &  106  & 0.2  &  $910 \pm 4$  &   $104 \pm 4$  &   $61 \pm 5$  &    $27.4 \pm 4.0$  &    $13.49 \pm 0.15$  &  [9.0, 78.2]  &     [24.1, 103.8]   \\
29  &       NGC 7817  &       MRK 335  &                  343 &  87 &  80 &  168  & 0.8  &  $2309 \pm 4$  &  $-178 \pm 10$  & $-360 \pm 4$  &  $29.3 \pm 0.9$  &    $13.8 \pm 0.01$  &   [-0.7, 0.8]  &     [-151.7, 204.4]   \\
29  &       NGC 7817  &       MRK 335  &                  343 &  87 &  80 &  168  & 0.8  &  $2309 \pm 4$  &  $-178 \pm 10$  & $-36 \pm 4$  &  $43.3 \pm 3.3$  &    $13.32 \pm 0.03$  &  [-0.7, 0.8]  &     [-151.7, 204.4]   \\
30  &       UGC 04238  &      PG0804+761  &              148 &  62 &  75 &  151  & 0.6  &  $1544 \pm 7$  &  $89 \pm 12$  &   $-21 \pm 7$  &  $36.7 \pm 4.3$  &    $13.04 \pm 0.04$  &  [7.0, 85.5]  &     [12.7, 120.5]   \\
30  &       UGC 04238  &      PG0804+761  &              148 &  62 &  75 &  151  & 0.6  &  $1544 \pm 7$  &  $89 \pm 12$  &   $67 \pm 8$  &   $23.4 \pm 7.1$  &    $12.6 \pm 0.08$  &   [7.0, 85.5]  &     [12.7, 120.5]   \\
31  &       UGC 06446  &      SDSSJ112448.30+531818.0  & 143 &  19 &  52 &   98  & 0.1  &   $645 \pm 1$  &   $62 \pm 5$  &    $16 \pm 3$  &  $39.8 \pm 4.0$  &    $14.07 \pm 0.04$  &  [6.0, 50.0]  &     [18.7, 53.4]   \\
32  &       UGC 08146  &      PG1259+593  &              114 &  52 &  78 &  123  & 0.3  &   $670 \pm 1$  &   $80 \pm 3$  &    $-49 \pm 8$ &  $19.6 \pm 8.4$  &    $13.04 \pm 0.14$  &  [2.9, 29.1]  &     [14.9, 101.7]   \\
32  &       UGC 08146  &      PG1259+593  &              114 &  52 &  78 &  123  & 0.3  &   $670 \pm 1$  &   $80 \pm 3$  &    $23 \pm 4$  &  $39.7 \pm 4.8$  &    $13.72 \pm 0.04$  &  [2.9, 29.1]  &     [14.9, 101.7]   \\
33  &       UGC 09760  &      SDSSJ151237.15+012846.0  & 123 &  87 &  90 &  156  & 0.6  &  $2094 \pm 16$  & $-54 \pm 16$  &  $-69 \pm 16$ &  $43.2 \pm 7.1$  &    $14.5 \pm 0.15$  &   [-0.0, 0.0]  &     [-44.9, 42.3]   \\
\enddata
\tablecomments{Non-detections toward SBS1116+523 and RBS1503 are marked with elipses.
\tablenotetext{a}{Galaxy rotation velocity in the direction of the target in column 3.} 
\tablenotetext{b}{Range of heliocentric velocities consistent with co-rotation from the \cite{steidel2002} model} 
\tablenotetext{c}{Range of heliocentric velocities consistent with co-rotation from the NFW halo model}}
\end{deluxetable*}
\end{longrotatetable}

\section{Discussion} \label{discussion}
We present data on 33 galaxy-QSO systems, representing 47 individual Ly$\rm \alpha$ component-galaxy pairs and two non-detections, for which we have galaxy information including kinematics, inclination, size and luminosity. This is the largest sample of its kind to date and provides the best yet opportunity to study the kinematic connection between galaxies and their neutral \HI~halos.

We designate each system as co-rotating or anti-rotating by comparing the absorption velocity difference, $\Delta v = v_{\rm absorber} - v_{\rm galaxy}$, to both the orientation of each galaxy and the model results. For example, the galaxy CGCG039-137 is probed on the receding side by the sightline RX\_J1121.2+0326, with a $\Delta v_{\rm Lya} = 53$ \kms~component detected. This $\Delta v$ lies within the NFW model range ([1.6, 210.4]) as well as the  Steidel ([1.1, 137.2]), and so is marked as "co-rotating" in all cases.

Table \ref{models} summarizes our galaxy-absorber sample and includes the rotation velocity and associated error for each galaxy in Column (7), with the resulting co-rotation velocity ranges predicted from the \cite{steidel2002} and NFW models given in Columns (10) and (11). In order to broadly account for velocity uncertainties we have calculated our model ranges to include the $1\sigma$ rotation velocity errors. We note that the majority of our ``co-rotating" sample fall well within the model ranges, so few would be thrown into uncertainty based on the relative size of included errors.

\begin{figure*}[ht!]
\centering
\vspace{6pt}
 \includegraphics[width=0.484\linewidth]{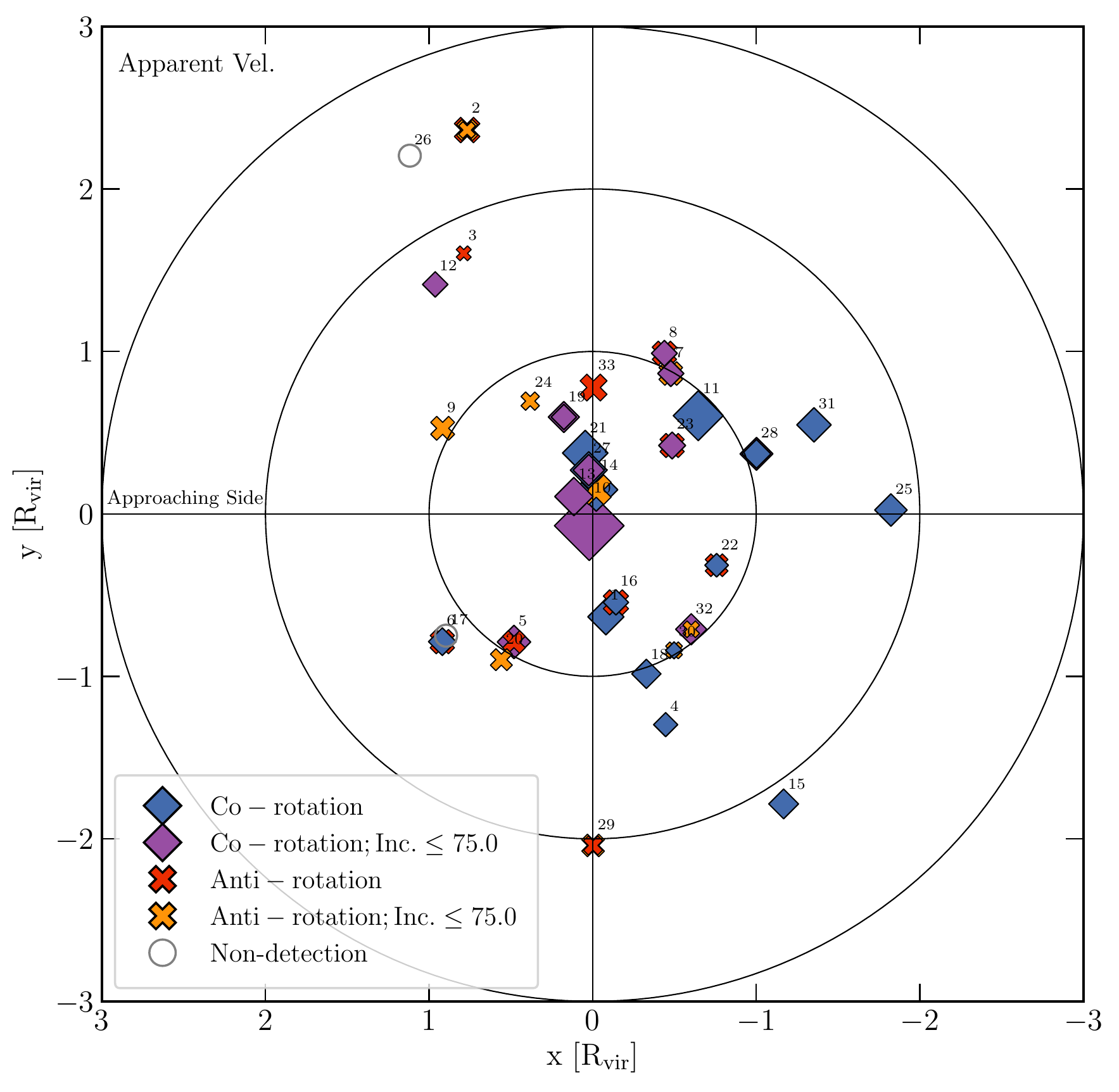}
  \includegraphics[width=0.505\linewidth]{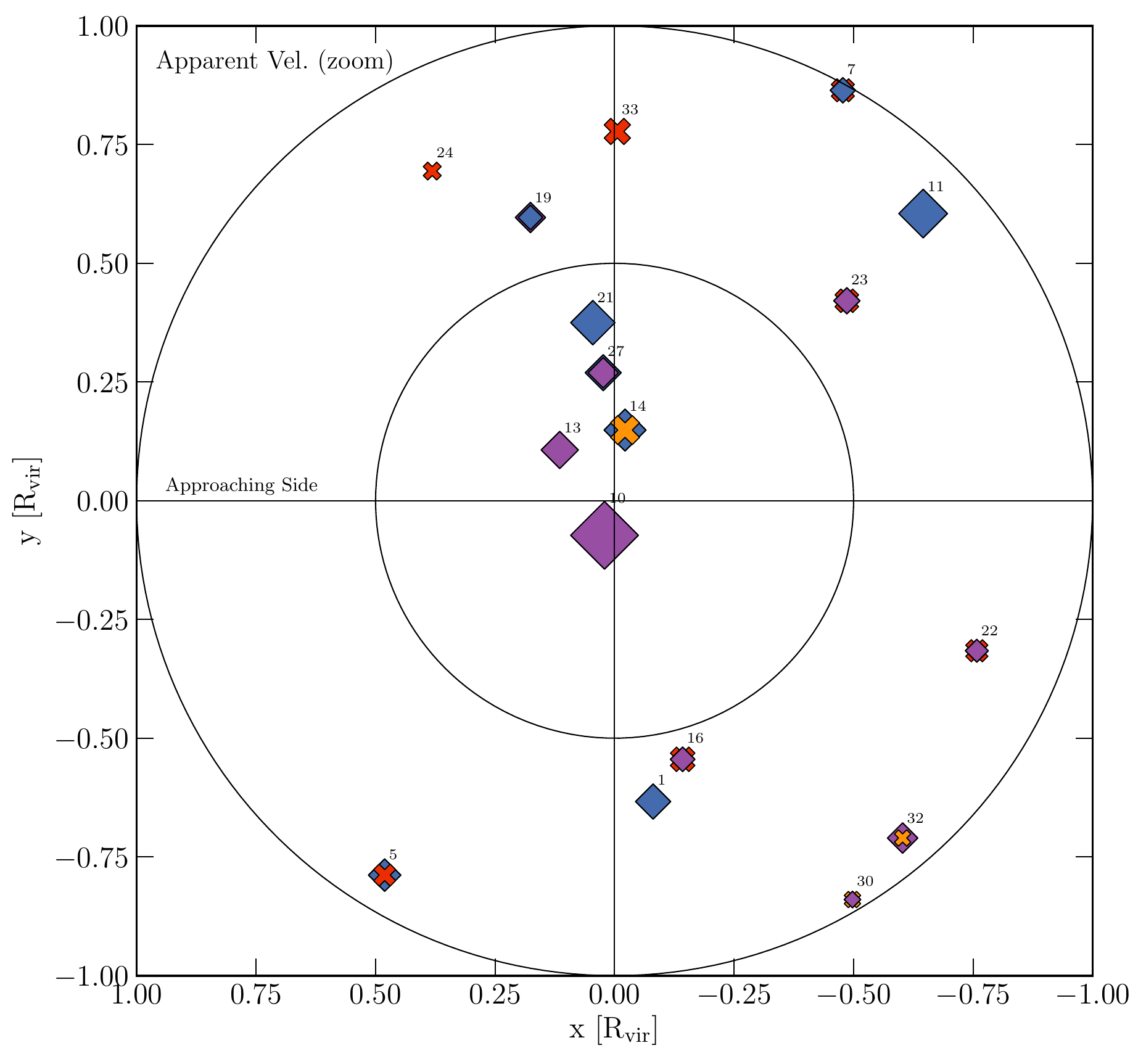}
  
   \includegraphics[width=0.485\linewidth]{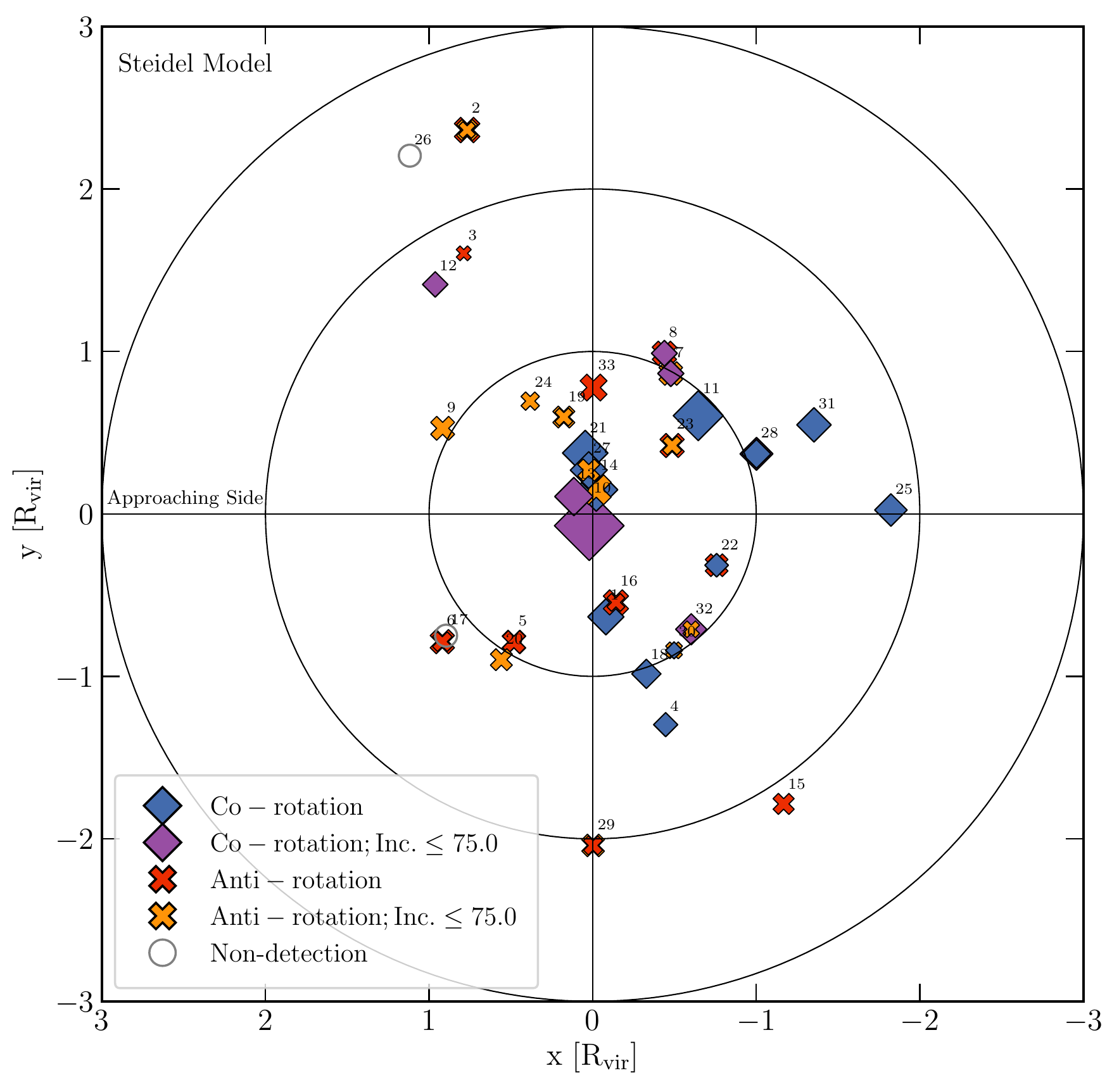}
  \includegraphics[width=0.485\linewidth]{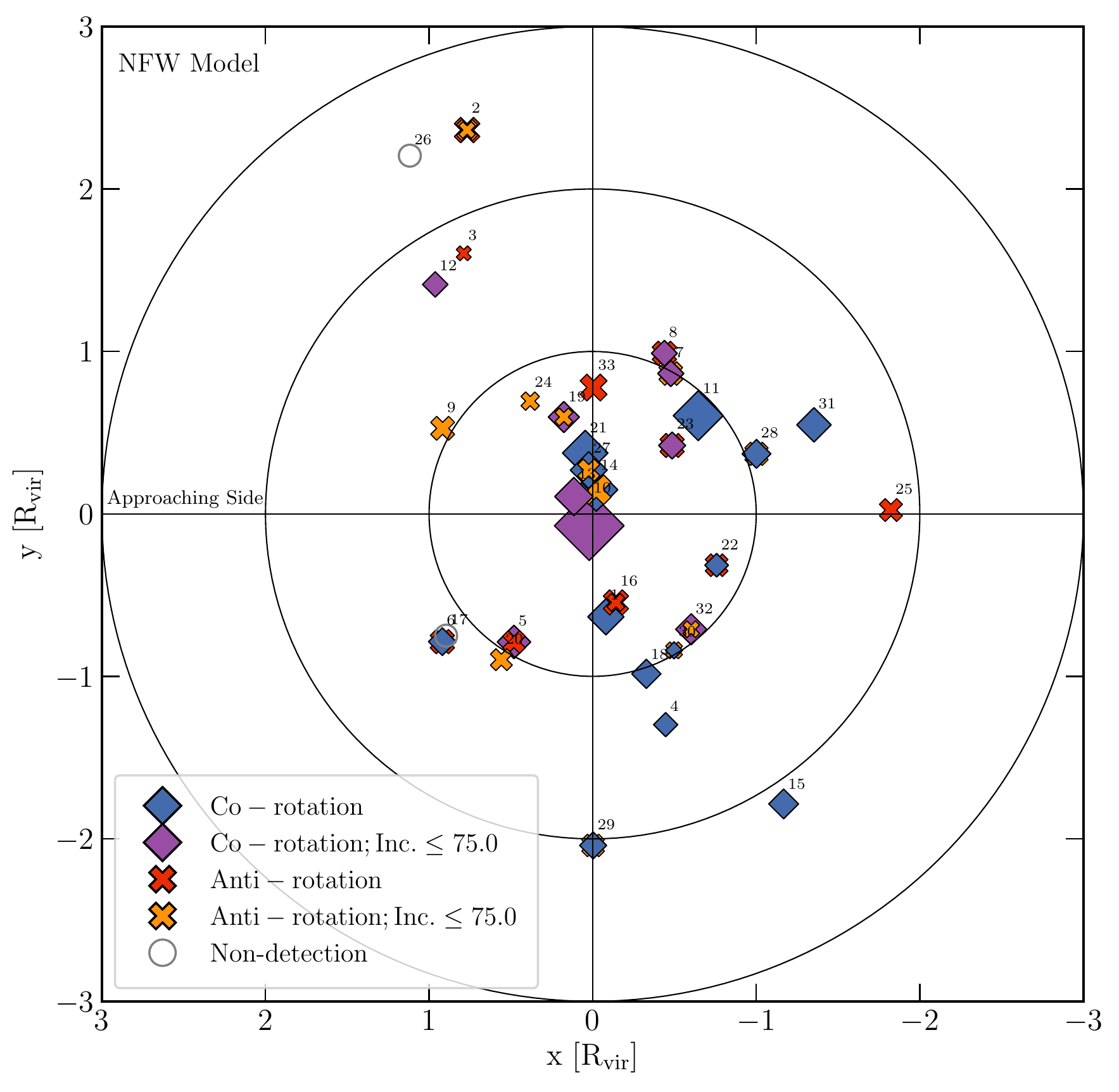}
  
  \caption{\small{\textbf{All:} Maps of the locations of each absorber normalized with respect to the galaxy virial radius. The color and style of each point indicates the line-of-sight velocity and inclination compared to that of the rotation of the nearby galaxy. The diamonds indicate absorbers whose $\Delta v$ sign aligns with the galaxy rotation (i.e., consistent with co-rotation), the crosses indicate anti-alignment, and the gray open circles indicate non-detections. The blue diamonds (red X's) indicate the associated galaxy has inclination $i > 75^{\circ}$ and co-rotating (anti-rotating) absorption, whereas the purple diamonds (orange X's) indicate less inclination ($i\leq 75^{\circ}$) and co-rotation (anti-rotation). The size of each point is scaled to reflect the column density of the absorber, and symbols are plotted in order of column density such that smaller symbols are always on top of larger symbols. All galaxies are rotated to PA = $90^{\circ}$ or $270^{\circ}$, such that their major axes are horizontal and their approaching side is on the left as indicated. The number identifiers correspond to the system number given in Column (1) of Table \ref{models}. The concentric rings indicate distances of 1, 2, and 3 $R_{\rm vir}$. \textbf{Top-Left:} Results based on apparent velocity on-sky only (model independent). \textbf{Top-Right:} A zoom-in of the apparent velocity map showing only those systems within $1R_{\rm vir}$. The concentric rings indicate distances of 0.5 and $1~R_{vir}$. \textbf{Bottom-Left:} Results based on the model from \cite{steidel2002}. \textbf{Bottom-Right:} Results based on our cylindrical NFW model.}}
  \vspace{10pt}
  \label{figure:full_map}
\end{figure*}

\subsection{Co-rotation Fraction}
Here we consider in aggregate our sample of Ly$\alpha$ absorbers, and the fraction consistent with co-rotation under various constraints.

To start we consider the fraction of absorbers whose $\Delta v$ velocity indicates they are on the ``correct" side of the galaxy to be consistent with co-rotation.
Based on their orientation and apparent velocities alone, we find 27/47 ($57\pm5\%$) of absorbers have velocities consistent with co-rotation with the nearby galaxy. The \cite{steidel2002} model produces 19/47 ($40\pm5\%$), and the NFW model produces 23/47 ($49\pm3\%$) consistent with co-rotation. All of our rotation fraction errors are calculated via a bootstrapping method which randomly resamples our dataset, thus robustly accounting for outliers.


Figure \ref{figure:full_map} presents maps of the locations of each absorber relative to it's assumed host galaxy. In Figure \ref{figure:full_map} we have rotated every system such that the galaxy major axes are horizontal with the approaching side on the left. The diamond symbols indicate the absorber velocity is consistent with co-rotation, the crosses indicate an anti-rotation, and the open gray circles indicate non-detections. They are further separated into two groups based on the associated galaxy inclination; the blue-diamonds and red-crosses represent $i> 75^{\circ}$ co- and anti-rotation, respectively, and the purple-diamonds and orange-crosses represent $i\leq75$ systems. We have also scaled the size of each marker according to its relative column density, and annotated each with a number corresponding to the appropriate system number given in Column (1) of Table \ref{models}. Figure \ref{figure:full_map} also includes the same map zoomed-in to a radius of $1 R_{\rm vir}$, as well as one each based on the \cite{steidel2002} and our cylindrical NFW model results.

\begin{figure*}[ht!]
\centering
 \subfigure[]{\includegraphics[width=0.495\linewidth]{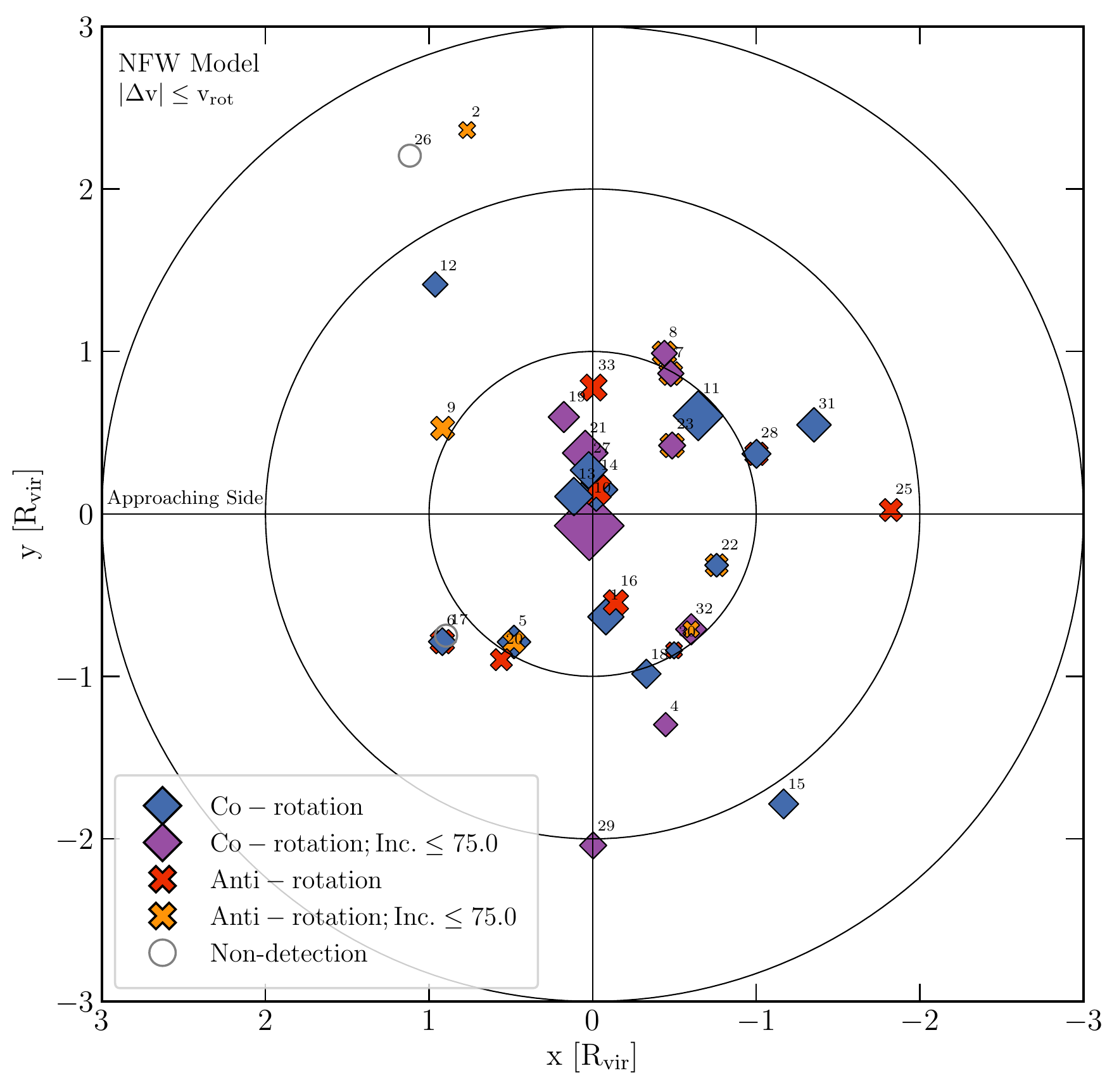}\label{figure:nfw_map_vconstraint}}
 \subfigure[]{\includegraphics[width=0.498\linewidth]{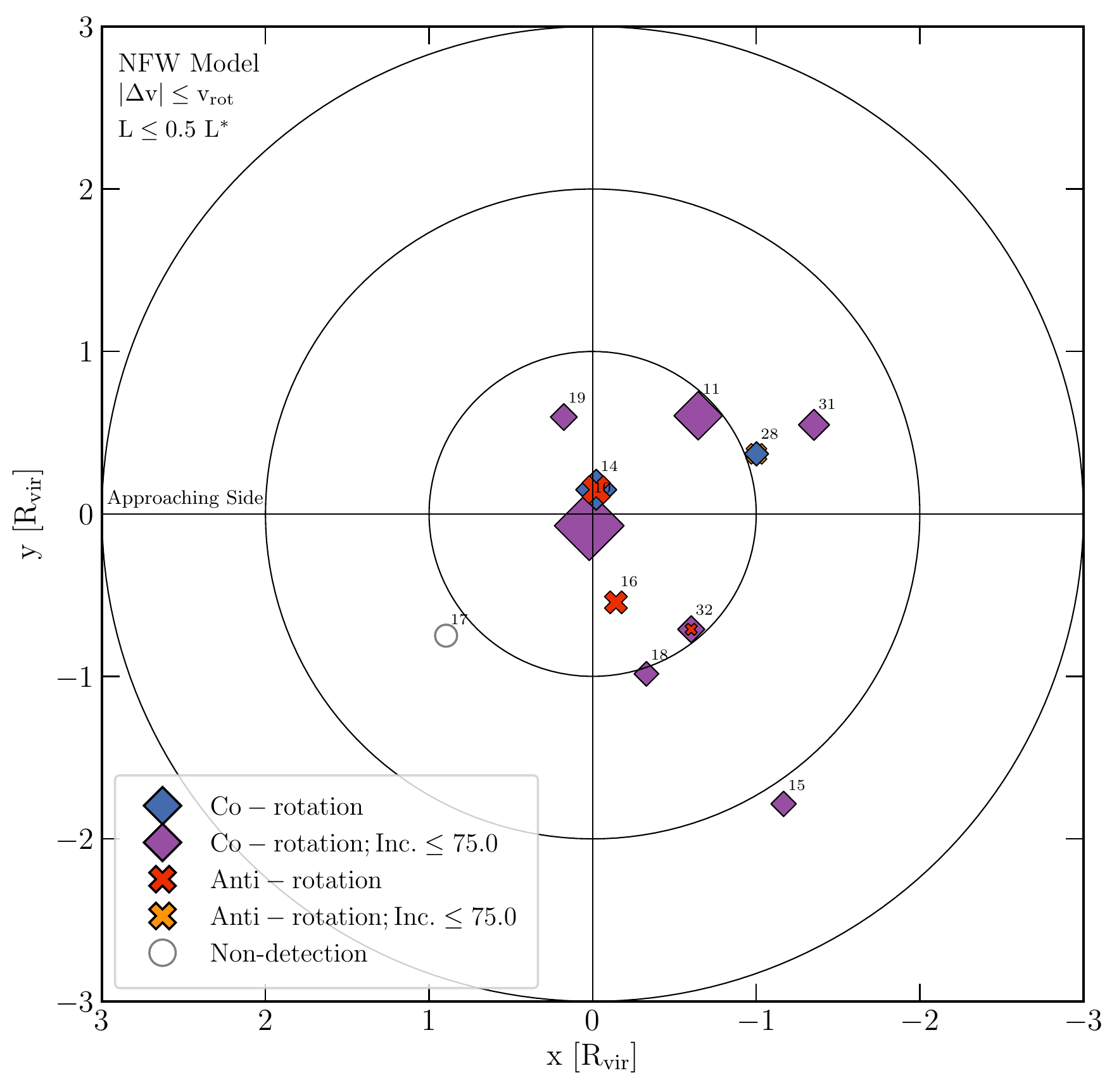}\label{figure:nfw_map_lstar}}
  \caption{\small{\textbf{Left:} Maps of the locations of each absorber normalized with respect to the galaxy virial radius. The color and style of each point indicates the NFW rotation model results and galaxy inclination with a $\lvert \Delta v \rvert \leq v_{\rm rot}$ constraint imposed. Diamonds indicate absorbers whose $\Delta v$ sign aligns with the galaxy rotation (i.e., consistent with co-rotation), crosses indicate anti-alignment, and grey open circles indicate non-detections. Blue and red symbols indicate the associated galaxy is highly inclined ($i > 75^{\circ}$), whereas purple and orange symbols indicate less inclination ($i\leq 75^{\circ}$). The size of each point is scaled to reflect the column density of the absorber, and symbols are plotted in order of column density such that smaller symbols are always on top of larger symbols. All galaxies are rotated to PA = $90^{\circ}$ or $270^{\circ}$, such that their major axis' are horizontal and their approaching side is on the left as indicated. The number identifiers correspond to the system number given in Column (1) of Table \ref{models}. Concentric rings indicate distances of 1, 2, and 3 $R_{vir}$. \textbf{Right:} The same map as \textbf{left}, but including only those absorbers associated with \Lstar $\leq 0.5$ galaxies.}}
  \vspace{15pt}
\end{figure*}


A cursory look at the maps from Figure \ref{figure:full_map} reveals several interesting results. First, the highest column density absorbers are all found within $1 R_{\rm vir}$. This is not surprising, given the results by numerous groups finding an impact parameter - column density anti-correlation (see, e.g., \citealt{french2017}, and references therein). Second, just under half of our absorbers lie beyond $1 R_{\rm vir}$. Previously, most groups have concentrated on studying the sub-$1 R_{\rm vir}$ regime, but doing so may artificially truncate the full extent of the CGM. Third, just over half (12/23) of galaxies have either multiple distinct velocity components in a single QSO sightline, or multiple sightlines containing absorbers. Of these 12, seven are oriented such that at least one component is co- and one anti-rotating with the galaxy.

However, a more in-depth look at the data reveals a number of these absorbers have $\Delta v$ much larger than the inclination-corrected galaxy rotation velocity ($v_{\rm rot}$). In other words, the velocity of the absorber relative to galaxy systemic is much greater than the rotation velocity of the galaxy disk. This results in a much smaller fraction of co-rotating absorbers when compared to our models, which will never output a velocity \emph{higher} than $v_{\rm rot}$. In undertaking this study we necessarily must begin by assuming that absorption within some velocity limit and impact parameter from a galaxy is likely associated with that galaxy. To start with we set these limits at $\Delta v_{\rm max} = 400$ \kms~and $\rho_{\rm max} = 3 R_{\rm vir}$, but now let us consider a stricter velocity range.

We now consider only absorbers with $\lvert \Delta v \rvert \leq v_{\rm rot}$, or absorbers with velocity differences no greater than the maximal galaxy rotation velocity (i.e., we are only considering absorbers where $|v_{\rm Ly\alpha} - v_{\rm sys}| \leq v_{\rm rot}$, which are those absorbers within the velocity range of $\pm$ rotation $-$ this constraint removes eight from our original sample of 47 absorbers). This constraint just means we only consider those absorbers for which it is a priori possible to be found as co-rotating. In the full sample, lines not meeting this criterion would \textit{always} be classified as non-rotating. Hence, we are effectively just setting the $\Delta v$ velocity separation limit on a case-by-case basis informed by each galaxy instead of globally given the additional information available. We could just as easily have started this study by looking for only absorbers within $\Delta v = 150$ \kms~of a galaxy instead of $\Delta v = 400$ \kms, which would have a similar overall effect. This criteria instead narrows the focus to only those absorbers kinematically close enough to a galaxy to test for a co-rotation fraction with minimal contamination from $\rm Ly\alpha$ forest lines.

With this rotation-based velocity constraint in place the co-rotating fractions for the \cite{steidel2002} and cylindrical NFW models increase to 19/39  ($49\pm5\%$) and 23/39 ($59\pm6\%$), while the apparent co-rotation fraction also increases slightly to 24/39 ($62\pm4\%$). Consistently, we find our NFW model to predict an $\sim 8-10 \%$ higher co-rotation fraction than the \cite{steidel2002} model, and similar to the apparent velocity results. This is a not wholly unexpected, and yet refreshing result; simulations have predicted that galaxies are strongly linked to their surroundings and share angular momentum, which should result in a higher than $50\%$ halo-gas co-rotation fraction. Figure \ref{figure:nfw_map_vconstraint} shows an absorber map based on our NFW model results with this velocity constraint in place.

\begin{figure*}[h!]
\label{lstar_impact_fig}
\centering
\subfigure[]{\includegraphics[width=0.49\linewidth]{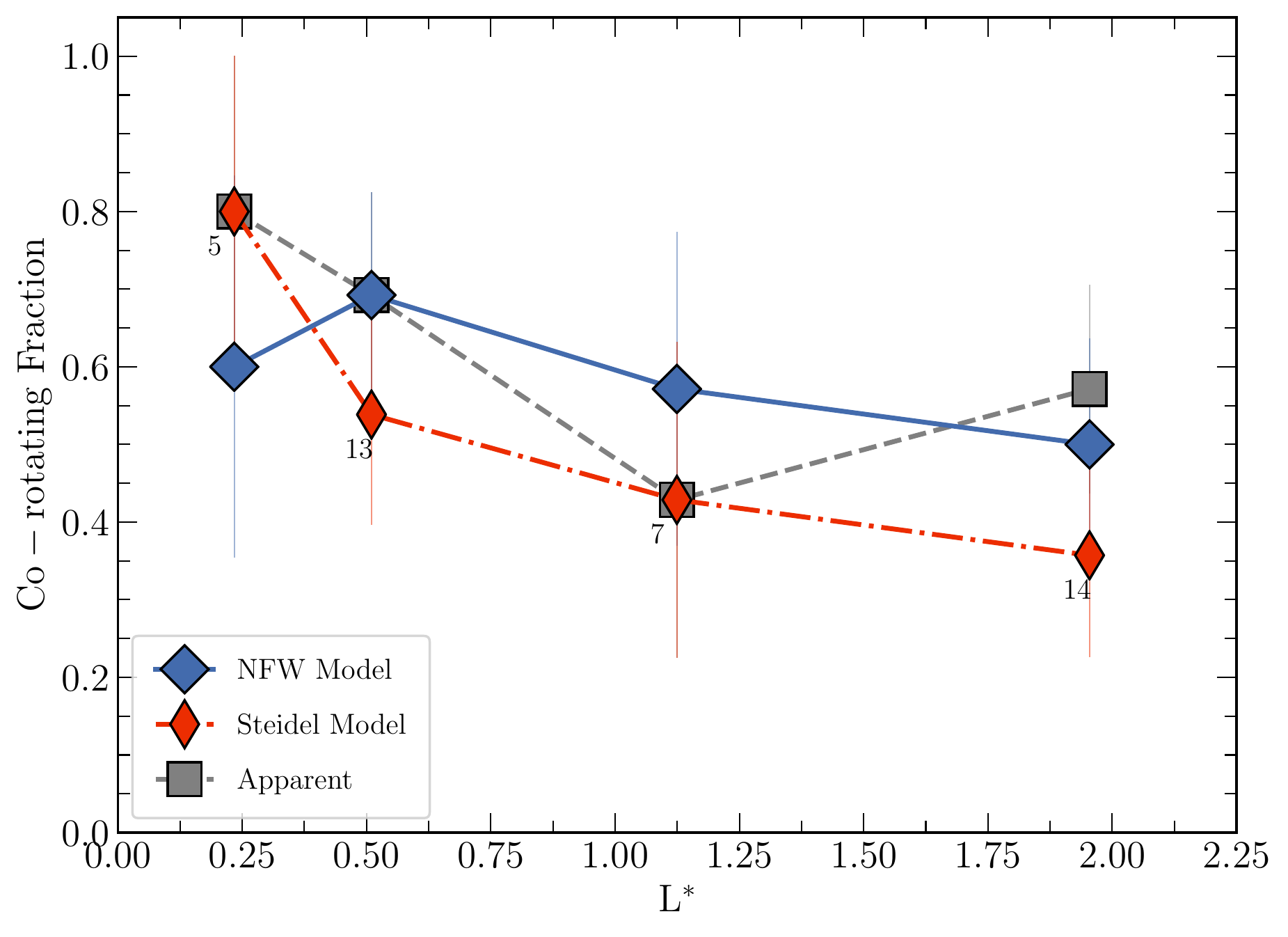}\label{figure:lstar_fraction}}
\subfigure[]{\includegraphics[width=0.49\linewidth]{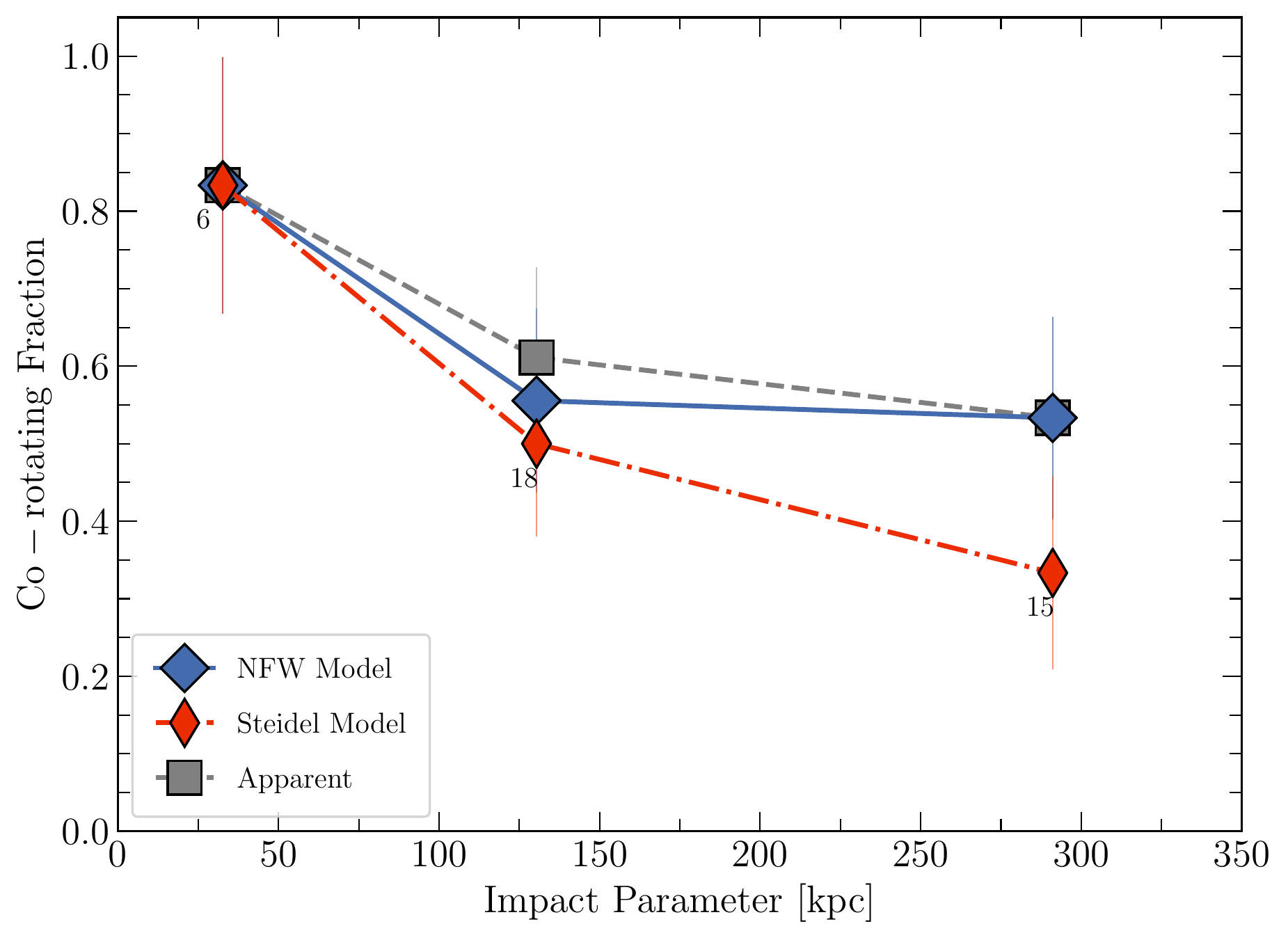}\label{impact_fraction}}
\caption{\small{\textbf{Left:} The fraction of co-rotating absorbers as a function of \Lstar. \textbf{Right:} The fraction of co-rotating absorbers as a function of impact parameter. \textbf{All:} Grey squares correspond to model-independent apparent velocity results, thin-red diamonds correspond to \cite{steidel2002} model results, and thick-blue diamonds correspond to our NFW model results. Bin edges are calculated to yield approximately equally sized bins. The numbers underneath each red-diamond show the number of absorber-galaxy systems in each bin.}}
\end{figure*}

\begin{figure*}[h!]
\centering
  \subfigure[]{\includegraphics[width=0.49\linewidth]{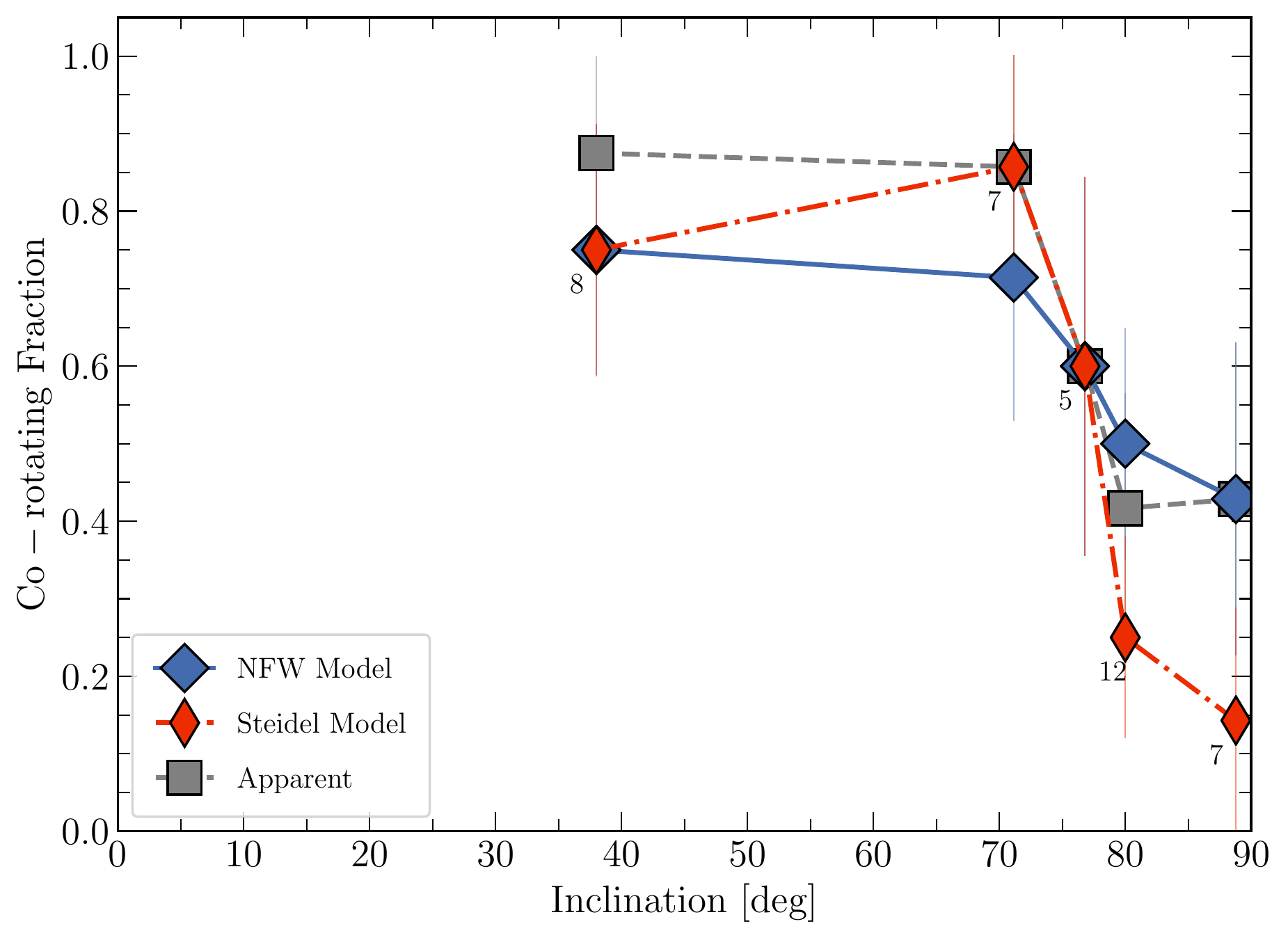}\label{figure:inclination}}
  \subfigure[]{\includegraphics[width=0.49\linewidth]{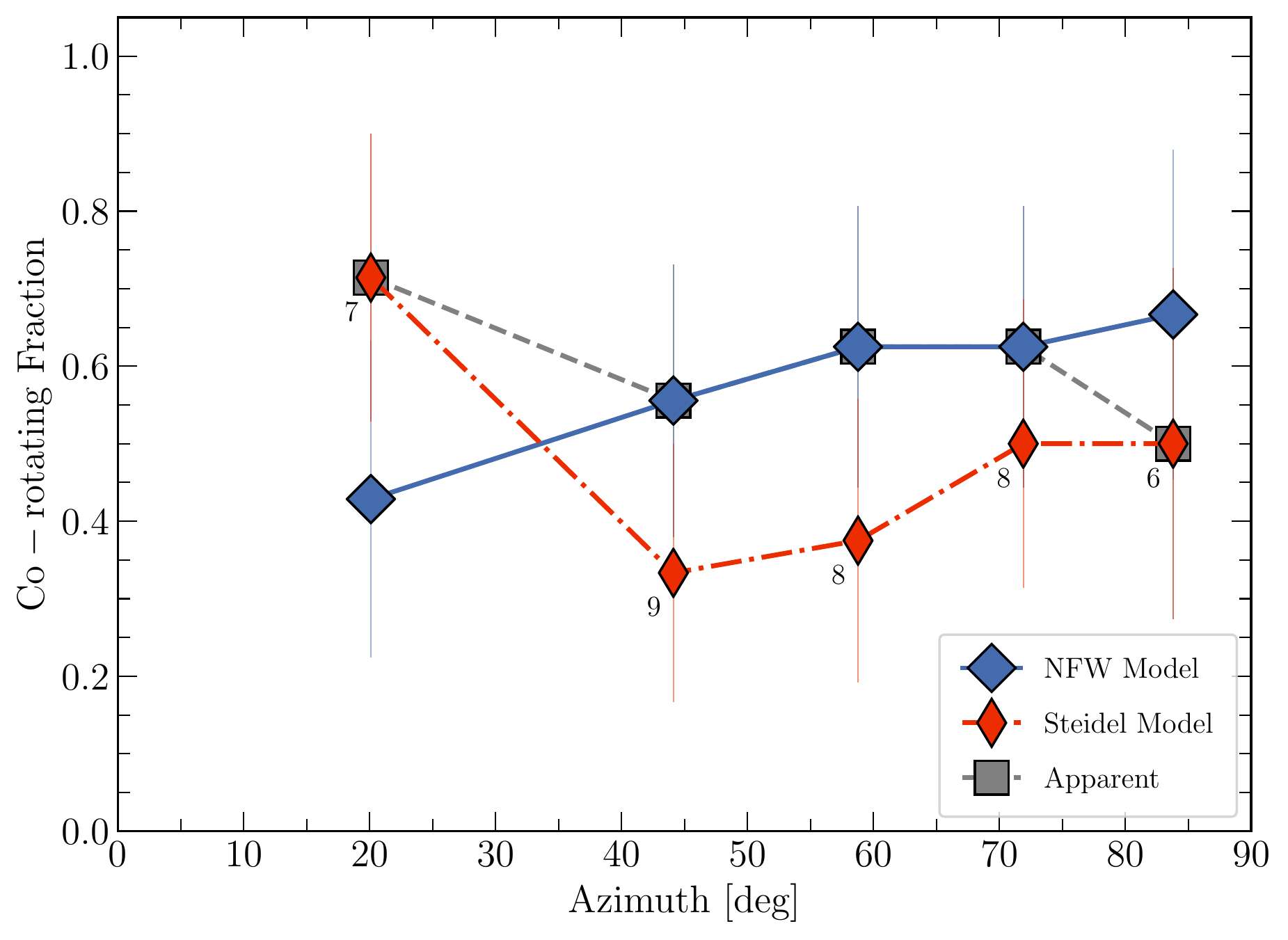}\label{figure:azimuth}}
  \caption{\small{The co-rotation fraction as a function of inclination angle (\textbf{left}) and azimuth angle (\textbf{right}) for $\lvert \Delta v \rvert \leq v_{\rm rot}$ absorbers. Results from apparent on-sky velocity are shown as grey-squares, \cite{steidel2002} model results as thin red-diamonds, and our NFW model results as thick blue-diamonds. Bins are chosen algorithmically to produce approximately even bins. Thus the first inclination bin appears at $\sim40$ degrees because we have mostly highly inclined galaxies. For reference, the mean azimuth angles within each inclination bin from left to right are 51, 57, 63, 50, and 65 degrees, with 90 degrees being along the galaxy minor axis.}}
\vspace{10pt}
\label{inc_az_fig}
\end{figure*}

\subsection{Co-rotation as a function of \Lstar}


For brevity's sake we will concentrate only on the NFW model results with our $\lvert \Delta v \rvert \leq v_{\rm rot}$ restriction from here on-wards. We now consider the effect of galaxy luminosity on co-rotation fraction by first separating our sample around $0.5$\Lstar. This results in 13 absorbers near $L \leq 0.5$\Lstar~galaxies and 26 around more luminous galaxies. The co-rotating fraction around luminous galaxies is then $54\pm5\%$, compared to $69\pm7\%$ around $L \leq 0.5$\Lstar~ galaxies. Figure \ref{figure:nfw_map_lstar} shows absorbers map for this $L \leq 0.5$\Lstar~ galaxy subsample (right) compared to the full sample (left).

Furthermore, we find this co-rotation fraction smoothly decreases as a function of \Lstar, as shown in Figure \ref{figure:lstar_fraction}. In this figure we have binned galaxy-absorber systems into four bins of luminosity, and are plotting the corresponding co-rotation fraction for each bin. This uneven bin spacing was chosen algorithmically to produce relatively evenly sized bins, and the exact binning does not affect the overall trend.
The bin sizes are labeled explicitly underneath each data point for clarity. We have used a bootstrapping routine to calculate errors for Figures \ref{lstar_impact_fig} and \ref{inc_az_fig}, which randomly resamples the selection of galaxies in each bin 10,000 times.


Given that recent simulation results suggest that co-rotating accretion gas is predominantly cold-mode for low-mass galaxies in the local universe, this may be a signature of this co-rotating, cold-mode accreting gas. Additionally, \cite{lutz2018} find that galaxies with high \HI mass compared to their stellar mass have higher halo angular momentum, which may be impeding their ability to efficiently form stars. While we do not have independent measures of \HI and stellar mass for our galaxies, it may not be unreasonable to think that such lower stellar-mass galaxies are analogous to the low luminosity galaxies in our sample.

Finally, we show in Figure \ref{impact_fraction} the co-rotation fraction as a function of impact parameter. For all three cases we consistently find a declining co-rotation fraction at greater physical distances from galaxies. Within $\sim 100$ kpc, nearly 85\% of Ly$\alpha$ absorbers have velocities consistent with co-rotation, compared to $\sim50$\% beyond this distance. As no correlation between disk gas and halo kinematics should result in a $\sim$ 50\% co-rotation fraction, this implies that the sub-100 kpc regime harbors the majority of coherent halo angular momentum. Moreover, we note that the \cite{steidel2002} model dives far below this 50\% co-rotation mark at large distances. This constant-velocity monolithic disk model is thus likely unphysical.


\subsection{Inclination \& Azimuth}
Here we investigate how the orientation of each galaxy affects the co-rotation rate. Figures \ref{figure:inclination} and \ref{figure:azimuth} show the co-rotation fraction for each model as a function of inclination and azimuth angle. We use the standard definitions wherein inclination ranges from $0^{\circ}$ (face-on) to $90^{\circ}$ (edge-on), and azimuth angle ranges from $0^{\circ}$ (major axis) to $90^{\circ}$ (minor axis). The co-rotation fraction is largely invariant as a function of azimuth angle but sharply drops off above inclinations of $\sim 70$ degrees, while remaining mostly flat below that. The \cite{steidel2002} model in particular predicts $\lesssim 20$\% co-rotation rate for highly inclined galaxies, which is less than half that predicted by our NFW model or apparent velocity ($\sim 40$\%). Hence, this simple monolithic disk model is likely not physical. 

\begin{figure}[ht!]
\centering
  \includegraphics[width=0.956\linewidth]{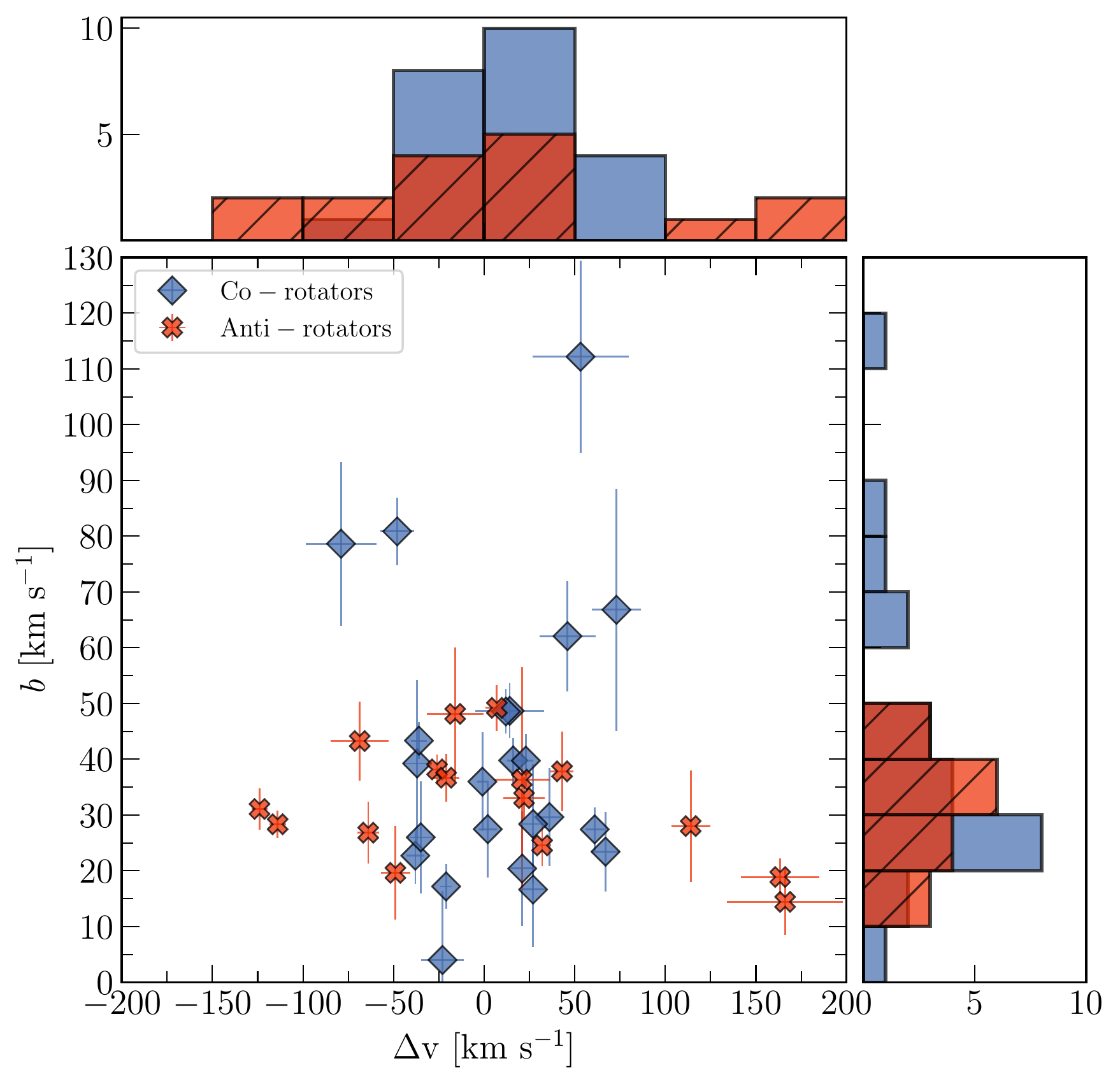}
  \caption{\small{The Doppler $b$-parameters for $\lvert \Delta v \rvert \leq v_{\rm rot}$ absorbers as a function of $\Delta v$. The co-rotating (blue diamonds) and anti-rotating (red crosses) designation is based on our NFW model results. The marginal histograms show the distributions of $b$ and $\Delta v$ similarly split into co-rotating (blue) and anti-rotating (red).}}
  \label{figure:b_vs_dv}
\end{figure}

\begin{figure}[ht!]
\centering
  \includegraphics[width=0.966\linewidth]{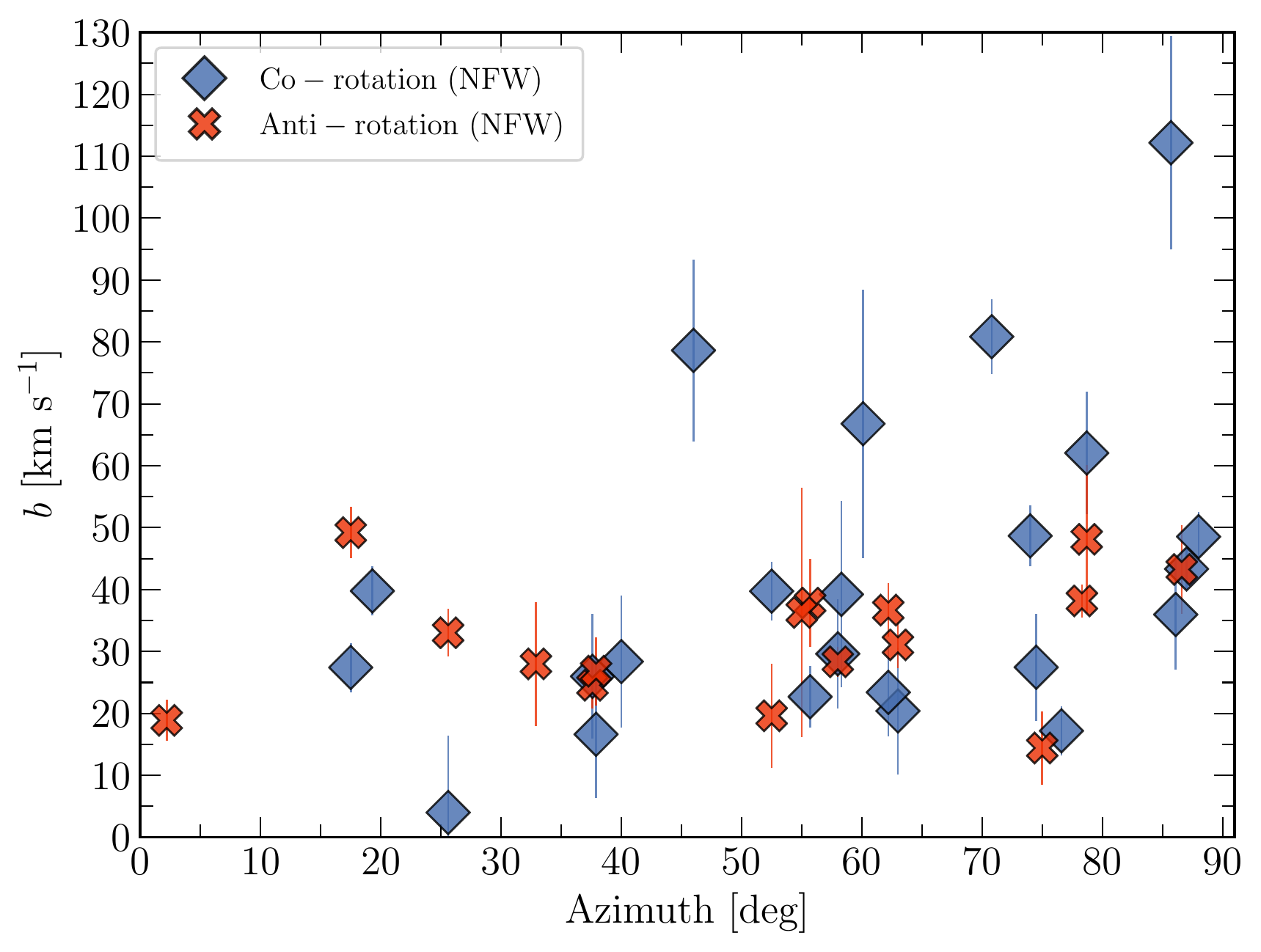}
  \caption{\small{The Doppler $b$-parameters for $\lvert \Delta v \rvert \leq v_{\rm rot}$ absorbers as a function of azimuth angle. The co-rotating (blue diamonds) and anti-rotating (red crosses) designation is based on our NFW model results.}}
  \label{figure:b_vs_az}
\end{figure}

\begin{figure*}[ht!]
\centering
  \subfigure[]{\includegraphics[width=0.476\linewidth]{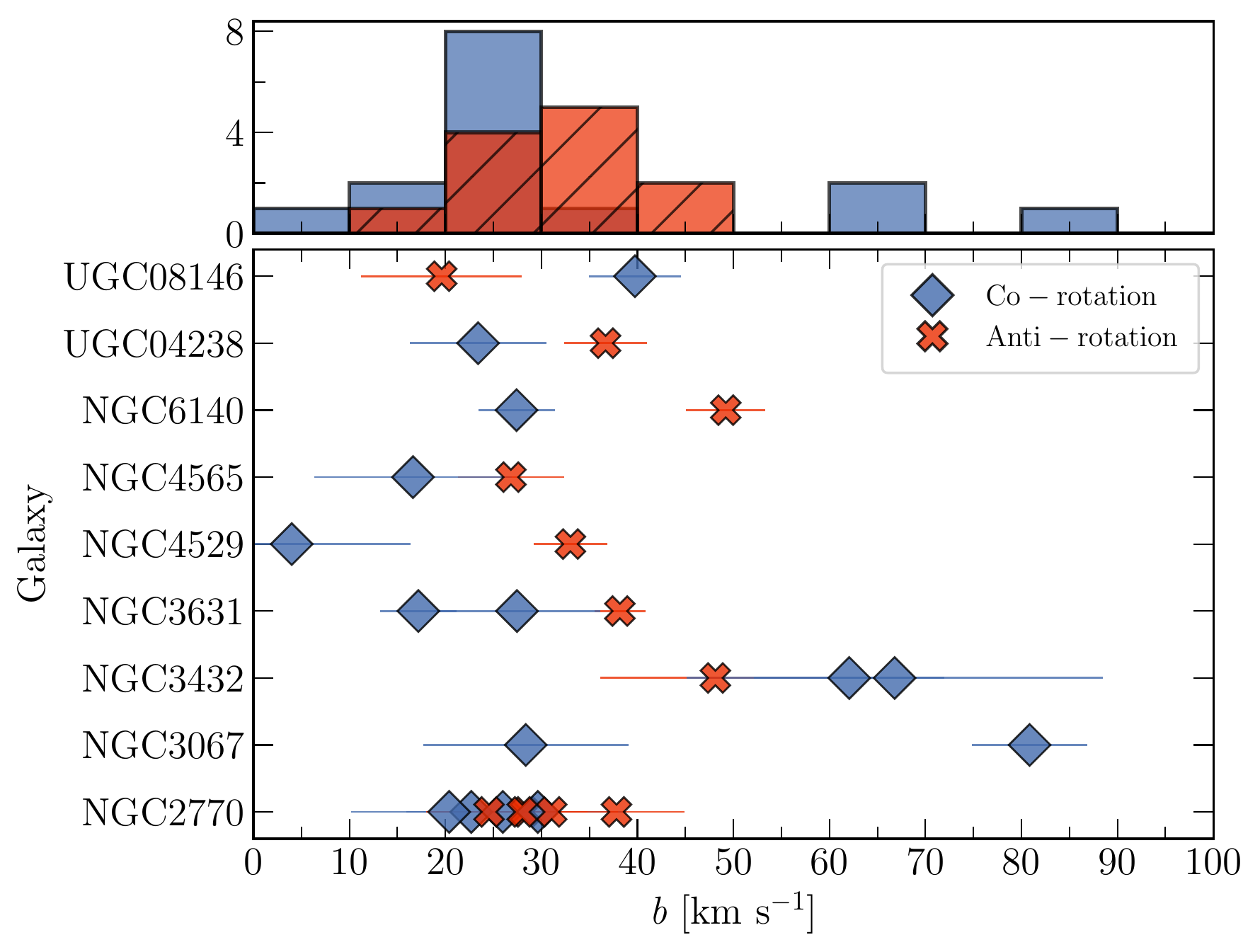}}
  \subfigure[]{\includegraphics[width=0.476\linewidth]{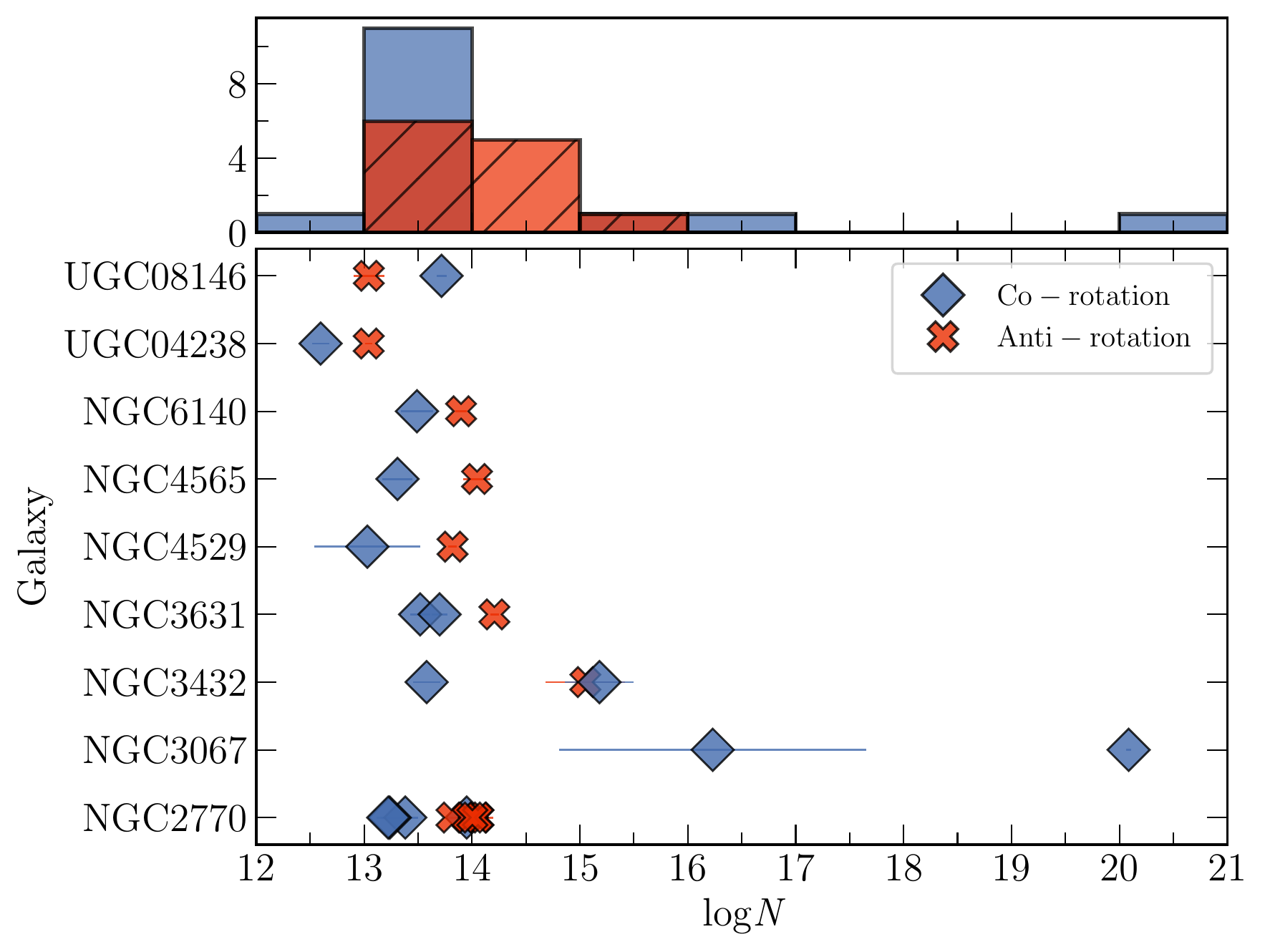}}
  \caption{\small{The Doppler $b$-parameters and column densities for $\lvert \Delta v \rvert \leq v_{\rm rot}$ absorbers where \emph{multiple components are associated with a single galaxy}. The co-rotating (blue diamonds) and anti-rotating (red crosses) designation is based on our NFW model results. The top histogram shows the distribution of $b$ similarly split into co-rotating (blue) and anti-rotating (red).}}
\label{figure:b_hist_multiple}
\vspace{5pt}
\end{figure*}

Numerous simulations (e.g., \citealt{stewart2011b} and \citealt{ho2019}) and metal-line absorption studies (e.g., \citealt{kacprzak2010},  \citealt{martin_2019}) suggest that co-rotation should instead \emph{peak} around the major axis of highly inclined galaxies (i.e., azimuth = $0^{\circ}$). Our results may differ for several reasons. First, we simply do not have many systems that are both highly inclined and aligned along the major axis. Of the 30 components associated with galaxies inclined $70^{\circ}$ or more, only eight are within $45^{\circ}$ of the major axis (and the co-rotation fraction of these eight does not differ significantly from the rest of the sample). We do not otherwise see a strong correlation with azimuth angle, but there may be a combination of azimuth and inclination effects present which are difficult to untangle. Second, our sample consists almost entirely of low column density gas, whereas the Mg\II~absorbers in other studies trace much denser material. As simulations suggest that cold, dense gas accretes mostly along an extended, warped disk, the more  diffuse gas we are tracing may be more geometrically and kinematically complex.




\subsection{Doppler $b$-parameters}
Finally, we consider the Doppler $b$-parameters of our absorber sample. In Figure \ref{figure:b_vs_dv} we show how the $b$-parameters vary as a function of $\Delta v$ for co-rotating versus anti-rotating absorbers (based on our NFW model results and limited to $\lvert \Delta v \rvert \leq v_{\rm rot}$ absorbers). Figure \ref{figure:b_vs_dv} also includes marginal histograms to show the distribution of $b$ and $\Delta v$. We might expect the co-rotating sample to occupy a narrower $\Delta v$ space as the NFW model predicts decreasing $\Delta v$ with distance, but the elevated $b$-parameters for these compared to the relatively flat distribution for anti-rotators is intriguing. All the anti-rotators have $b \lesssim 50$ \kms, leaving all five broader lines as co-rotating systems. 

As previously discussed however, the picture described by the simulations of \cite{stewart2011b} and others describes a scenario where co-rotating gas is predominately the product of cold-mode accretion. Hotter, and thus broader, outflowing gas would likely carry angular momentum from the disk with it, but this would be quickly lost as the outflows expand into the halo and result in negligible observable rotation signature. Indeed, as shown in Figure \ref{figure:b_vs_az}, there is a slight correlation between $b$ and azimuth angle, with the broadest absorbers within $\sim 50^{\circ}$ of the minor axis (azimuth = $90^{\circ}$).

As the broadest lines are higher than expected for purely thermal motions within a single $\rm Ly\alpha$ structure, we may be observing either a number of clouds that are close in velocity space or a filament with a range of turbulent, internal velocities. In this scenario these high $b$-parameters could be consistent with filamentary inflows versus around higher \Lstar~ galaxies where virial shocks are perhaps breaking larger structures into smaller, more isolated cloudlets and producing lower $b$-parameter absorption.

To further explore this, we now consider the nine galaxies for which we find multiple Ly$\alpha$ components. Figure \ref{figure:b_hist_multiple} shows the $b$ and $\log N(\HI)$ distributions for each of these galaxies. Intriguingly, 8/9 of these galaxies have at least one component consistent with both co-rotation and anti-rotation. Of these eight, six are oriented such the lower $b$ component is co-rotating, and 7/8 such that the lower $\log N(\HI)$ component is co-rotating. Hence, narrower, lower column density components appear more likely to be co-rotating. Once again these results differ from the finding of numerous metal-line kinematic studies that suggest the entirety of Mg\II~absorption tends to lie on one side of galaxy systemic velocities (e.g., \citealt{kacprzak2010}, \citealt{ho2017}). Galaxy halos are thus likely multiphase, with distributions of material complex both geometrically and kinematically.


\begin{deluxetable*}{l | c c | c c | c c}
\tabletypesize{\footnotesize}
\tablewidth{0pt}
\tablecaption{CGM Rotation: Summary of Results\label{results}}
\tablehead{
\colhead{Sub-sample} & \colhead{Co-rotating} & \colhead{Anti-rotating}  & \colhead{Co-rotating} & \colhead{Anti-rotating} & \colhead{Co-rotating} & \colhead{Anti-rotating}	\\
			  		&          			 	& 			 				& $\rho \leq 1$			& $\rho \leq 1$				& $\rho >$ 1		& $\rho > 1$	}
\startdata
Apparent Vel.			&	27				&	20						&		17				&		10				&	10					&	10		\\
\hline
Steidel Model			&	19				&	28						&		11				&		16				&	8					&	12		\\
\hline
NFW Model				&	23				&	24						&		10				&		12				&	15					&	19		\\
\hline
\hline
With Constraint:		&				    &					    	&					    &					    &					    &		    \\
$\lvert \Delta v \rvert \leq v_{\rm rot}$ 	&		&					&				    	&					    &			        	&			\\
\hline
Apparent Vel.	    	&	24				&	15				    	&	    14			    &		9			    &	10			        &	6		\\
\hline
Steidel Model           &	19				&	20						&		11				&		12				&	8					&	8		\\
\hline
NFW Model               &	23				&	16						&		14				&		9				&	9					&	7		\\
\hline
\hline
With Constraint:		&				    &					    	&					    &					    &					    &		    \\
$\lvert \Delta v \rvert \leq v_{\rm rot}$ 	&		&					&				    	&					    &			        	&			\\
$[0 \leq $\Lstar $ \leq 0.5]$               &		&					&				    	&					    &			        	&			\\
\hline
Apparent Vel.	    	&	9				&	4				    	&	    5			    &		3			    &	4			        &	1		\\
\hline
Steidel Model           &	8				&	5						&		4				&		4				&	4					&	1		\\
\hline
NFW Model               &	9				&	4						&		5				&		3				&	4					&	1		\\
\hline
\hline
With Constraint:		&				    &					    	&					    &					    &					    &		    \\
$\lvert \Delta v \rvert \leq v_{\rm rot}$ 	&		&					&				    	&					    &			        	&			\\
$[$\Lstar $ > 0.5]$     &		            &					        &				    	&					    &			        	&			\\
\hline
Apparent Vel.	    	&	14				&	12				    	&	    9			    &		6			    &	5			        &	6		\\
\hline
Steidel Model           &	11				&	15						&		7				&		8				&	4					&	7		\\
\hline
NFW Model               &	14				&	12						&		9				&		6				&	5					&	6		\\
\enddata
\tablecomments{Each column gives the number of galaxy-absorber systems that appear to be co- or anti-rotating for each model and with each of the constraints we placed in the text. The middle and right columns give the results inside and outside of $1 R_{\rm vir}$.}
\end{deluxetable*}

\vspace{5pt}

\section{Summary} \label{summary}
We have presented complimentary COS Ly$\alpha$ absorption-line and nearby galaxy rotation curve analysis for a sample of 33 galaxy-QSO pairs, including 47 Ly$\alpha$ components, resulting in the largest yet sample of its kind. Table \ref{results} provides a summary of the absorber co-rotation fractions resulting from the various models and constraints we have explored. Overall, our findings suggest that galaxy halo rotation is only one of many factors contributing to the velocity distribution of Ly$\alpha$ clouds near galaxies. The fact that we see coherent co-rotation trends with galaxy luminosity, impact parameter, and orientation suggests that these absorbers are kinematically connected to the nearby galaxy disks.

These findings contrast slightly with recent results concerning the kinematics of metal tracers such as Mg\II~(e.g., \citealt{kacprzak2010}, \citealt{ho2017}, \citealt{martin_2019}), which find clear and ubiquitous co-rotation with nearby galaxy disks. Likely, the neutral \HI~studied here is tracing more complex structures, with contributions from both extended, co-rotating disks as well as infalling IGM material. The effect of outflow, inflow, and turbulent velocities certainly also plays an important role, and one that is extremely difficult to untangle. Metals mostly originate within already co-rotating disks however, which should account for their higher co-rotation rates. Our main conclusions are the summarized below:
\vspace{10pt}

1. We have tested two halo rotation models against a simple on-sky velocity designation of Ly$\alpha$ absorber velocities. We find a $62 \pm 4$\% apparent on-sky Ly$\alpha$ co-rotation fraction, while the \cite{steidel2002} model results in $49 \pm 5$\% and our new cylindrical NFW model results in $59 \pm 6$\% fractions. We find the constant-velocity, monolithic disk \cite{steidel2002} model to be a poor descriptor of Ly$\alpha$ CGM kinematics, as it predicts unphysically low co-rotation fractions for higher-impact parameter systems (Figure \ref{impact_fraction}), as well as for those at high-azimuth and inclination angles (Figure \ref{inc_az_fig}).


\vspace{10pt}

2. The fraction of Ly$\alpha$ absorbers appearing to co-rotate with the nearby galaxy declines as a function galaxy luminosity (\Lstar). Based on the predicted velocity of our NFW halo model, $69\pm7\%$ of absorbers co-rotate around $\rm \leq 0.5 $\Lstar~ galaxies, which falls down to $54\pm5\%$ around more luminous galaxies at $z \sim 0$. Our overall co-rotation fraction is broadly consistent with the simulation results of \cite{stewart2011b, stewart2013}, and the effect of galaxy luminosity on halo gas co-rotation is consistent with predicted cold-mode filamentary accretion schemes.

\vspace{10pt}

3. Over 80\% of Ly$\alpha$ absorbers appear to co-rotate with galaxies within 100 kpc. Beyond this the co-rotation fraction returns to approximately 50\%, consistent with no correlation, according to our NFW model and apparent on-sky velocities.


\vspace{10pt}

4. The Ly$\alpha$ co-rotation fraction is mostly inclination-independent below $\sim 70$ degrees, but sharply declines at higher inclinations. This is likely due to a combination of azimuth and inclination effects which are difficult to untangle. If gas is primarily accreting along galaxies' major axes, the infall velocity may further complicate the projected velocity profile. A larger sample encompassing a wide range of azimuth and inclination angles will be needed to make further progress here.  






\vspace{10pt}

5. Co-rotating absorbers (when chosen from the sample restricted to $\lvert \Delta v \rvert \leq v_{\rm rot}$) occupy a wide range in Doppler $b$-parameter, while anti-rotators have mostly $b \leq 50$ \kms. However, for galaxies with multiple sightlines, 8/9 have both a co-rotating and anti-rotating absorber. Within this subsample, the co-rotating absorbers are both narrower (i.e., lower $b$ values) and have lower column densities in 6/8 and 7/8 cases, respectively.


\vspace{-10pt}
\acknowledgements

D.M.F. thanks Claire Murray for useful insights, particularly related to our halo model, and Julie Davis for invaluable SALT data reduction pointers. This research has made use of the NASA/IPAC Extragalactic Database (NED), which is operated by the Jet Propulsion Laboratory, California Institute of Technology, under contract with the National Aeronautics and Space Administration. Based on observations with the NASA/ESA \textit{Hubble Space Telescope}, obtained at the Space Telescope Science Institute (STScI), which is operated by the Association of Universities for Research in Astronomy, Inc., under NASA contract NAS 5-26555. Spectra were retrieved from the Barbara A. Mikulski Archive for Space Telescopes (MAST) at STScI. Some of the observations reported in this paper were obtained with the Southern African Large Telescope (SALT) under program 2016-1-SCI-062 (PI: Wakker). Over the course of this study, D.M.F. and B.P.W. were supported by grant AST-1108913 from the National Science Foundation and GO-13444.01-A, GO-14240.01-A, and AR-14577.01-A from STScI.



\facility{HST (COS), SALT (RSS)}
\clearpage

\bibliography{rotation_paper_submitted}

\bibliographystyle{aasjournal}

\clearpage

\appendix
\label{galaxy_sample}

\section{SALT Galaxies} \label{SALT_sample}
In this section we summarize each galaxy-QSO system observed by SALT. Here we provide rotation curves and finder chart images for the subsample of galaxies with newly observed SALT data. Please see Table \ref{models} for further details. Each rotation curve figure includes a legend with the galaxy name, $v_{\rm sys}$, and $v_{\rm rot}$. Each associated finder chart includes the position of the SALT-RSS slit in red, an arrow toward each nearby QSO with QSO name and impact parameter labels in blue (or a red circle around the QSO for CGCG039-137, the only case where the QSO is within the image), and a physical scale legend in the lower-left corner. 



\subsection{CGCG039-137}

CGCG039-137 is an isolated Scd type galaxy with a measured systemic velocity $v_{\rm sys} = 6918 \pm 24$ \kms~and inclination of $i = 72^{\circ}$. The QSO RX\_J1121.2+0326 is located nearby at an impact parameter of 99 kpc and azimuth angle of $86^{\circ}$ on the receding side. The data for  RX\_J1121.2+0326 has low signal-to-noise ($\sim 4.2$), but we are able to detect Ly$\alpha$ at 6971 \kms.

\begin{figure}[hb]
\centering
  \subfigure[]{\includegraphics[width=0.5\linewidth]{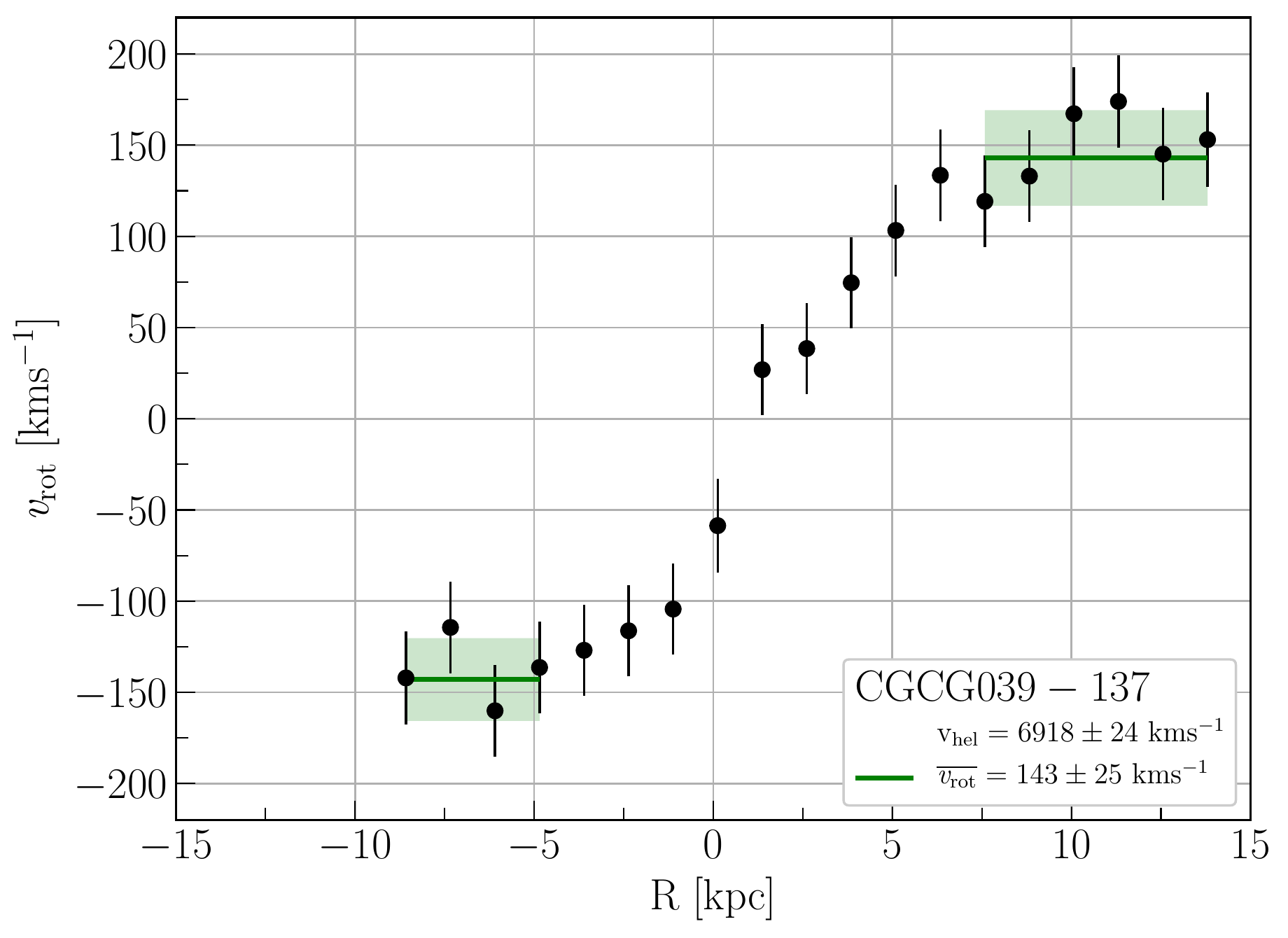}}{\label{rotationcurve_CGCG039-137}}
  \subfigure[]{\includegraphics[width=0.49\linewidth]{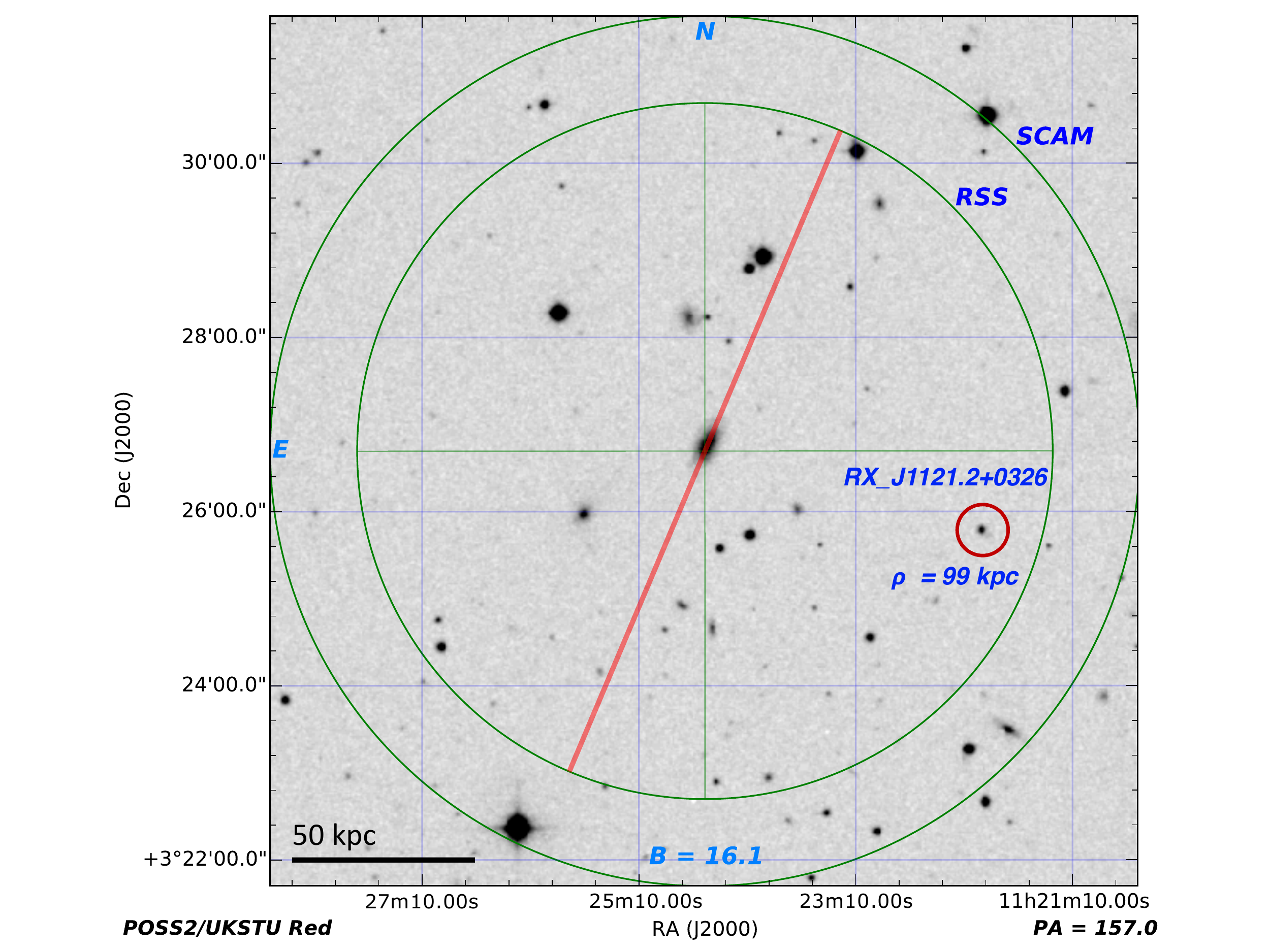}\label{finderchart_CGCG039-137}}
  \caption{\small{a) Rotation curve of CGCG039-137. The solid green line indicates the weighted mean velocity over the corresponding x-axis region, and the shaded green indicates the 1$\sigma$ error in the mean. b) SALT finder chart for CGCG039-137 showing the position of the slit in red.}}
\vspace{0pt}
\end{figure}

\subsection{ESO343-G014}
ESO343-G014 is an edge-on ($i = 90^{\circ}$) spiral galaxy with a measured systemic velocity $v_{\rm sys} = 9139 \pm$ 32 \kms. It has a smaller neighboring galaxy, 2MASXJ21372816-3824412, located north of its major axis at a projected distance of 216 kpc and velocity of 9129 \kms. The nearest sightline is towards RBS1768 at $\rho = 466$ kpc and $75^{\circ}$ azimuth angle on the approaching side. We detect 3 blended Ly$\rm \alpha$ absorption components toward RBS1768 at $v_{\rm Ly\alpha} = 9305, 9383, 9447$ \kms.


\begin{figure}[ht]
\centering
  \subfigure[]{\includegraphics[width=0.5\linewidth]{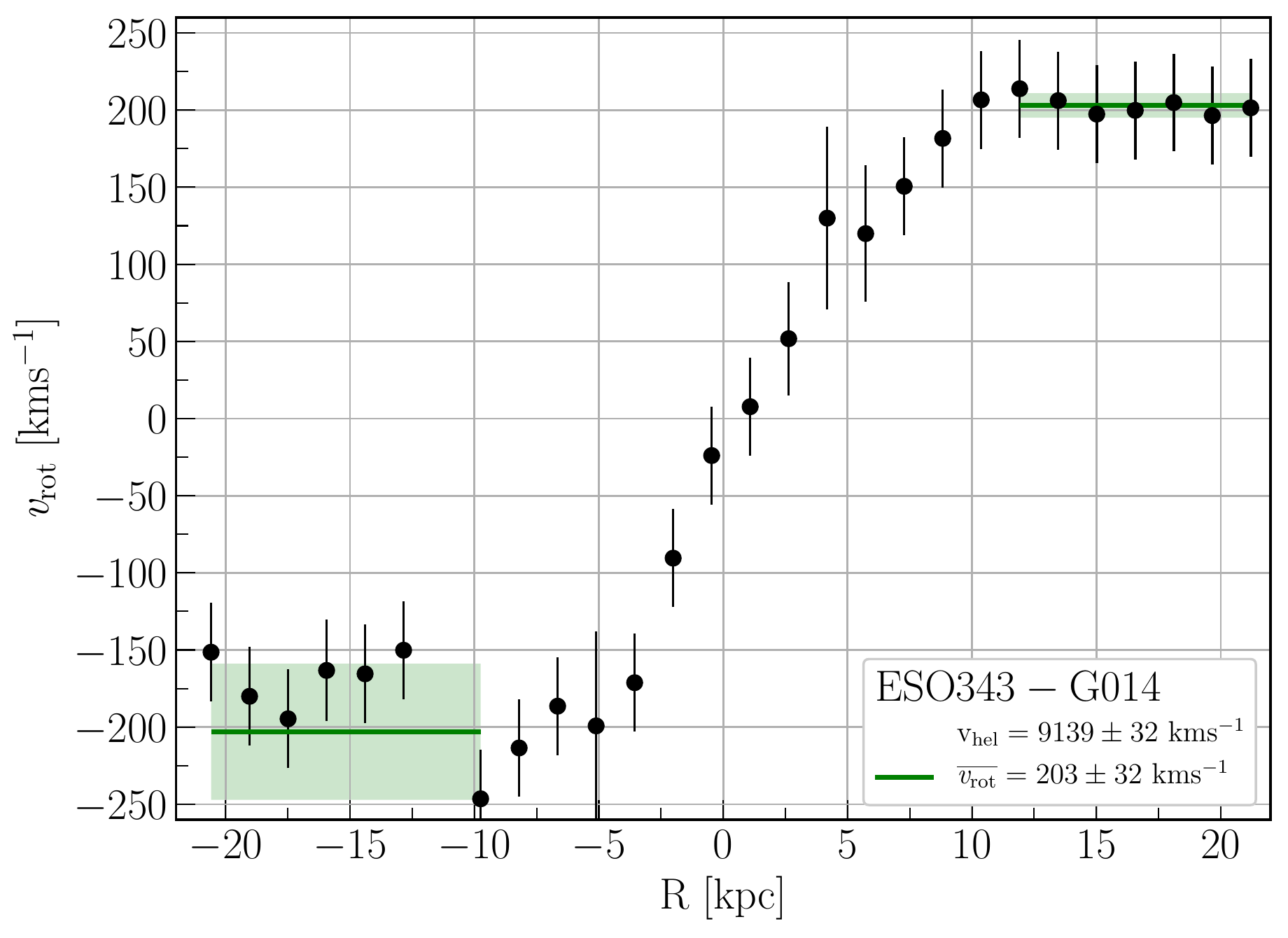}}{\label{rotationcurve_ESO343-G014}}
  \subfigure[]{\includegraphics[width=0.49\linewidth]{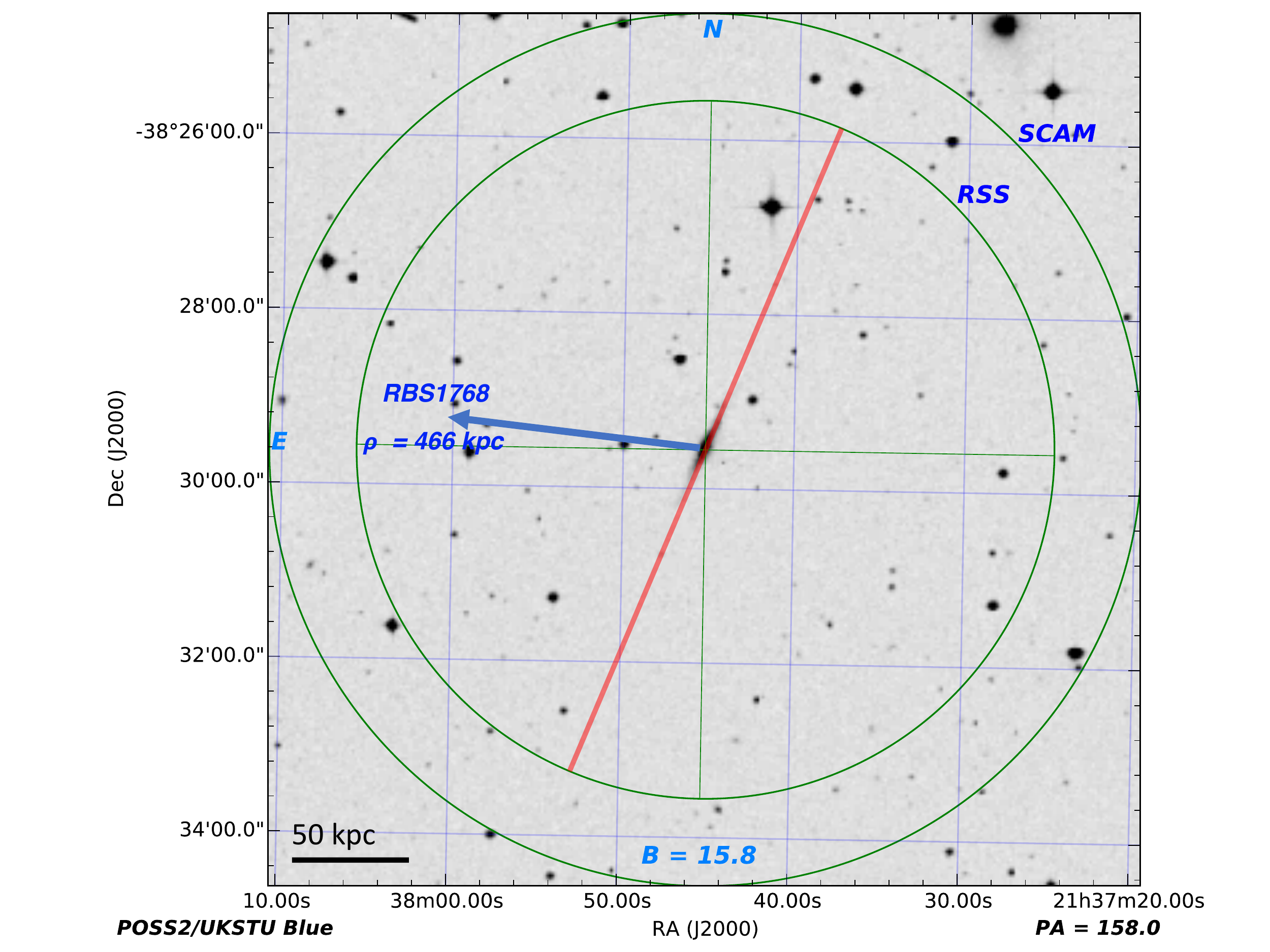}\label{finderchart_ESO343-G014}}
  \caption{\small{a) Rotation curve of ESO343-G014. The solid green line indicates the weighted mean velocity over the corresponding x-axis region, and the shaded green indicates the 1$\sigma$ error in the mean. b) SALT finder chart for ESO343-G014 showing the position of the slit in red.}}
\vspace{0pt}
\end{figure}

\subsection{IC5325}
IC5325 is a mostly face-on ($i = 25^{\circ}$) SAB(rs)bc type galaxy with a measured systemic velocity $v_{\rm sys} = 1512 \pm 8$ \kms. It's inclination is just high enough to obtain a reasonable rotation curve. The background QSO RBS2000 is located northeast at $\rho = 314$ kpc and  $67^{\circ}$ azimuth angle on the approaching side of IC5325. We detect Ly$\alpha$ at $v_{\rm Ly\alpha} = 1596$ \kms~towards RBS2000.



\begin{figure}[ht]
\centering
  \subfigure[]{\includegraphics[width=0.5\linewidth]{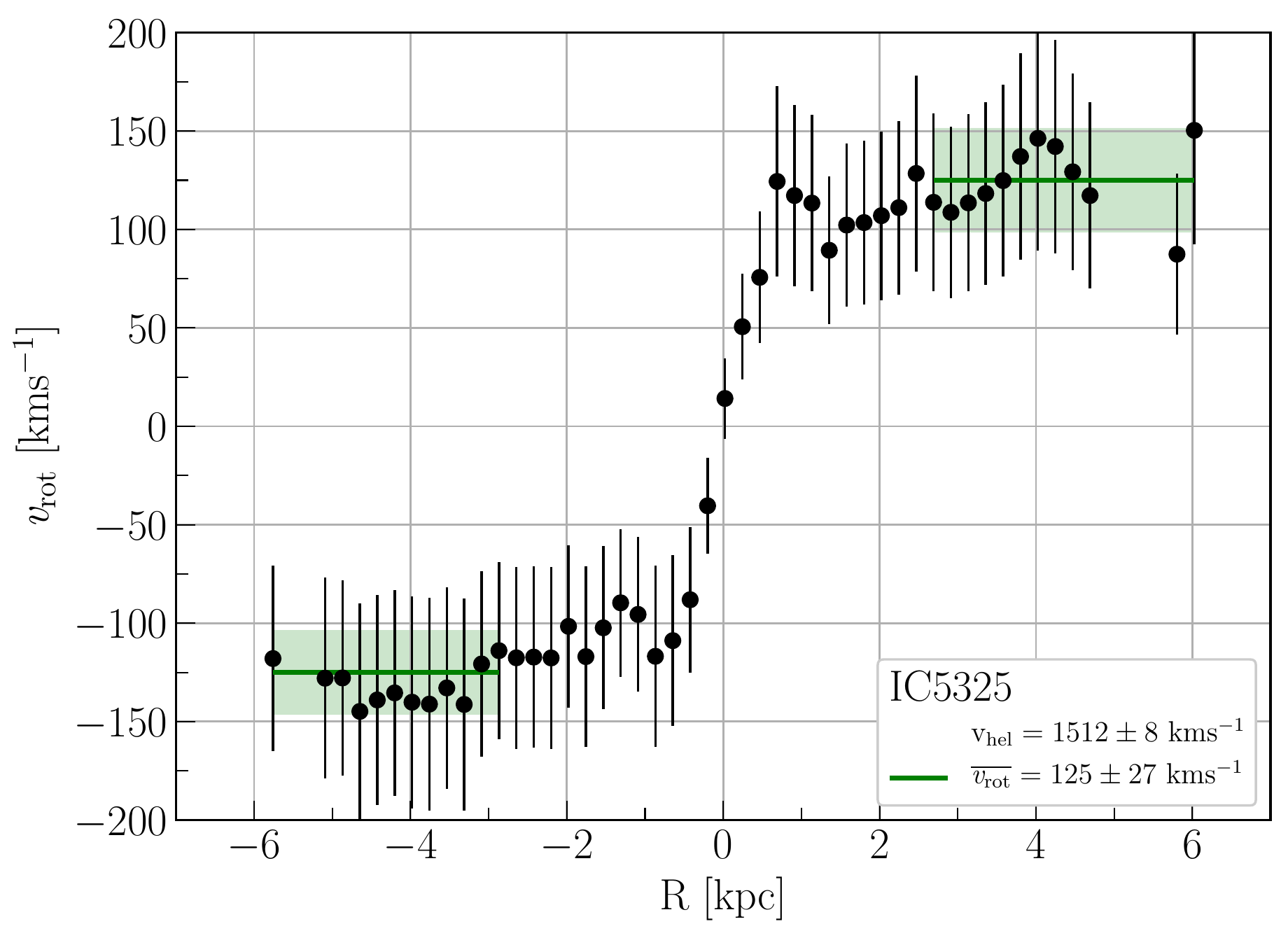}}{\label{rotationcurve_IC5325}}
  \subfigure[]{\includegraphics[width=0.49\linewidth]{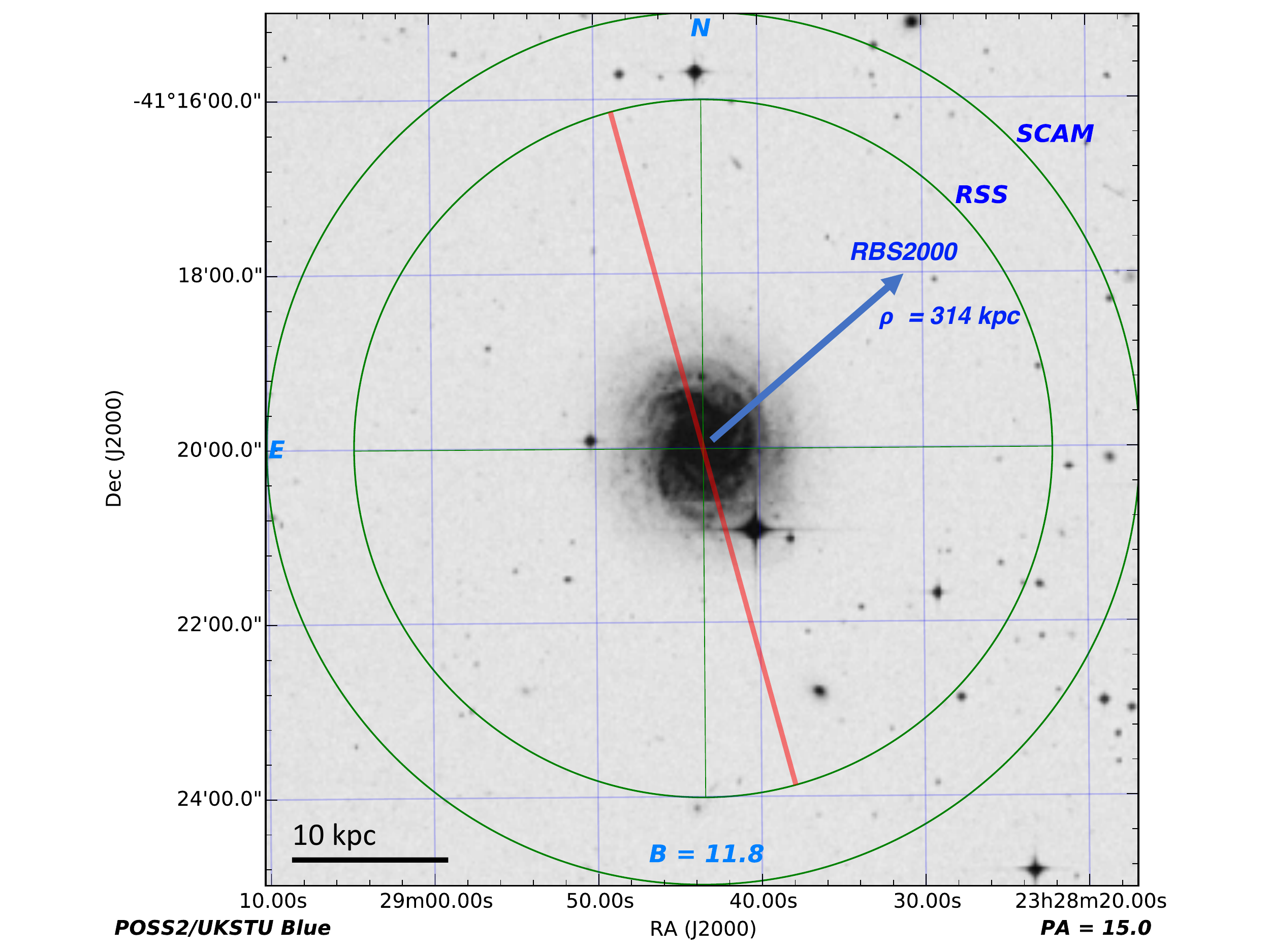}\label{finderchart_IC5325}}
  \caption{\small{a) Rotation curve of IC5325. The solid green line indicates the weighted mean velocity over the corresponding x-axis region, and the shaded green indicates the 1$\sigma$ error in the mean. b) SALT finder chart for IC5325 showing the position of the slit in red.}}
\vspace{0pt}
\end{figure}

\subsection{MCG-03-58-009}
MCG-03-58-009 is a massive and very isolated Sc type galaxy at a measured systemic velocity of $v_{\rm sys} = 9015 \pm 19$ \kms~and inclination angle of $i = 61^{\circ}$. The background QSO MRC2251-178 is located southeast at $\rho = 355$ kpc at an azimuth angle of $74^{\circ}$ on the receding side. We detect a weak Ly$\rm \alpha$ absorber at $v_{\rm Ly\alpha} = 9029$ \kms~ towards MRC2251-178. 


\begin{figure}[ht]
\centering
  \subfigure[]{\includegraphics[width=0.5\linewidth]{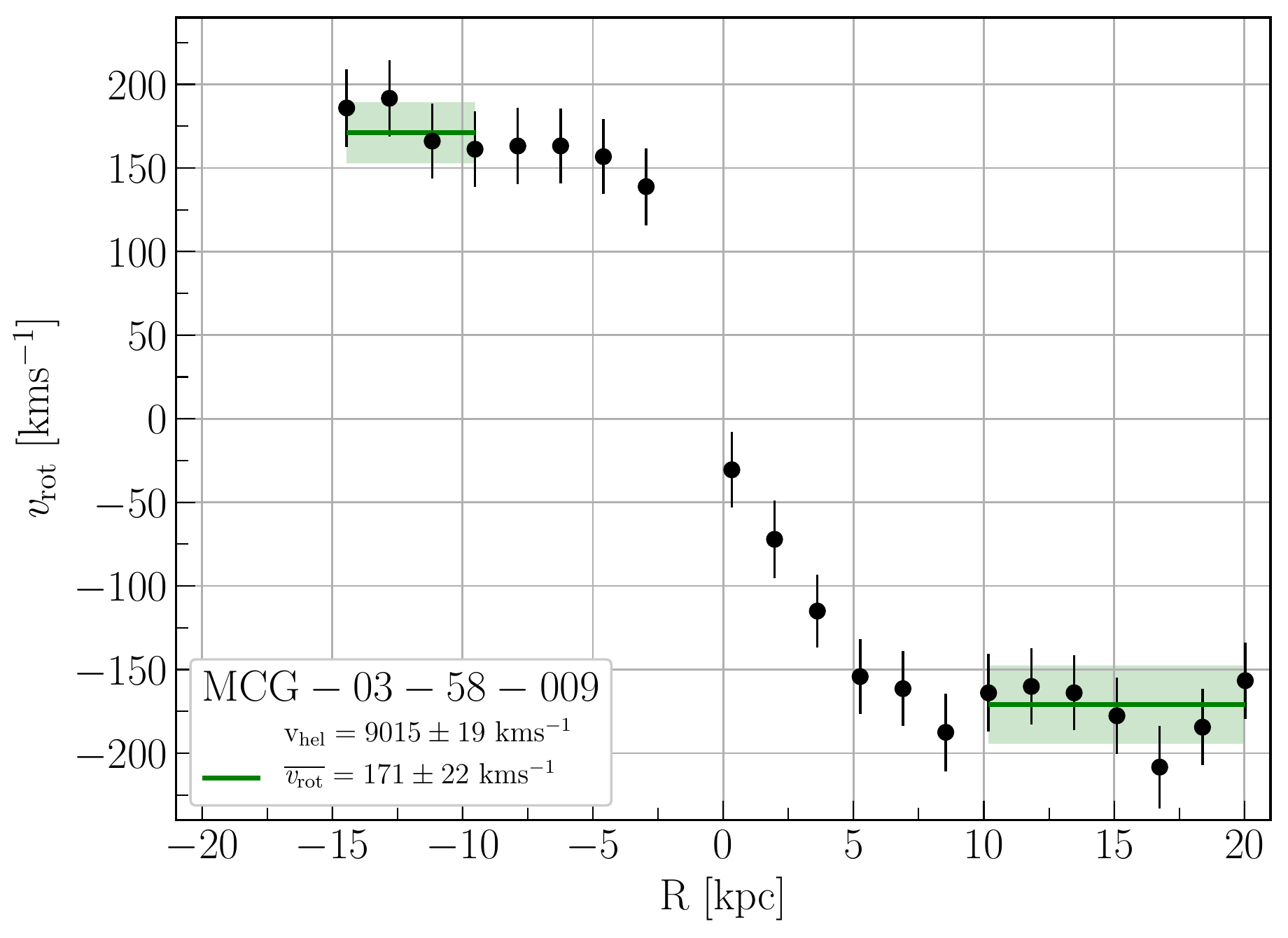}}{\label{rotationcurve_MCG-03-58-009}}
  \subfigure[]{\includegraphics[width=0.49\linewidth]{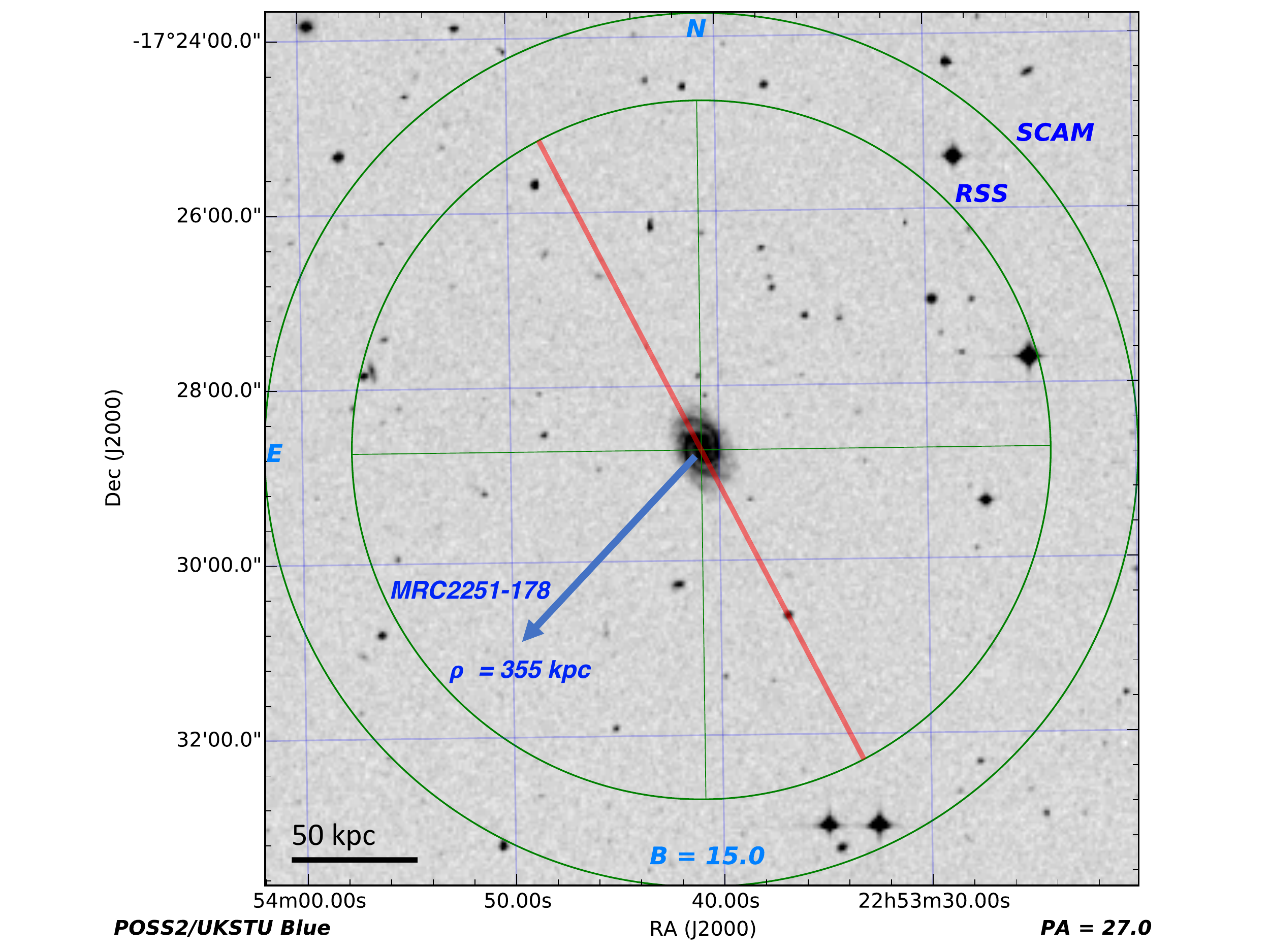}\label{finderchart_MCG-03-58-009}}
  \caption{\small{a) Rotation curve of MCG-03-58-009. The solid green line indicates the weighted mean velocity over the corresponding x-axis region, and the shaded green indicates the 1$\sigma$ error in the mean. b) SALT finder chart for MCG-03-58-009 showing the position of the slit in red.}}
\vspace{0pt}
\end{figure}

\subsection{NGC3633}
NGC3633 is an isolated, edge-on ($i = 72^{\circ}$) SAa type galaxy with a measured systemic velocity $v_{\rm sys} = 2587 \pm 7$ \kms. Several locations along the disk of NGC3633 show two velocities for emission. We have combined these into a single velocity measurement via a weighted average. 

The background QSO RX\_J1121.2+0326 is located southeast at $\rho = 184$ kpc and $55^{\circ}$ azimuth on the approaching side of NGC3633. We detect Ly$\rm \alpha$ at $v_{\rm Ly\alpha} = 2608$ \kms~toward RX\_J1121.2+0326.


\begin{figure}[ht]
\centering
  \subfigure[]{\includegraphics[width=0.5\linewidth]{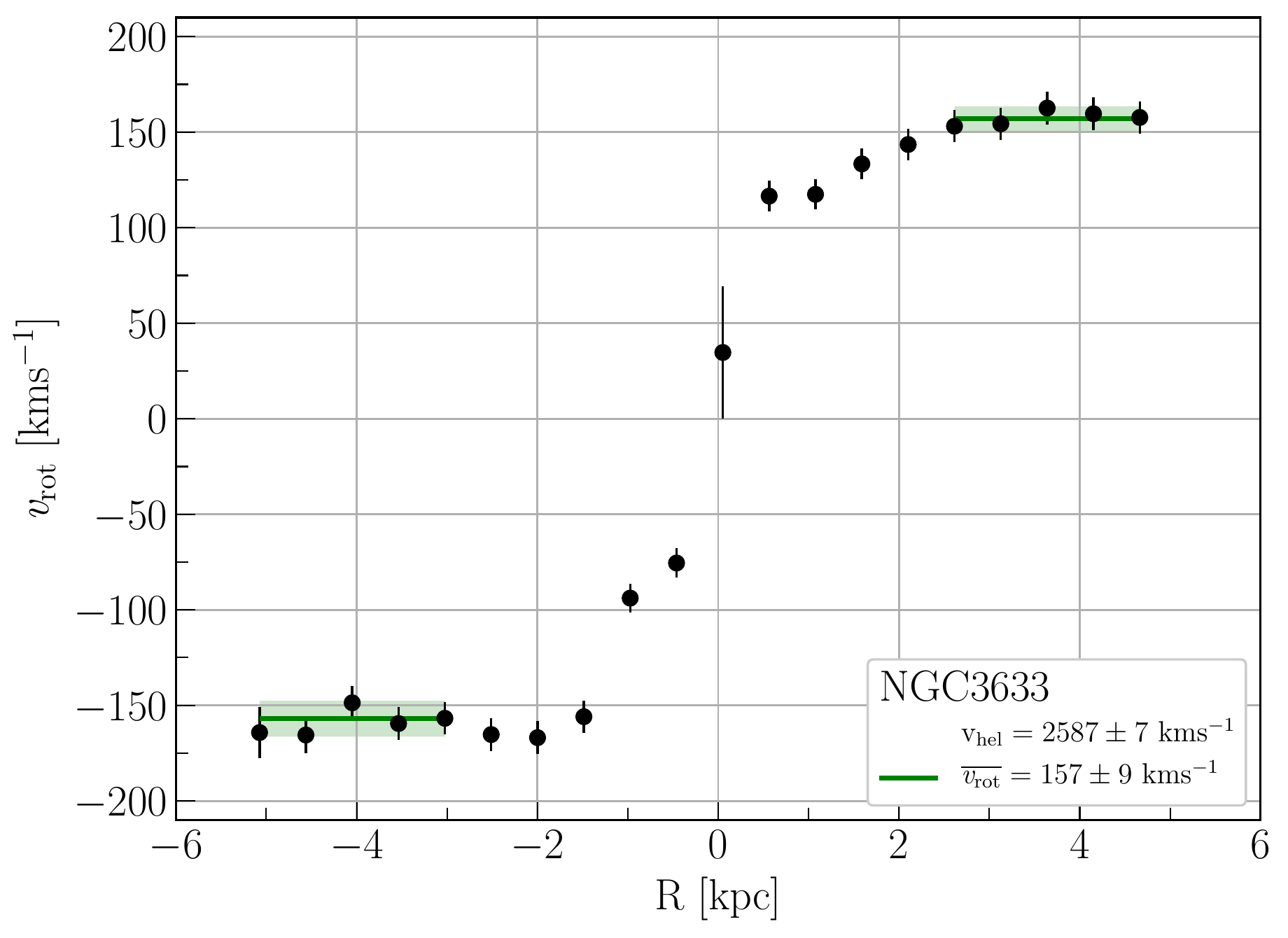}}{\label{rotationcurve_NGC3633}}
  \subfigure[]{\includegraphics[width=0.49\linewidth]{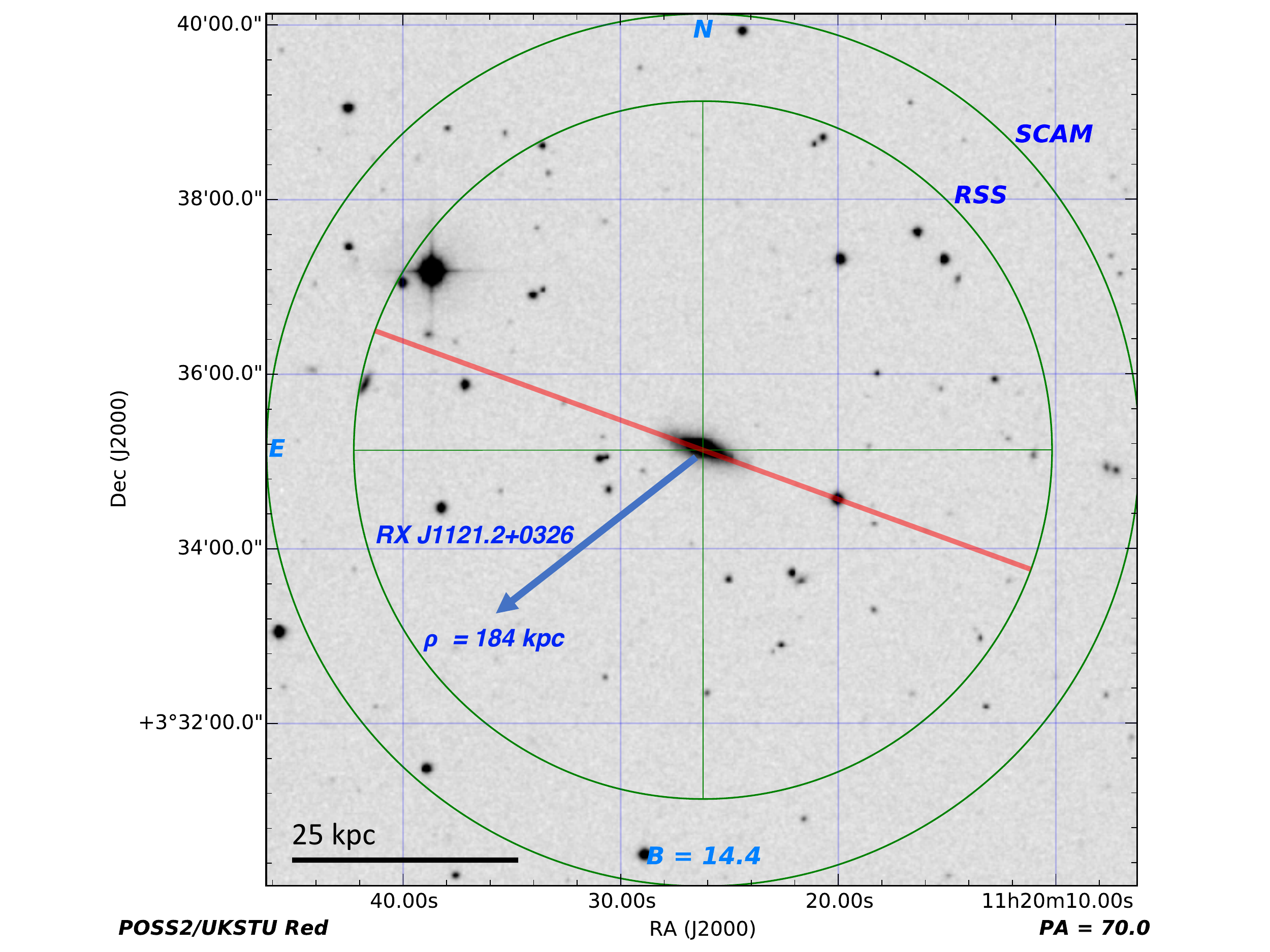}\label{finderchart_NGC3633}}
  \caption{\small{a) Rotation curve of NGC3633. The solid green line indicates the weighted mean velocity over the corresponding x-axis region, and the shaded green indicates the 1$\sigma$ error in the mean. b) SALT finder chart for NGC3633 showing the position of the slit in red.}}
\vspace{0pt}
\end{figure}

\subsection{NGC4939}
NGC4939 is a large SA(s)bc type galaxy with measured systemic velocity $v_{\rm sys} = 3093 \pm 33$ \kms~and inclination $i = 61^{\circ}$. The background QSO PG1302-102 is located southeast at $\rho = 254$ kpc and $64^{\circ}$ azimuth angle on the approaching side of NGC4939. We detect a Ly$\rm \alpha$ absorber at $v_{\rm Ly\alpha} = 3449$ \kms~towards PG1302-102.


\begin{figure}[ht!]
\centering
  \subfigure[]{\includegraphics[width=0.5\linewidth]{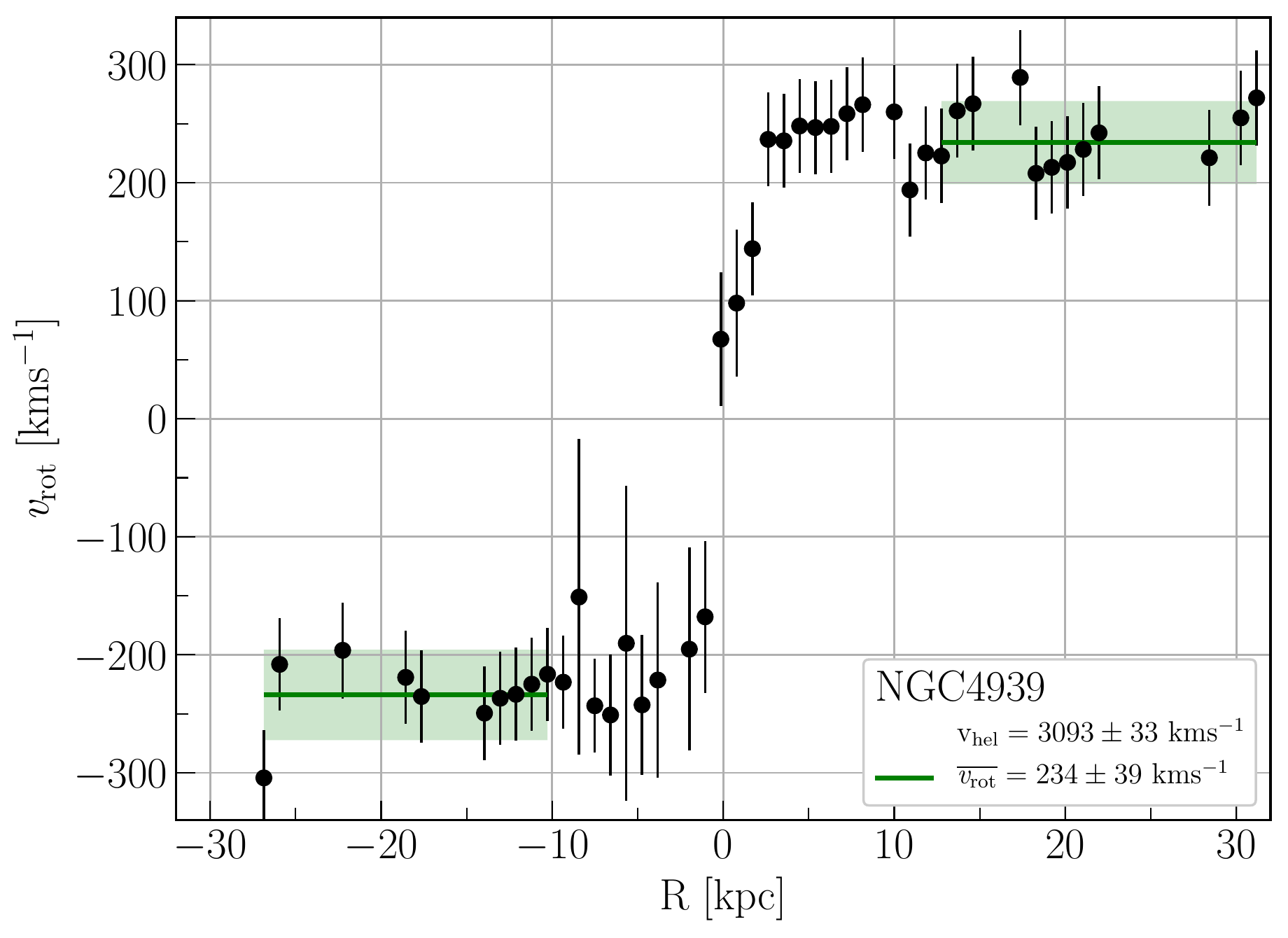}}{\label{rotationcurve_NGC4939}}
  \subfigure[]{\includegraphics[width=0.49\linewidth]{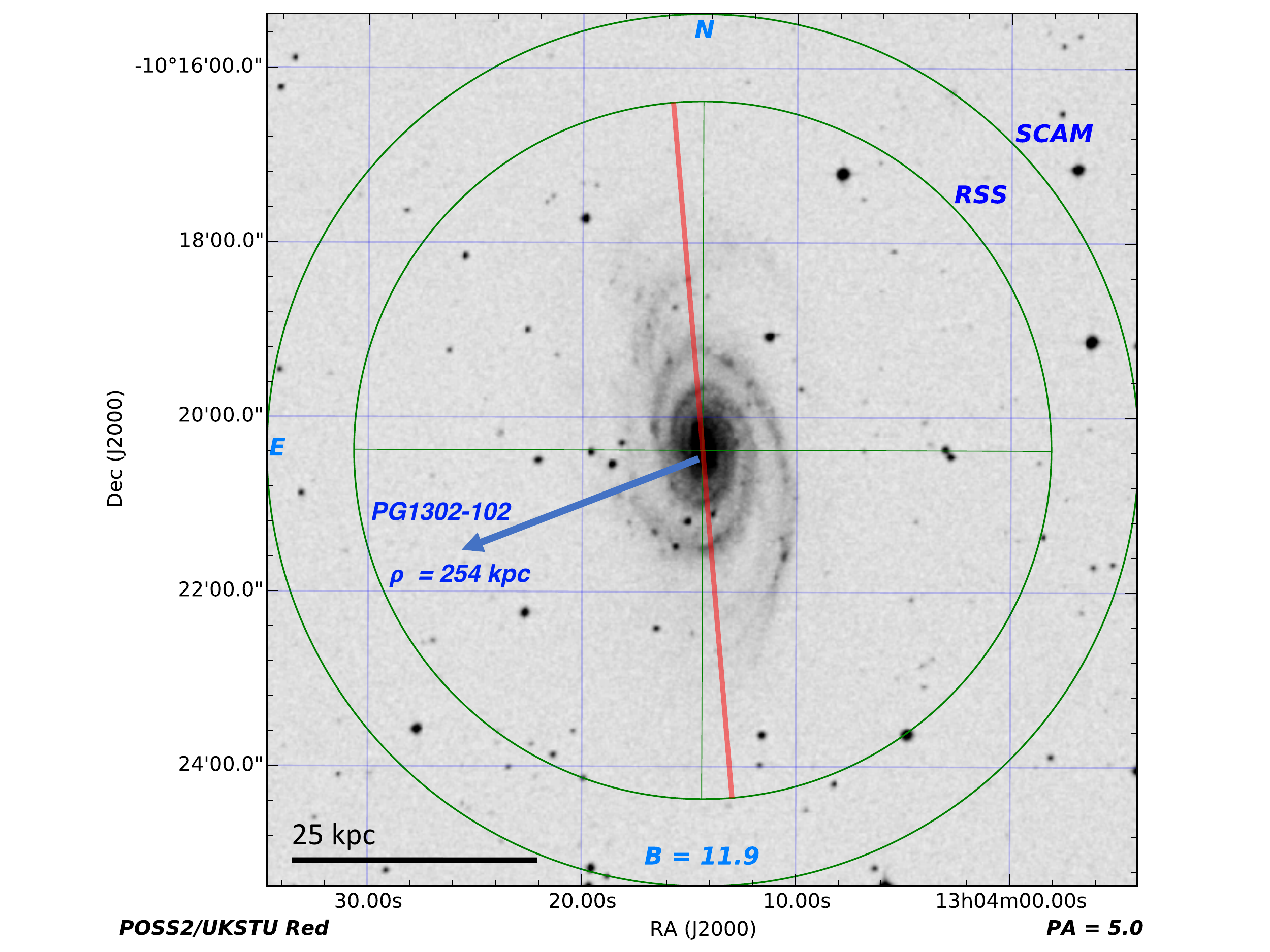}\label{finderchart_NGC4939}}
  \caption{\small{a) Rotation curve of NGC4939. The solid green line indicates the weighted mean velocity over the corresponding x-axis region, and the shaded green indicates the 1$\sigma$ error in the mean. b) SALT finder chart for NGC4939 showing the position of the slit in red.}}
\vspace{0pt}
\end{figure}

\subsection{NGC5786}
NGC5786 is a large, strongly-barred spiral galaxy with measured systemic velocity $v_{\rm sys} = 2975 \pm 22$ \kms~and inclination $i = 65^{\circ}$. The background QSO QSO1500-4140 is located directly east at $\rho = 453$ kpc and $2^{\circ}$ azimuth angle on the receding side of NGC5786. We detect Ly$\rm \alpha$ at $v_{\rm Ly\alpha} = 3138$ \kms~ toward QSO1500-4140.


\begin{figure}[ht]
\centering
  \subfigure[]{\includegraphics[width=0.5\linewidth]{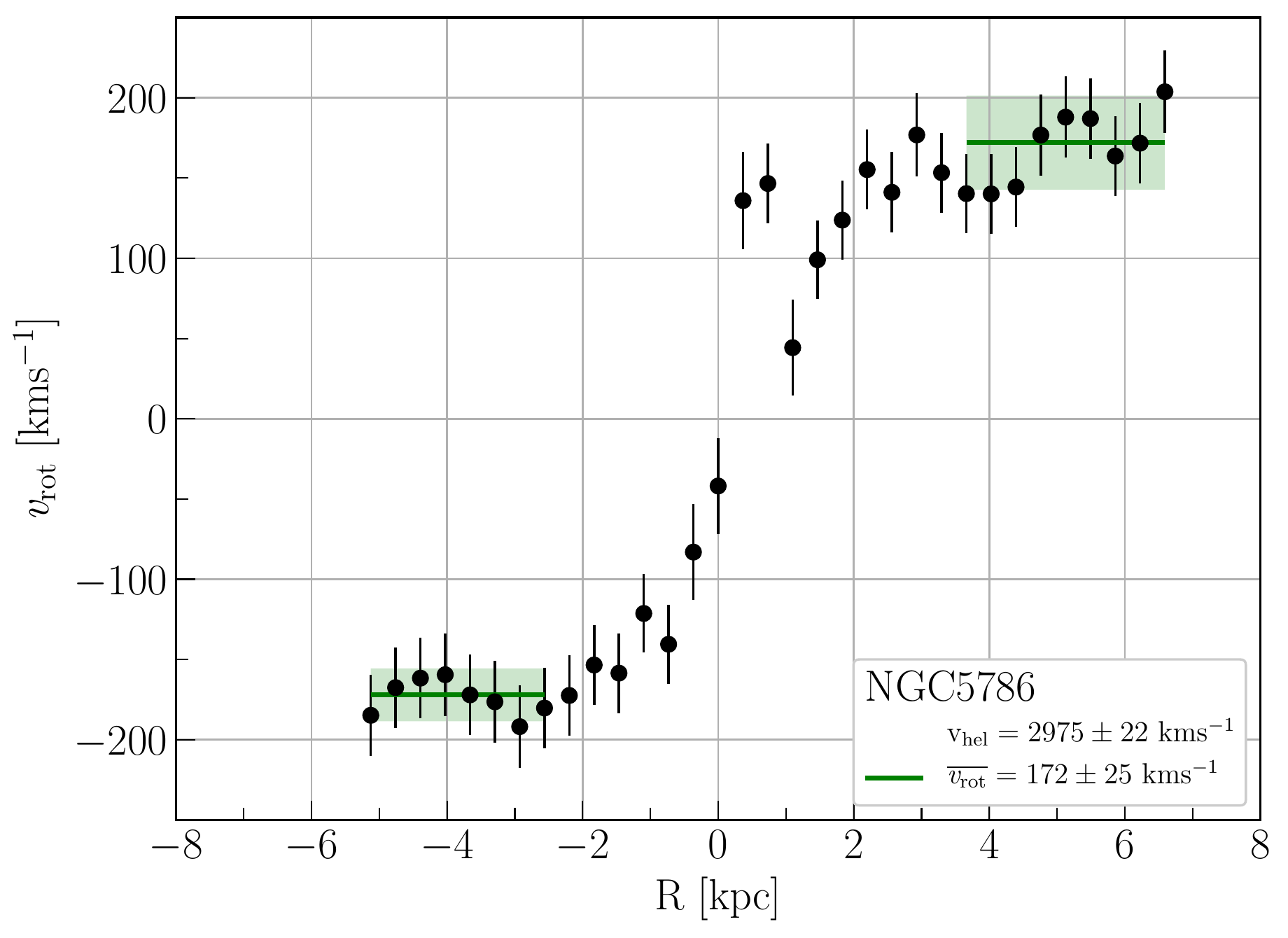}}{\label{rotationcurve_NGC5786}}
  \subfigure[]{\includegraphics[width=0.49\linewidth]{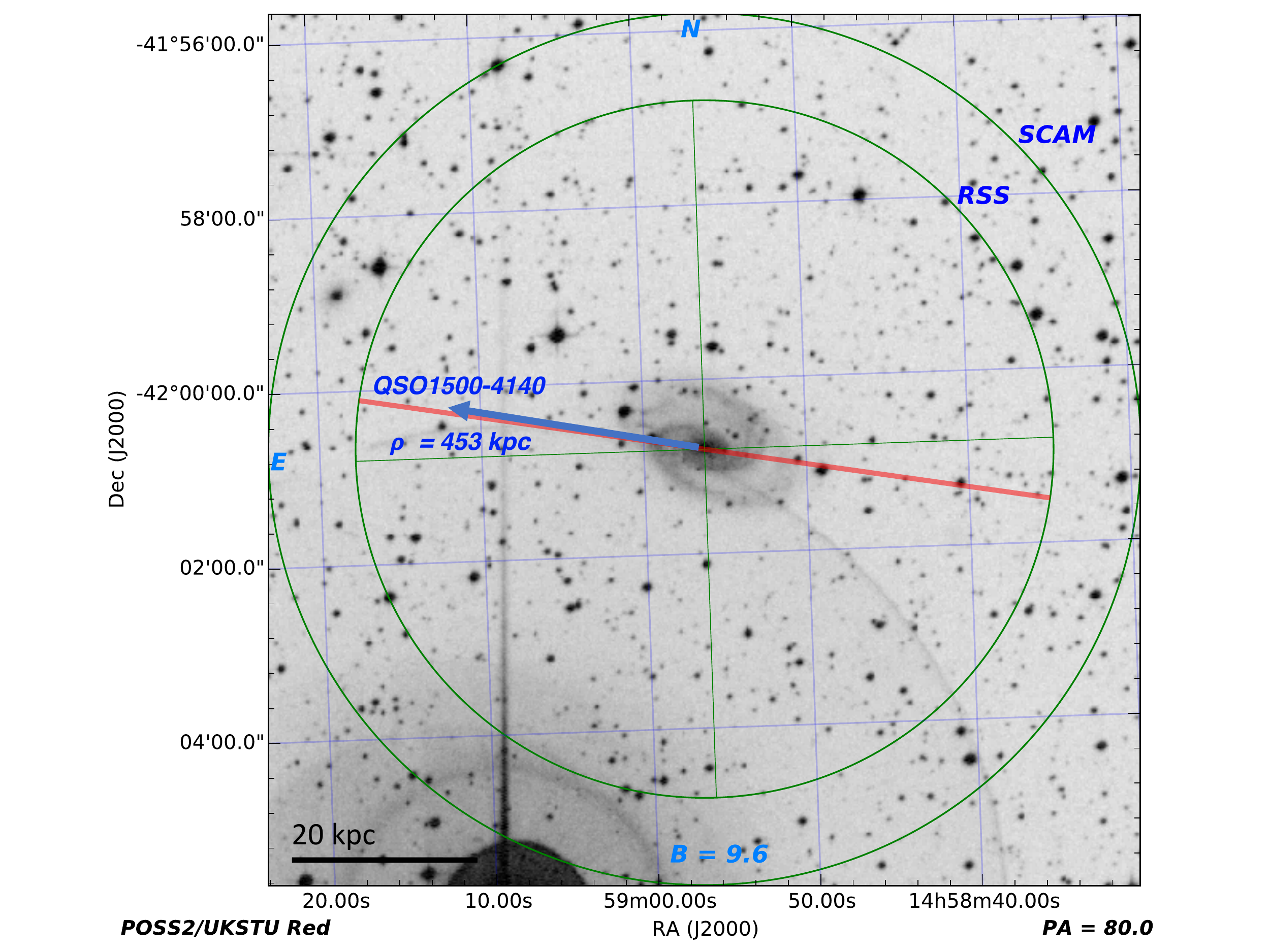}\label{finderchart_NGC5786}}
  \caption{\small{a) Rotation curve of NGC5786. The solid green line indicates the weighted mean velocity over the corresponding x-axis region, and the shaded green indicates the 1$\sigma$ error in the mean. b) SALT finder chart for NGC5786 showing the position of the slit in red.}}
\vspace{0pt}
\end{figure}

\subsection{UGC09760}
UGC09760 is an edge-on ($i = 90^{\circ}$), slow-rotating Sd galaxy with measured systemic velocity $v_{\rm sys} =  2094 \pm 16$ \kms. This systemic velocity deviates slightly from other published redshifts, such as the The Updated Zwicky Catalog value of $v_{\rm sys} = 2023 \pm 2$ \kms~\citep{falco1999}. This is likely due to our method of imposing rotation symmetry and averaging the approaching and receding velocities to derive $v_{\rm sys}$. If we do not sample the rotation curve far enough out, a systematic offset is not unreasonable. Indeed, we do not detect the rotation curve turnover or flattening point.

The background QSO SDSSJ151237.15+012846.0 is located southeast at $\rho = 123$ kpc and $87^{\circ}$ azimuth angle. We detect Ly$\rm \alpha$ absorption at $v_{\rm Ly\alpha} = 2025$ \kms~toward SDSSJ151237.15+012846.0. It is worth noting that there are several small satellite galaxies nearby, including SDSSJ151208.16+013508.5, SDSSJ151121.63+013637.6, SDSSJ151241.38+013723.7 and UGC09746 (impact parameters $\rho = 53, 88, 82, 230$ kpc respectively). All of these galaxies lie slightly blue-ward of UGC09760, and thus \emph{farther} away in velocity from the Ly$\rm \alpha$ absorber at 2025 \kms.


\begin{figure}[ht]
\centering
  \subfigure[]{\includegraphics[width=0.5\linewidth]{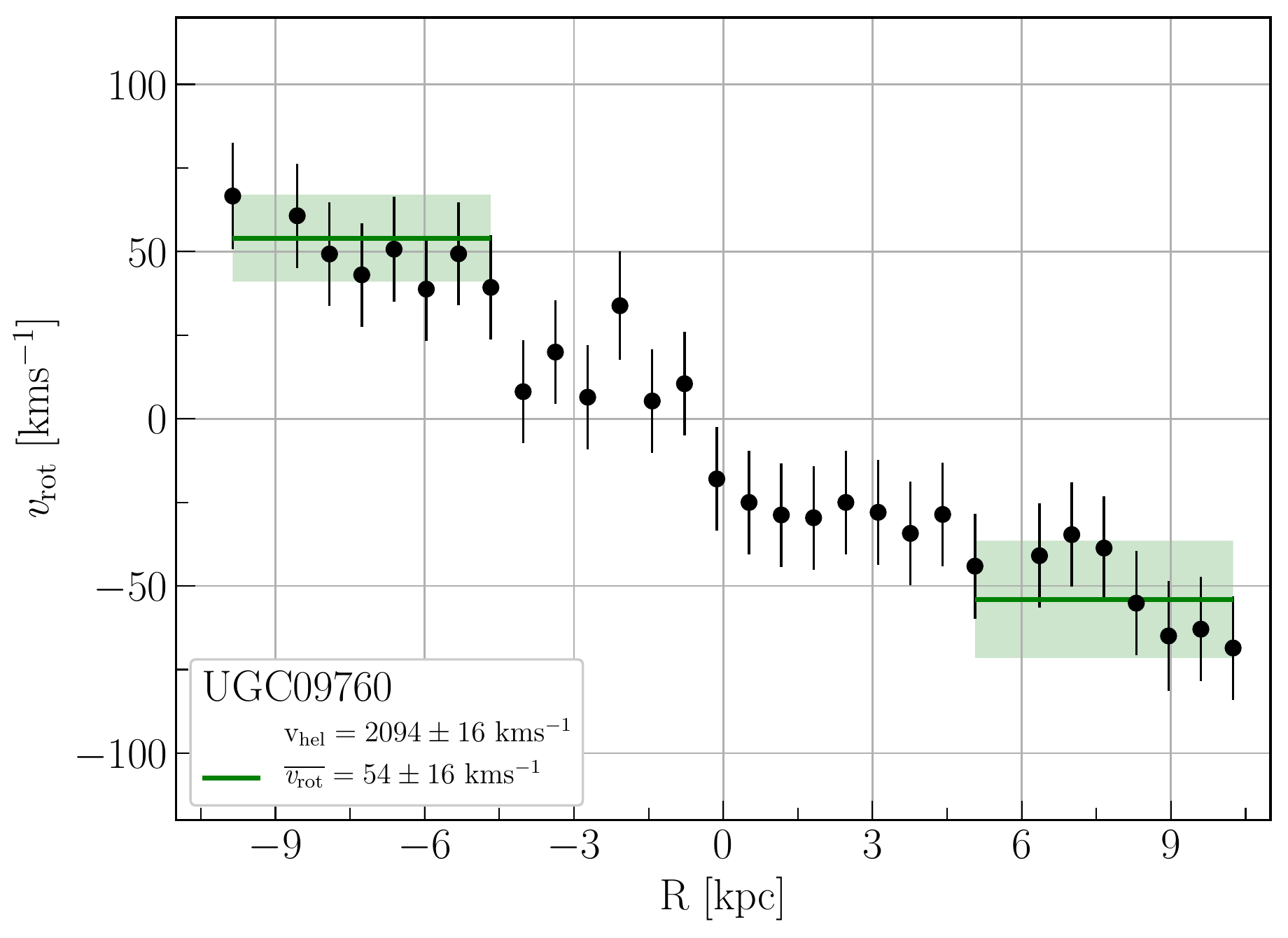}}{\label{rotationcurve_UGC09760}}
  \subfigure[]{\includegraphics[width=0.49\linewidth]{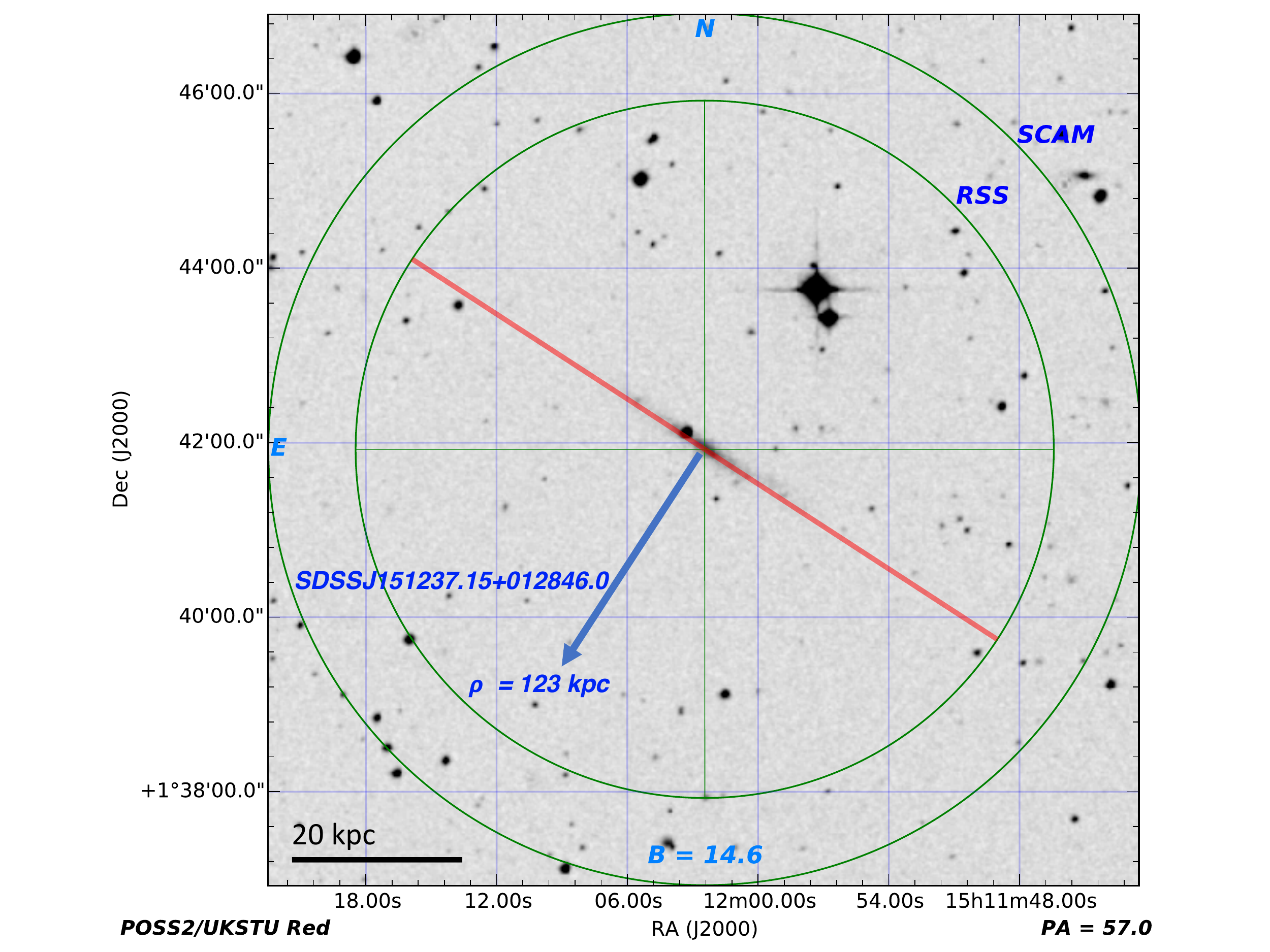}\label{finderchart_UGC09760}}
  \caption{\small{a) Rotation curve of UGC09760. The solid green line indicates the weighted mean velocity over the corresponding x-axis region, and the shaded green indicates the 1$\sigma$ error in the mean. b) SALT finder chart for UGC09760 showing the position of the slit in red.}}
\vspace{0pt}
\end{figure}

\section{Ancillary Data} \label{ancillary_data}
To increase our sample size we have also searched the literature for galaxies with published rotation curves and orientations. Unfortunately, while the rotation velocity is available for thousands of galaxies, only a handful of publications also include the \emph{orientation} of the rotation on the sky. Of these, we were able to find 16 additional galaxies which have a systemic velocity greater than $\sim 500$ \kms, and are near to a COS sightline with available data. We have included 3 of the galaxy-QSO systems analyzed by \cite{cote2005}. We briefly summarize each of these systems here, and refer the reader to \cite{cote2005} for a more complete discussion. As new spectra and redshift-independent distances are available for these systems our results, while similar, are not identical.

\subsection{NGC2770}
NGC2770 is a large, edge-on ($i = 80^{\circ}$) Sc type galaxy with systemic velocity $v_{\rm sys} = 1948 \pm 2$ \kms. It is mostly isolated except for two nearby small dwarfs MCG+06-20-036NED02 and GALEXASCJ090946.88+330840.4 (both 25 kpc away, on opposite sides of NGC2770). We take the rotation curve and orientation information produced by \cite{rhee1996}.  There are five nearby QSOs, which we present in order of increasing impact parameter. 

First, the QSO FBQSJ0908+3246 is located south at $\rho = 204$ kpc and $56^{\circ}$ azimuth angle on the approaching side of NGC2770. We detect Ly$\rm \alpha$ at $v_{\rm Ly\alpha} = 1910, 1991$ \kms~toward FBQSJ0908+3246.


Second, the QSO TON1015 is located northeast at $\rho = 218$ kpc and $58^{\circ}$ azimuth angle on the receding side of NGC2770. We detect Ly$\rm \alpha$ at $v_{\rm Ly\alpha} = 1834, 1984$ \kms~toward TON1015.


Third the QSO SDSSJ091127.30+325337.0 is located southeast at $\rho = 234$ kpc and $33^{\circ}$ azimuth angle on the approaching side of NGC2770. We detect Ly$\rm \alpha$ at $v_{\rm Ly\alpha} = 2062$ \kms~toward SDSSJ091127.30+325337.0.


Fourth, the QSO SDSSJ091052.80+333008.0 at is located northeast at $\rho = 239$ kpc and $63^{\circ}$ azimuth angle on the receding side of NGC2770. We detect Ly$\rm \alpha$ at $v_{\rm sys} = 1824, 1969$ \kms~toward SDSSJ091052.80+333008.0.


Finally, the QSO TON1009 is located south at $\rho = 267$ kpc and $38^{\circ}$ azimuth angle on the approaching side of NGC2770. We detect Ly$\rm \alpha$ at $v_{\rm sys} = 1913, 1980$ \kms~toward TON1009.



\subsection{NGC3067}
NGC3067 is a mostly edge-on ($i = 71^{\circ}$) SAB(s)ab type galaxy with systemic velocity $v_{\rm sys} = 1465 \pm 5$ \kms. This galaxy and the nearby QSO sightline toward 3C232 is a particularly well studied system. They are separated by only $\rho = 11$ kpc ($71^{\circ}$ azimuth angle on the northwest, receding side) and a Lyman Limit System (LLS) with column density $N_{\scriptsize \HI} = 1 \times 10^{20}$ $\rm cm^{-2}$ is detected toward 3C232 at $v_{\rm Ly\alpha} = 1417$ \kms, which has been postulated as a high velocity cloud (HVC) orbiting NGC3067 \citep{carilli1989, keeney2005}.

We obtained the rotation curve for NGC3067 from \cite{rubin1982} and the orientation from \cite{carilli1989}. We fit a single component to the Lyman Limit System with $v_{\rm Ly\alpha} = 1417$ \kms $\rm log \emph{N}_{\scriptsize \HI} = 20.086$ $\rm cm^{-2}$. Some studies suggest this system could have 2 or more components (e.g., \citealt{keeney2005} and \citealt{stocke2010}), but we do not see strong evidence for multiple components here.


A second QSO SDSSJ095914.80+320357.0 is located farther away, to the southeast at $\rho = 128$ kpc and $40^{\circ}$ azimuth angle on the receding side of NGC3067. We detect Ly$\rm \alpha$ at v$_{\rm Ly\alpha} = 1492$ \kms~toward SDSSJ095914.80+320357.0.


\subsection{NGC3198}
NGC3198 is a SB(rs)c type galaxy with systemic velocity $v_{\rm sys} = 660 \pm 1$ \kms and inclination = $i = 73^{\circ}$. It is a well studied galaxy, and is included in the detailed THINGS rotation curve study of \cite{deblok2008}. NGC3198 has an even and flat rotation curve, with an average velocity of $v_{\rm rot} = 152$ \kms. The background QSO RX\_1017.5+4702 is located northeast at $\rho = 370$ kpc and $58^{\circ}$ azimuth angle on the approaching side of NGC3198. We detect Ly$\rm \alpha$ toward RX\_1017.5+4702 at $v_{\rm Ly\alpha} = 623$. We note that the small dwarf galaxy SDSSJ101848.77+452137.0 is located 65 kpc away from NGC3198 toward the southwest.



\subsection{NGC3351}
NGC3351 is a mostly face-on ($i = 42^{\circ}$) SB(r)b type galaxy with systemic velocity  $v_{\rm sys} = 778 \pm 4$ \kms. It is located $\sim200$ kpc southwest of the core of the Leo I group. We take the rotation curve and orientation produced by \cite{dicaire2008}. While we expect any extended disk rotation to be quickly disrupted due to the complex Leo I environment, this galaxy also has one of the closest sightlines in our sample with SDSSJ104335.90+115129.0 at $\rho = 31$ kpc and $46^{\circ}$ azimuth on the northwest, approaching side. We detect Ly$\rm \alpha$ at $v_{\rm Ly\alpha} = 699, 862, 1036$ \kms~toward SDSSJ104335.90+115129.0. We note also that there are multiple metal ions associated with the $v_{\rm Ly\alpha} = 699$ \kms~line, including C\,{\sc ii}, N\,{\sc i}, N\,{\sc v}, O\,{\sc i}, Si\,{\sc ii}, Si\,{\sc iii}, Si\,{\sc iv}, S\,{\sc ii}, and Fe\,{\sc ii}.


\subsection{NGC3432}
NGC3432 is an edge-on ($i = 90^{\circ}$) SB(s)m type galaxy with systemic velocity $v_{\rm sys} = 616 \pm 4$ \kms. It is interacting with the nearby dwarf galaxy UGC05983 located 11 kpc away and at $v_{\rm sys} = 765$ \kms. We take a rotation curve and orientation for NGC3432 from \cite{rhee1996}. The QSO CSO295 is located just 20 kpc away and just to the receding side of the minor axis ($79^{\circ}$ azimuth angle). This is the second closest pair in our sample, after the 11 kpc separated NGC3067-3C232 system. We detect Ly$\rm \alpha$ at $v_{\rm Ly\alpha} = 600, 662$ \kms~toward CSO295. We also detect C\,{\sc ii}, Si\,{\sc ii}, Si\,{\sc iii}, and Si\,{\sc iv} associated with this absorption system.


A second QSO RX\_J1054.2+3511 is located south at $\rho = 290$ kpc and $60^{\circ}$ azimuth angle on the receding side of NGC3432. We detect Ly$\rm \alpha$ at $v_{\rm sys} = 689$ \kms~toward RX\_J1054.2+3511. 

	

\subsection{NGC3631}
NGC3631 is a mostly face-on ($i = 17^{\circ}$) SA(s)c type galaxy with systemic velocity $v_{\rm sys} = 1156 \pm 1$ \kms. We take the rotation curve and orientation information produced by \cite{knapen1997}. There are 4 nearby QSOs, which we will present in order of increasing impact parameter.

First, the closest background QSO RX\_J1117.6+5301 is located southwest at $\rho = 78$ kpc and $78^{\circ}$ azimuth angle on the receding side of NGC3631. We detect Ly$\rm \alpha$ at $v_{\rm Ly\alpha} = 1130, 1265$ \kms~toward RX\_J1117.6+5301.


Second, background QSO SDSSJ112448.30+531818.0 is located northeast at $\rho = 86$ kpc and $77^{\circ}$ azimuth angle on the approaching side of NGC3631. We detect Ly$\rm \alpha$ at $v_{\rm Ly\alpha} = 1018, 1135$ \kms~ toward SDSSJ112448.30+531818.0.


Third, the background QSO SDSSJ111443.70+525834.0 is located in the same direction but farther than RX\_J1117.6+5301, at $\rho = 145$ kpc and $74^{\circ}$ azimuth angle on the receding side of NGC3631. We detect Ly$\rm \alpha$ at $v_{\rm Ly\alpha} = 1158$ \kms~toward SDSSJ111443.70+525834.0.


Finally, the background QSO SBS1116+523 is located south at $\rho = 163$ kpc and $37^{\circ}$ azimuth angle on the approaching side of NGC3631, but we do not detect any Ly$\rm \alpha$ within $\pm400$ \kms~of NGC3631.

\subsection{NGC3666}
NGC3666 is a mostly isolated and edge-on ($i = 78^{\circ}$) SA(rs)c type galaxy with systemic velocity $v_{\rm sys}=1060 \pm 1$ \kms. We take the rotation curve and orientation information produced by \cite{rhee1996}. The QSO SDSSJ112439.50+113117.0 is located north at $\rho = 58$ kpc and $86^{\circ}$ azimuth angle on the approaching side of NGC3666. We detect Ly$\rm \alpha$ at $v_{\rm sys} = 1062$ \kms~toward SDSSJ112439.50+113117.0.

\subsection{NGC4529} \label{NGC4529}
NGC4529 is an edge-on ($i = 80^{\circ}$) and isolated Scd type galaxy with systemic velocity $v_{\rm sys} = 2536 \pm 11$ \kms. We take the rotation curve and orientation information produced by \cite{cote2005}. The QSO MRK771 is located west at $\rho = 158$ kpc and $26^{\circ}$ azimuth angle on the approaching side of NGC4529. We detect Ly$\rm \alpha$ at $v_{\rm sys} = 2558, 2513$ \kms~toward MRK771.


\subsection{NGC4565}
NGC4565 is an edge-on ($i = 86^{\circ}$) SA(s)b type galaxy with systemic velocity $v_{\rm sys} = 1230 \pm 5$ \kms. We take the rotation curve and orientation produced by \cite{sofue1996}. The background QSO RX\_J1236.0+2641 is located directly north at $\rho = 147$ kpc and $38^{\circ}$ azimuth angle on receding side of NGC4565. We detect Ly$\rm \alpha$ absorption at $v_{\rm Ly\alpha} = 1166, 1257$ \kms~toward RX\_J1236.0+2641. 


\subsection{NGC5907}
NGC5907 is a large, edge-on ($i = 90^{\circ}$) SA(s)c type galaxy with systemic velocity $v_{\rm sys} = 667 \pm 3$ \kms. We take the rotation curve and orientation produced by \cite{yim2014}. The background QSO RBS1503 is located southeast at $\rho = 478$ kpc and $66^{\circ}$ azimuth angle on the receding side of NGC5907. We do not detect Ly$\rm \alpha$ toward RBS1503 within $\pm 400$ \kms~of NGC5907.


\subsection{NGC5951}
NGC5951 is a large, edge-on ($i = 74^{\circ}$) SBc type galaxy with systemic velocity $v_{\rm sys} = 1780 \pm 1$ \kms. We take the rotation curve and orientation for NGC5951 from \cite{rhee1996}. The QSO 2E1530+1511 is located east at $\rho = 55$ kpc and $88^{\circ}$ azimuth angle on the receding side of NGC5951. We detect Ly$\rm \alpha$ at $v_{\rm Ly\alpha} = 1792, 1959$ \kms~toward 2E1530+1511. 


\subsection{NGC6140} \label{NGC6140}
NGC6140 is a small SB(s)cd type galaxy with systemic velocity $v_{\rm sys} = 910 \pm 4$ \kms~and inclination $i = 49^{\circ}$. We take the rotation curve and orientation information produced by \cite{cote2005}. A background QSO Mrk876 is located northwest at $\rho = 113$ kpc and azimuth angle $18^{\circ}$. We detect Ly$\rm \alpha$ at $v_{\rm Ly\alpha} = 917, 971$ \kms~toward MRK876. Additionally, we detect Ly$\rm \beta$ and O\,{\sc vi} associated with this system  (see \citealt{narayanan2010}).



\subsection{NGC7817}
NGC7817 is an edge-on ($i = 80^{\circ}$) SAbc type galaxy with systemic velocity $v_{\rm sys} = 2309 \pm 4$ \kms. We take the rotation curve and orientation information produced by \cite{rhee1996}. The background QSO MRK335 is located southeast at $\rho = 343$ kpc and almost directly along the minor axis of NGC7817 ($87^{\circ}$ azimuth angle). We detect Ly$\rm \alpha$ at $v_{\rm sys} = 1949, 2273$ \kms~toward MRK335. 


\subsection{UGC04238} \label{UGC04238}
UGC04238 is an isolated, mostly edge-on ($i = 62^{\circ}$) SBd type galaxy with systemic velocity $v_{\rm sys} = 1544 \pm 7$ \kms. We take the rotation curve and orientation information produced by \cite{cote2005}. The background QSO PG0804+761 is located directly south at $\rho = 148$ kpc and $62^{\circ}$ azimuth on the receding side of UGC04238. We detect Ly$\rm \alpha$ at $v_{\rm Ly\alpha} = 1523, 1611$ \kms~toward PG0804+761. 


\subsection{UGC06446} \label{UGC06446}
UGC06446 is a Sd type galaxy with systemic velocity $v_{\rm sys} = 645 \pm 1$ \kms~and inclination $i = 52^{\circ}$ on the far northwest edge of the Ursa Major cluster of galaxies. We take the rotation curve and orientation information produced by \cite{verheijen2001} and  \cite{swaters2009}. The background QSO SDSSJ112448.30+531818.0 is located southwest at $\rho = 143$ kpc and $19^{\circ}$ azimuth angle on the receding side of UGC06446. We detect Ly$\rm \alpha$ at $v_{\rm Ly\alpha} = 661$ \kms~toward SDSSJ112448.30+531818.0.



\subsection{UGC08146} \label{UGC08146}
UGC08146 is an isolated and edge-on ($i = 78^{\circ}$) Sd type galaxy with systemic velocity $v_{\rm sys} = 670 \pm 1$ \kms. This galaxy (and the nearby QSO PG1259+593) are included in the \cite{cote2005} sample also, but we have taken the rotation curve and orientation information from \cite{rhee1996}. The QSO PG1259+593 is located northwest at $\rho = 114$ kpc at $52^{\circ}$ azimuth angle on the receding side of UGC08146. While \cite{cote2005} cite a single Ly$\rm \alpha$ component at $v_{\rm Ly\alpha} = 679$ \kms, we detect two components at $v_{\rm Ly\alpha} = 621, 693$ \kms~in the higher signal-to-noise COS data now available for PG1259+593.


\end{document}